\documentclass[aps,prb,twocolumn,superscriptaddress,10pt]{revtex4-2}

\usepackage{graphicx}
\usepackage{amssymb, amsmath}
\usepackage[utf8]{inputenc}
\usepackage[T1]{fontenc}
\usepackage{enumerate}
\usepackage{hyperref}
\usepackage{braket}
\usepackage{amsmath}
\usepackage{mathdots}

\newcommand{\bonnpi}{Physikalisches Institut, Universität Bonn, Nussallee 12, 53115 Bonn, Germany}
\newcommand{\Salernodf}{Dipartimento di Fisica ‘‘E.R. Caianiello’’, Università degli Studi di Salerno and INFN, Sezione di Napoli, Gruppo collegato di Salerno, Via Giovanni Paolo II, 132, I-84084 Fisciano (Sa), Italy}
\newcommand{\cnr}{CNR/SPIN, Fisciano (Sa), 84084, Italy}

\usepackage{xcolor}

\begin{document}
\title{Floquet Engineering of Quantum Transport through two Driven Impurities}

\author{Vincenzo Bruno}
\affiliation{\bonnpi} 
\affiliation{\Salernodf}

\author{Corinna Kollath}
\affiliation{\bonnpi}

\author{Roberta Citro}
\affiliation{\Salernodf}
\affiliation{\cnr}

\author{Ameneh Sheikhan}
\affiliation{\bonnpi}

\date{\today}

\begin{abstract}
Floquet engineering offers powerful tools to manipulate quantum states by periodically driving physical parameters. In this work, we investigate the quantum transport through two periodically driven impurities in a mesoscopic one-dimensional channel. By mapping the time-dependent Hamiltonian into an effective multichannel scattering problem, we unveil a rich landscape of transport phenomena arising from the interplay between Fabry-Pérot cavity modes and Fano interference. We demonstrate that the inter-impurity distance acts as a critical control parameter, allowing for the formation of  Bound States in the Continuum (BICs). Furthermore, we identify Quasi-BICs---extremely narrow resonances with finite lifetimes---that can be dynamically tuned by the drive amplitude. We show that these states enable a robust coherent trapping mechanism, allowing the system to switch from perfect transparency or reflection to strong localization with giant Wigner time delays. Our results suggest possible applications for tunable delay lines and quantum memories, with feasible experimental realizations in the context of cold atoms.
\end{abstract}

\maketitle

\section{Introduction}
\label{Intro}

Floquet engineering, which refers to the manipulation of quantum systems using periodic drives, has attracted considerable attention in recent years due to its ability to coherently control quantum states \cite{d2016quantum,mankowsky2016non,zhang2017manipulating,eckardt2017colloquium,aidelsburger2018artificial,oka2019floquet, zhou2023pseudospin}. The introduction of local periodic drives in quantum systems has attracted significant interest recently, as it allows for the control of transport properties. Transport through a single oscillating impurity has been studied in both discrete lattices \cite{thuberg2016quantum,reyes2017transport,hubner2022floquet,hubner2023momentum,bruno2025transport,ahmadi2025pair} and continuum systems \cite{kim1998coherent,martinez2001transmission, citro_2020}. The latter are particularly relevant for describing transport in mesoscopic systems, such as ballistic one-dimensional semiconductors \cite{Datta1995transport}, ballistic wires for electron optics \cite{bocquillon2014electron}, and ultracold atoms in ballistic channels \cite{brantut2012conduction,krinner2015observation,lebrat2018band}.

A particularly intriguing aspect of periodically driven systems is that one can engineer bound states in the continuum (BICs) \cite{longhi2013floquet}. These are peculiar states that remain localized despite being embedded in the continuum spectrum, yet are completely decoupled from it \cite{hsu2016bound,kang2023applications}. The theoretical existence of such states was first demonstrated by von Neumann and Wigner \cite{vonNeumann1929bic}, and they can be described as resonant states with zero width \cite{friedrich1985interfering}. Experimental detection of these states typically requires a perturbation to the system configuration, allowing them to couple weakly to the continuum and manifest as narrow resonances \cite{plotnik2011experimental,liu2025high}. In such cases, these  states partially coupled to the continuum spectrum are  referred to as quasi-BICs \cite{yang2021nanoparticle}. Quasi-BICs are closely related to Fano resonances \cite{melik2021fano}, which arise from the Fano interference between discrete and continuum scattering pathways \cite{Fano1961resonance}.
Floquet engineering provides a powerful mean to induce such resonances, making it a promising route to realize quasi-BICs in novel platforms. In systems with local periodic drives, different Floquet sidebands at positive and negative energies can act as effective discrete and continuum spectra.
The coherent exchange of energy quanta with the drives allows an incident particle to dynamically couple these pathways, and the resulting Fano interference between the propagating continuum and the localized sidebands manifests as Fano resonances in the transport properties.

In this work, we investigate the transport of a quantum particle in a one-dimensional continuum system containing two periodically driven impurities, oscillating at the same frequency and in-phase. Previous analyses of this configuration  focused primarily on the regime of small driving amplitudes \cite{kim1999coherent, kim1998dynamic}. Our first objective is to revisit and extend this weak-driving analysis. We show how BICs can be engineered, and how they can be related to the Fano resonances. Furthermore, we perform an exploration of the strong-driving regime, where analytical treatments are no longer feasible. To do so, we carry out a comprehensive numerical study of the transport properties. In this regime we show how the transport properties of the system are determined by the interplay between Fano interference and cavity effects.
The paper is organized as follows. In Sec.~\ref{2-model}, we introduce the model and apply Floquet theory to derive a time-independent effective Hamiltonian. In Sec.~\ref{3-T}, we calculate the transmission probability for an incoming free particle transmitted beyond the second impurity. In Sec.~\ref{4-single}, we briefly review the simpler case of a single oscillating impurity to introduce the basic mechanism of Fano interference in locally periodically driven systems before adding the complexity of the second impurity. Section~\ref{5-overview} provides an overview of the results, highlighting the different regimes and physical phenomena discussed in the following sections. Section~\ref{6-wcr} presents the analysis of the weak-coupling regime, where we demonstrate coherent control of Fano resonances by tuning the driving parameters and impurity separation, and engineer BICs. Section~\ref{7-bwcr} addresses the physics beyond the weak-coupling regime, in particular how most of the features of the transmission spectrum can be expalined in terms of the interplay between Fano interference and cavity effects. Section~\ref{8-localization} discusses the coherent control of the system's transport properties. Specifically, we show that by changing the drive amplitude, the system can be tuned between regimes of perfect transmission, perfect reflection and localization. Finally, we give conclusions in Sec.~\ref{sec:conclusion}.

\section{The Model and Floquet Analysis}
\label{2-model}
\begin{figure}[h]
    \centering
    \includegraphics[width=0.45\textwidth]{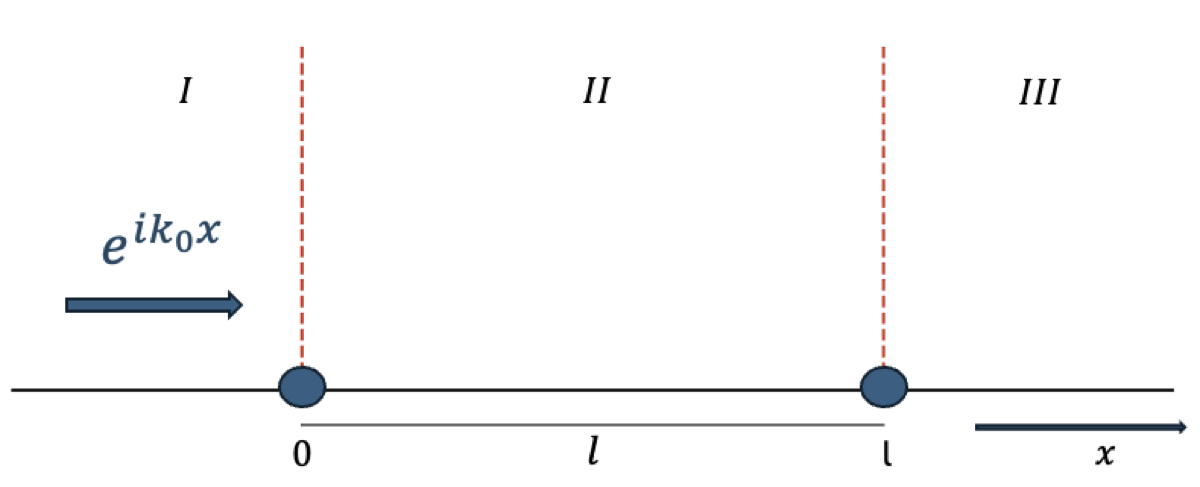}
    \caption{Sketch of the one-dimensional system with two impurities. The arrow indicates the incoming particle with momentum $k_{0}$, while the two vertical red dashed lines represent the localized, periodically driven impurities.}
    \label{fig:1}
\end{figure}
The system consists of a one-dimensional wire with two driven impurities as sketched in Fig. \ref{fig:1}. The impurities are represented by two Dirac delta functions located at $x=0$ and $x=l$ and the amplitude of these functions is modulated in time. In order to probe the transport properties of the system, an incoming free particle with mass $m$ and energy $E=\frac{\hbar^{2}k^{2}_{0}}{2m}$ is prepared and scatters with this two-impurity system.
The Hamiltonian of the system is
\begin{equation}
    \label{Hamiltonian}
    \hat{H}(t) = -\frac{\hbar^{2}}{2m}\frac{d^2}{dx^2} + V(x,t),
\end{equation}
where the potential takes the form
\begin{equation}
\label{potential_model}
\begin{split}
V(x,t) = & -[\nu_{1} + \gamma_{1}\cos(\omega t)]\delta(x) + \\
         & -[\nu_{2} + \gamma_{2}\cos(\omega t)]\delta(x-l).
\end{split}
\end{equation}
Here the impurities have  static potential contributions $-\nu_{1}$ and $-\nu_{2}$ and a driven part  with amplitudes $\gamma_{1}$ and $\gamma_{2}$ which are modulated in time by two cosine terms with the same frequency $\omega$.  If we consider the static limit ($\gamma_1=\gamma_2=0$), the system reduces to the well-known double-delta problem in quantum mechanics~\cite{book,griffiths2018introduction}. This configuration admits up to two bound states depending on the system parameters. The lowest bound state, which is the ground state, exists for any attractive potential strength $\nu_{1,2} > 0$, regardless of how weak it is. In contrast, the other bound state which is the first excited state  exists only when $l > \frac{\hbar^2}{2m}\left(\frac{1}{\nu_1}+\frac{1}{\nu_2}\right)$; specifically, this state disappears below a critical distance (at fixed $\nu_{1,2}$) or below critical potential strengths (at fixed $l$).  

\begin{figure}[h]
    \centering
    \includegraphics[width=0.45\textwidth]{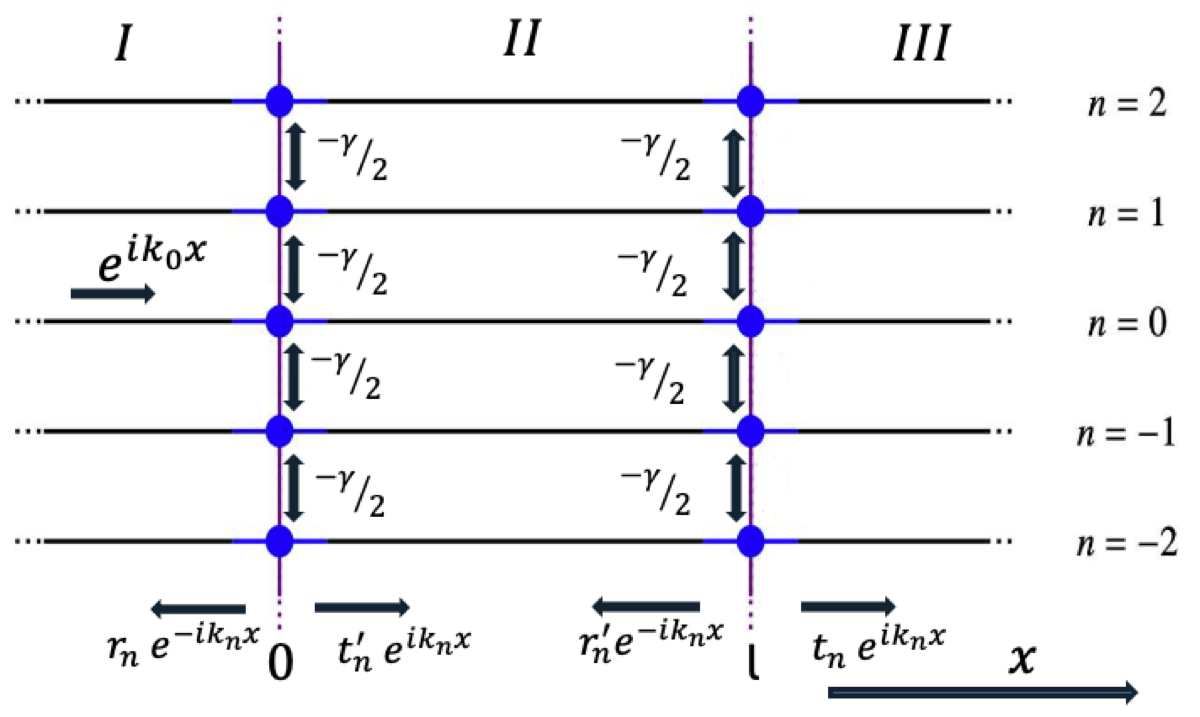}
    \caption{Sketch of the multi-channel setup arising from the Floquet mapping of the time-dependent Hamiltonian $H(t)$ onto the extended Hilbert space. The coupling $\gamma_{1}=\gamma_{2}=\gamma$ correspond to the drive amplitudes of the time-dependent model and $\nu_{1}=\nu_{2}=\nu$ are the amplitudes of the static potential. The incoming plane wave $e^{ik_0 x}$
 , along with the possible reflection and transmission pathways, are depicted by arrows. The reflection and transmission amplitudes are denoted by $r^{(')}_n$ and $t^{(')}_n$, respectively. }
    \label{fig:2}
\end{figure}

As shown in Fig.~\ref{fig:1}, we can identify three distinct spatial regions: region I ($x<0$), from which the particle is incident and to where it can be reflected; region II ($0<x<l$), between the two impurities, which acts as a cavity where the particle can be reflected and transmitted multiple times and may resonate; and region III ($x>l$), into which the particle can be transmitted and propagate away from the second impurity.
This quantum transport problem involving a periodic drive can be effectively tackled using Floquet theory. Analogous to Bloch's theorem for spatially periodic potentials, this formalism exploits the temporal periodicity of the system to map the original time-dependent Hamiltonian, acting on the physical Hilbert space $\mathcal{H}$, onto a time-independent effective Hamiltonian acting on an extended Hilbert space $\mathcal{F}$~\cite{santoro2019introduction,eckardt2015high}.
This enlarged space is constructed as the tensor product $\mathcal{F} = \mathcal{H} \otimes \mathcal{T}$, where $\mathcal{T}$ is the space of periodic functions spanned by the orthonormal basis of Fourier harmonics $\{ e^{in\omega t} \mid n \in \mathbb{Z} \}$. In this framework, the time variable is treated as an additional coordinate, and the Floquet index $n \in \mathbb{Z}$ labels the discrete sites of a synthetic frequency lattice, allowing the dynamical problem to be solved as a stationary multichannel scattering problem.
According to the Floquet theorem~\cite{floquet1883equations}, the solution to the time-dependent Schrödinger equation,
\begin{equation}
    \label{TDSE}
    -i\hbar\frac{\partial \ket{\psi(t)}}{\partial t} = \hat{H}(t)\ket{\psi(t)}
\end{equation}
has the form
\begin{equation}
    \label{Floquet_solution}
    \ket{\psi(t)} = e^{-i\frac{E}{\hbar} t}\ket{\phi(t)}=\sum_{n\in \mathbb{Z}}e^{-i\frac{(E+n\hbar\omega)}{\hbar}t}\ket{\psi_{n}}.
\end{equation}
The Floquet mode $\ket{\phi(t)}$ is periodic in time with period $\frac{2\pi}{\omega}$ and can be written as a Fourier series, whose $n$-th coefficient is $\ket{\psi_{n}}$, with $n \in \mathbb{Z}$. 
The Fourier components of the Floquet mode $\ket{\psi_{n}}$ satisfy the Floquet equation
\begin{equation}
    \label{Quasienergy}
    \hat{Q}\ket{\psi_{n}}=E_{n}\ket{\psi_{n}}
\end{equation}
where $\hat{Q}=\hat{H}(t)-i\hbar\frac{\partial}{\partial t}$ is the Floquet operator (or quasi-energy operator) and $E_{n} = E + n\hbar\omega$ are the quasi-energies. In the extended Hilbert space $\mathcal{F}$, the Floquet operator $\hat{Q}$ is characterized by a block structure~\cite{eckardt2015high}, explained in Appendix~\ref{App:Floquet_Structure}, defined by the matrix elements:
\begin{equation}
    \label{Q_block}
    \hat{Q}_{n,n^{'}} = \hat{H}_{n-n^{'}} + \delta_{n,n^{'}}n\hbar\omega,
\end{equation}
where $\hat{H}_{j}$ is the $j$-th coefficient of the Fourier expansion of the time-periodic Hamiltonian $H(t)$, with $n,n^{'} \in \mathbb{Z}$, and $j = n-n^{'}$. 
Crucially, while the operator $\hat{Q}$ acts on the extended space $\mathcal{F}$, the Fourier components $\hat{H}_j$, the state vectors $\ket{\psi_n}$, and the block components $\hat{Q}_{n,n^{'}}$ are all defined on the physical Hilbert space $\mathcal{H}$. Consequently, the eigenvalue problem in $\mathcal{F}$, given by Eq. (\ref{Quasienergy}), decomposes into an infinite set of coupled time-independent equations in $\mathcal{H}$:
\begin{equation}
    \label{Floquet_equation_in_F}
    \hat{H}_{0}\ket{\psi_{n}} + \hat{H}_{-1}\ket{\psi_{n-1}} + \hat{H}_{1}\ket{\psi_{n+1}}=E_{n}\ket{\psi_{n}}.
\end{equation}
where
\begin{equation}
    \hat{H}_0 = -\frac{\hbar^2}{2m}\frac{d^2}{d x^2} - \nu_1 \delta(x) - \nu_2\delta(x-l).
\end{equation}
and
\begin{equation}
    \hat{H}_{1} = \hat{H}_{-1} = -\frac{\gamma_1}{2} \delta(x)-\frac{\gamma_2}{2} \delta(x-l).
\end{equation}
In this framework, we have mapped the original time-dependent problem of two periodically driven impurities in a single channel onto an equivalent time-independent problem involving an infinite number of coupled channels, each corresponding to a Fourier component of the Floquet mode $\ket{\psi_{n}}$. This results in a stationary multi-channel scattering problem, as depicted in Fig.~\ref{fig:2}.
At the impurity locations $x=0$ and $x=l$, the channels are locally coupled by the external drive. This coupling is governed by the off-diagonal terms $\hat{H}_{\pm 1}$ in Eq.~\eqref{Floquet_equation_in_F}, which connect the $n$-th channel $\ket{\psi_n}$ to its nearest neighbors $\ket{\psi_{n\pm 1}}$. Physically, this implies that the oscillating potential induces transitions between Floquet sidebands, allowing the particle to exchange energy quanta $\hbar\omega$ with the drive at the impurity sites. The strength of this inter-channel coupling is directly controlled by the driving amplitudes $\gamma_{1,2}$, as the terms $\hat{H}_{\pm 1}$ are proportional to it.

We can describe the scattering process in this multichannel picture. An incoming particle from region I in the channel ($n=0$), with energy $E=\hbar^{2}k_{0}^{2}/2m$, impinges onto the first impurity at $x=0$. Here, the wavefunction can split: it can be reflected back into region I into a generic $n$-th channel with amplitude $r_{n}$, or transmitted into the $n$-th channel of the central region II with amplitude $t^{'}_{n}$. Inside region II, the particle propagates as a superposition of channel states until it encounters the second impurity at $x=l$. At this second interface, the particle undergoes further scattering, being either reflected back into any channel in region II (with amplitudes $r^{'}_{n}$ for all $n\in\mathbb{Z}$) or transmitted into any channel in region III with amplitude $t_{n}$, for all $n\in\mathbb{Z}$.
Ultimately, this mapping reveals that instead of a single dynamical cavity, the system behaves as an infinite set of static cavities coupled at the boundaries, where the particle undergoes multiple reflections and transmissions involving different energy sidebands. In this multi-channel framework, the quasi-energy $E_{n}$ corresponds to the physical energy of the particle propagating in the $n$-th channel. Therefore, throughout this paper, we will refer to $E_n$ simply as the channel energy, implying that a particle scattered into the $n$-th sideband carries a total energy $E_n = E + n\hbar\omega$.

\section{Transport Properties}
\label{3-T}
Let $\psi_{n}(x) = \braket{x|\psi_{n}}$ denote the $n$-th Fourier component of the Floquet mode in the position representation \cite{book}. This function represents the spatial wavefunction associated with the particle in the $n$-th channel. For the multichannel scattering problem depicted in Fig.~\ref{fig:2}, we adopt the following piecewise ansatz:

\begin{equation}
    \label{Floquet_modes}
    \psi_n(x) =
\begin{cases}
\delta_{0n}e^{ik_nx} + r_ne^{-ik_nx} & \text{if } x < 0 \\
t_n'e^{ik_nx} + r_n'e^{-ik_nx} & \text{if } 0 \le x \le l \\
t_ne^{ik_nx} & \text{if } x > l,
\end{cases}
\end{equation}
where $k_n$ is the wavevector associated with the energy $E_n$. 
In the same representation, the time-dependent wavefunction, $\psi(x,t)=\braket{x|\psi(t)}$, is given by the Floquet expansion (\ref{Floquet_solution}):
\begin{equation}
    \label{psi_tot}
    \Psi(x, t) = \sum_{n=-\infty}^{+\infty} e^{-\frac{i}{\hbar}E_nt}\psi_n(x).  
\end{equation}
 The wave number $k_{n}$ for each channel can be derived from the free particle dispersion relation  $E_{n} = \frac{\hbar^2 k_{n}^2}{2m}$ as

\begin{equation}
\label{k_n}
    k_{n}=\frac{1}{\hbar}\sqrt{2m(E+n\hbar\omega)}.
\end{equation}
The propagation behavior of the wavefunction depends fundamentally on whether the wavevector $k_{n}$ is real or imaginary.
When the channel energy is positive ($E+n\hbar\omega>0$), $k_{n}$ is real, and the mode $\psi_{n}(x)$ behaves as a propagating plane wave in regions I and III. Such a channel is referred to as an {open channel and contributes to the transport.
Conversely, if the energy is negative ($E+n\hbar\omega<0$), $k_{n}$ becomes purely imaginary. In this scenario, the mode $\psi_{n}(x)$ corresponds to an exponentially decaying wave in regions I and III. These are known as closed or evanescent channels, as they do not carry any current away from the scattering region.
To determine the unknown scattering amplitudes, we apply the boundary conditions to the wavefunction ansatz in Eq.~\eqref{Floquet_modes}. These conditions require the continuity of the wavefunction at the impurity locations ($x=0$ and $x=l$) and a discontinuity in its first derivative. Due to the   properties of the Dirac delta potential the latter is derived explicitly by integrating Eq.~\eqref{Floquet_equation_in_F} in small intervals around the two impurities.
\begin{widetext}
 At $x=0$, the boundary conditions impose, for each channel $n\in \mathbb{Z}$:
\begin{subequations}
\label{BC_Explicit_0}
\begin{align}
    \psi_n(0^+) &= \psi_n(0^-), \\
    \left. \frac{d\psi_n}{dx} \right|_{0^+} - \left. \frac{d\psi_n}{dx} \right|_{0^-} &= -\frac{2m}{\hbar^2} \{ \nu_1 \psi_n(0) + \frac{\gamma_1}{2} \left[ \psi_{n-1}(0) + \psi_{n+1}(0) \right] \}.
\end{align}
\end{subequations}

Similarly, at $x=l$, the boundary conditions for each channel $n\in \mathbb{Z}$ read:
\begin{subequations}
\label{BC_Explicit_l}
\begin{align}
    \psi_n(l^+) &= \psi_n(l^-), \\
    \left. \frac{d\psi_n}{dx} \right|_{l^+} - \left. \frac{d\psi_n}{dx} \right|_{l^-} &= -\frac{2m}{\hbar^2} \{ \nu_2 \psi_n(l) + \frac{\gamma_2}{2} \left[ \psi_{n-1}(l) + \psi_{n+1}(l) \right] \}.
\end{align}
\end{subequations}
In the preceding equations, we adopted the notation $\displaystyle\lim_{x \to x_{0}^{\pm}}f(x)=f(x_{0}^{\pm})$ and $\displaystyle\lim_{x \to x_{0}^{\pm}}\frac{df(x)}{dx}=\left. \frac{df}{dx} \right|_{x_{0}^{\pm}}$.
Substituting the piecewise ansatz from Eq.~\eqref{Floquet_modes} into these boundary conditions yields the following system of coupled linear equations for the scattering amplitudes in each channel $n$:
\begin{subequations}
\label{eq:LinearSystem}
\begin{align}
    \label{eq:sys_cont_0}
    \delta_{0n} + r_n &= t_n' + r_n', \\
    \label{eq:sys_jump_0}
    ik_n(t_n' - r_n' - \delta_{0n} + r_n) &= -\frac{2m}{\hbar^2} \left[ \nu_1 (t_n' + r_n') + \frac{\gamma_1}{2} \sum_{\sigma=\pm 1}(t_{n+\sigma}' + r_{n+\sigma}') \right], \\
    \label{eq:sys_cont_l}
    t_n'e^{ik_nl} + r_n'e^{-ik_nl} &= t_ne^{ik_nl}, \\
    \label{eq:sys_jump_l}
    ik_n \left[ t_n e^{ik_nl} - (t_n' e^{ik_nl} - r_n' e^{-ik_nl}) \right] &= -\frac{2m}{\hbar^2} \left[ \nu_2 t_n e^{ik_nl} + \frac{\gamma_2}{2} \sum_{\sigma=\pm 1} t_{n+\sigma} e^{ik_{n+\sigma}l} \right].
\end{align}
\end{subequations}
\end{widetext}
Equations~\eqref{eq:sys_cont_0}-\eqref{eq:sys_jump_l} constitute an infinite set of coupled linear equations for unknown amplitudes (as $n \in \mathbb{Z}$), reflecting the infinite dimensions of the extended Hilbert space.
In order to obtain information about the transport properties of the system one has to consider total transmission probability $T$. This is defined as the probability that a particle, incident from the fundamental channel ($n=0$), is transmitted into any open channel in region III. Based on the conservation of the probability current, $T$ is calculated as the sum of transmitted fluxes over all open channels:
\begin{equation}
    \label{Transmission_probability}
    T = \sum_{n \in \mathcal{S}_{\text{open}}} |t_{n}|^{2}\frac{k_{n}}{k_{0}}.
\end{equation}
The sum runs over the set of open channels $\mathcal{S}_{\text{open}} = \{n \in \mathbb{Z} \mid k_n \in \mathbb{R}\}$, since evanescent modes do not contribute to the net current. Further details on this derivation are provided in Appendix~\ref{Transmission}.

Before discussing the numerical solution of this system, it is convenient to render the equations dimensionless. We define the dimensionless energy $\tilde{E}$ as
\begin{equation}
\tilde{E} = \frac{E}{\hbar\omega}
\end{equation}
and the dimensionless wave number as:
\begin{equation}
    \tilde{k}=k\sqrt{\frac{\hbar}{2m\omega}}.
\end{equation}
All the other dimensionless physical quantities that we are going to use are
\begin{equation}
\begin{split}
& \tilde{l} = l\sqrt{\frac{2m\omega}{\hbar}},  \quad
\tilde{t} = t\omega, \\
&\tilde{\gamma}_{1,2} = \gamma_{1,2}\sqrt{\frac{2m}{\hbar^3\omega}},  \quad \tilde{\nu}_{1,2} = \nu_{1,2}\sqrt{\frac{2m}{\hbar^3\omega}} .
\end{split}
\end{equation}
 All subsequent numerical results in the plots, as well as the equations from Sec.~\ref{4-single}(B) are presented using these dimensionless variables.
To obtain a tractable numerical problem, the infinite number of equations must be truncated. If not stated otherwise, we  introduce a cutoff $n_{\text{max}} \in \mathbb{N}$ and retain only the channels with index $|n| \le n_{\text{max}}$. This approximation reduces the problem to a finite system of equations involving $N = 2n_{\text{max}} + 1$ coupled channels.
Since there are 4 unknown amplitudes ($r_n, t_n, t_n', r_n'$) for each channel, this results in a closed linear algebraic system of dimension $4N \times 4N$, which is inverted numerically.
The choice of the cutoff $n_{\text{max}}$ is crucial for the accuracy of the results. We determine the necessary cutoff by checking for convergence, increasing $n_{\text{max}}$ until the calculated quantities stabilize. We show that different channel coupling regimes (corresponding to different values of $\gamma_{1,2}$) require different cutoffs $n_{max}$. 
In general, the stronger the coupling between the channels, the larger the required $n_{max}$. For sufficiently low values of the coupling also an approximate analytical solution can be derived. This allows us to introduce a weak-coupling regime, where the physics of the system is accurately described by retaining only two channels ($n=0$ and $n=-1$).

\section{Single Impurity}
\label{4-single}

In this section, we review the transport properties of a system featuring a periodically driven single impurity, obtained from the general model in Eq.~\eqref{potential_model} by setting $\gamma_2=0$ and $\nu_2=0$ ($\gamma_{1}=\gamma\ne0$ and $\nu_{1}=\nu\ne0$). In previous work~\cite{kim1998coherent,martinez2001transmission}, it has been shown that transport through a periodically driven impurity leads to the formation of a Fano resonance, making it possible to engineer a Fano mirror that reflects particles with specific energy or momentum. We extend previous works by the derivation of approximate analytical expressions for the transmission poles and zeros associated with the Fano resonance in the weak-coupling regime, relying on a minimal two-channel model ($n=0, -1$). Remarkably, in this regime, we recover the results of Ref.~\cite{kim1998coherent}, which were originally obtained using a three-channel truncation ($n=0, \pm 1$). Furthermore, we extend the analysis to the case of a vanishing static potential ($\nu=0$). This specific limit requires a distinct theoretical treatment: while it can be addressed using a continued fraction approach~\cite{martinez2001transmission}, we employ an aymptotic expansion in the drive amplitude $\gamma$, which yields explicit analytical expressions describing the weak-coupling physics.
This analysis serves as a benchmark for understanding the fundamental mechanisms underlying Fano resonances, BICs and quasi-BICs, while establishing the scattering formalism that will be extended to the two-impurity system. 
Finally, we provide insights into relevant experimental platforms, focusing specifically on ultracold atoms and semiconductive nanoscale channels. By analyzing the typical localization times associated with Fano resonances in these systems, we introduce the concept of quasi-BICs.

\subsection{General Properties of Fano Resonances and Bound States in the Continuum}

A Fano resonance arises from the quantum interference between two scattering pathways~\cite{Fano1961resonance,iizawa2021quantum,miroshnichenko2010fano}: (i) a direct, non-resonant path through a continuum, and (ii) a resonant path involving a discrete quasi-bound state coupled to the continuum. 
Due to this interference, the transmission probability can vary drastically, potentially leading to complete suppression. The transmission profile is described by the Fano resonance formula \cite{miroshnichenko2010fano}:
\begin{equation} \label{Fano_profile}
    T(\epsilon) = \frac{(\epsilon+q)^{2}}{\epsilon^{2} + 1}.
\end{equation}
Here, we introduce the reduced energy $\epsilon = (E-E_r)/(\Gamma/2)$, where $E_r$ is the energy of the resonant state and $\Gamma$ represents its spectral width arising from the coupling to the continuum. Physically, the resonant state acts as a quasi-bound state with energy $E_{r}$ and a lifetime $\tau = \hbar/\Gamma$. The Fano asymmetry parameter, $q$, quantifies the interference; its modulus $|q|$ measures the relative weight between the resonant and direct channels, while its sign determines their phase relationship.
From Eq.~\eqref{Fano_profile}, several key spectral characteristics of the Fano resonance can be identified.

\begin{figure}[h!]
    \centering
    \includegraphics[width=0.45\textwidth]{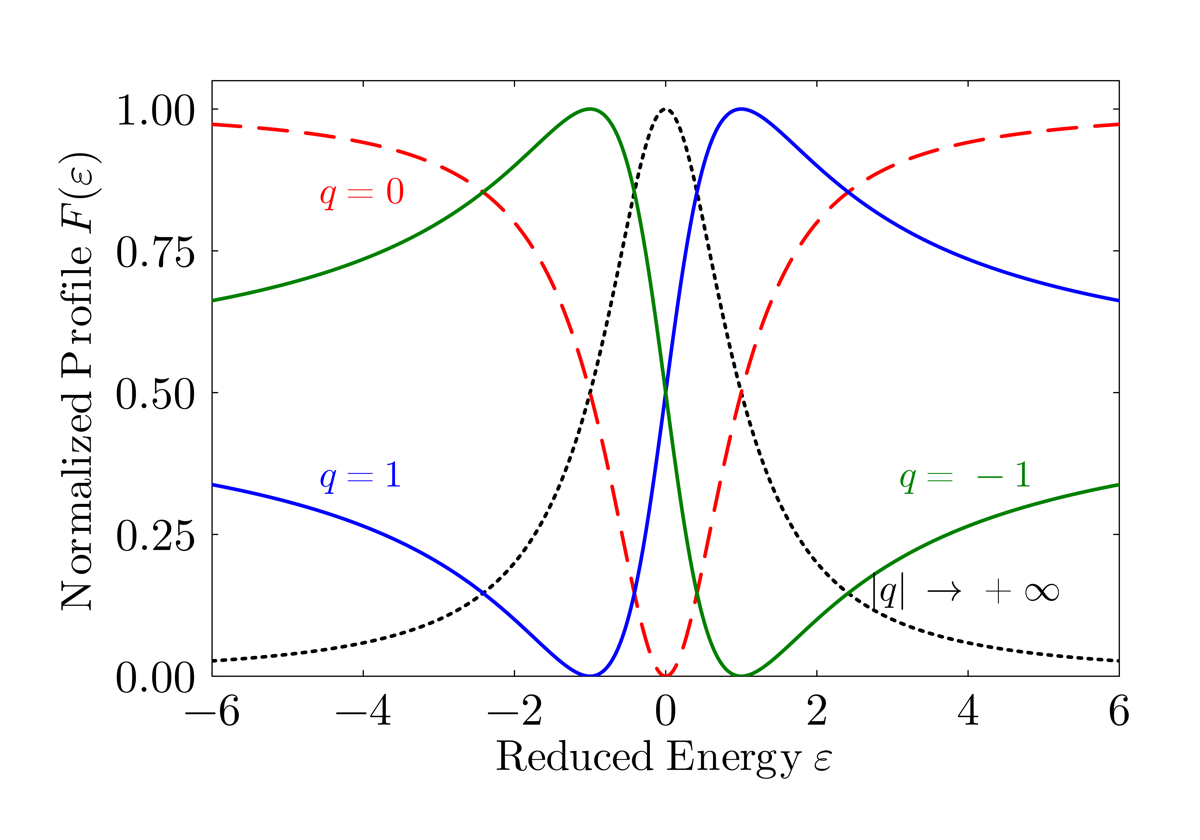}
    \caption{Normalized Fano profile, $F(\epsilon)=T(\epsilon)/(1+q^{2})$, with $T$ given by Eq.~\eqref{Fano_profile}. For $q>0$, the profile exhibits a dip-peak structure, while for $q<0$ it shows a peak-dip structure. The Lorentzian limits correspond to $q \rightarrow \pm \infty$ (peak) and $q=0$ (dip).}
    \label{fig:fano_pedagogical}
\end{figure}
By imposing the condition $T=0$ in Eq.~\eqref{Fano_profile}, we find that the transmission zero occurs at the energy:
\begin{equation} \label{zero_Fano}
    E_{zt} = E_{r} - \frac{\Gamma}{2}q.
\end{equation}
Regarding the resonance pole, the vanishing denominator of the Fano profile mathematically yields two solutions, $\epsilon = \pm i$. However, the solution corresponding to $\epsilon = +i$ is unphysical, as it would imply a resonance with a probability amplitude that diverges in time. Consequently, the physical pole is given by the root with the negative imaginary part:
\begin{equation} \label{pole_Fano}
    E_{p} = E_{r} - i \frac{\Gamma}{2}.
\end{equation}
Finally, by analyzing Eq.~\eqref{Fano_profile}, one finds that the maximum transmission probability occurs at:
\begin{equation} \label{maximum_fano}
    E_{m} = E_{r} + \frac{\Gamma}{2q}.
\end{equation}
The width $\Gamma$ determines the spectral distance between the zero and the maximum of the Fano profile. Depending on the sign of $q$, distinct spectral shapes emerge. For $q>0$, the energies follow the order $E_{zt} < E_r < E_m$, which corresponds to a ``dip-peak'' structure. Conversely, for $q<0$, the ordering is inverted ($E_m < E_r < E_{zt}$), resulting in a ``peak-dip'' structure, as shown in Fig.~\ref{fig:fano_pedagogical}.
Considering the limiting cases, for $q=0$, Eqs.~\eqref{zero_Fano} and \eqref{pole_Fano} imply that the zero coincides with the resonance energy ($E_{zt} = E_r$). This manifests as a symmetric, dip-like resonance where the transmission vanishes exactly at $E_r$. On the other hand, in the limit $|q| \rightarrow \infty$, the transmission profile $T$ evolves into a Lorentzian resonance where $E_{r} \approx E_{m} = \Re\{E_{p}\}$ and the zero is pushed to $E_{zt} \rightarrow \pm\infty$. In this limit, $\Gamma$ corresponds to the full width at half maximum  of the Lorentzian peak.
Finally, if the coupling to the continuum vanishes ($\Gamma \to 0$) for a state at energy $E_r$ embedded in the continuous spectrum, the complex pole becomes real ($E_p \to E_r$). Such a state, known as a BIC, is completely decoupled from the scattering states and becomes invisible in the transmission spectrum, $T \rightarrow1$. A BIC, therefore, manifests as the complete disappearance of the Fano resonance linewidth.

\subsection{Driven Single Impurity Analysis}

\begin{figure*}[t]
    \centering
    % --- Pannello (a) ---
    \begin{minipage}[b]{\textwidth}
        \centering
         \includegraphics[width=\textwidth]{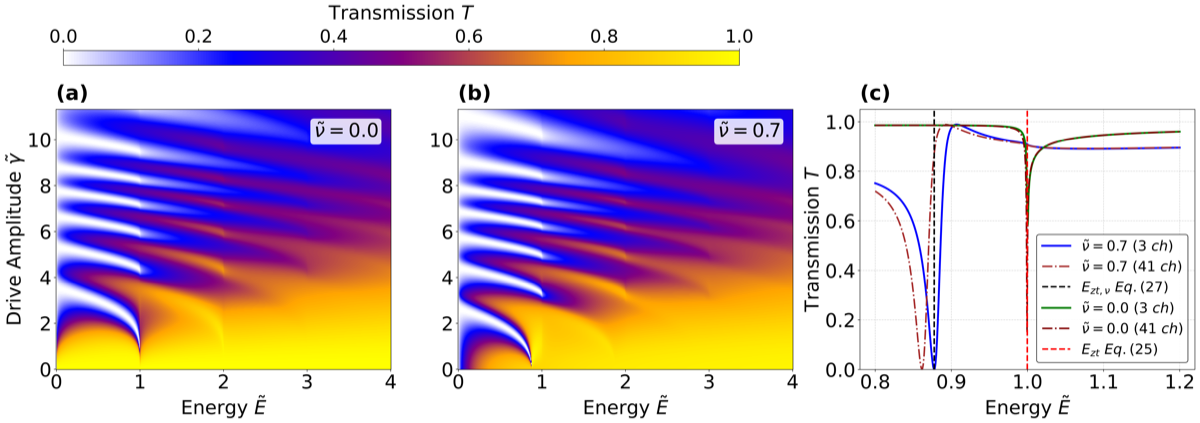}
    \end{minipage}

    \caption{Transport properties of a single periodically driven impurity . Transmission probability 
 as a function of incident energy $\tilde{E}$ and driving amplitude $\tilde{\gamma}$ are displayed  for \textbf{(a)}  the case $\tilde{\nu}=0$ and  \textbf{(b)} with an additional static potential ($\tilde{\nu}=0.7$). For both numerical solutions 41 channels have been taken into account. Panel \textbf{(c)} shows the energy profiles of the transmission probability at fixed drive amplitude ($\tilde{\gamma}=0.5$).
The solid curves represent the results obtained within the 2-channel approximation, while the dotted curves correspond to the 41-channel case.
For both $\tilde{\nu}=0$ and $\tilde{\nu}=0.7$,
the vertical dotted lines mark the theoretical predictions derived from Eq. \eqref{one_zero_nostatic} for $\tilde{\nu}=0$ and Eq. \eqref{zero_1_imp} for $\tilde{\nu}=0.7$. The values of the zeros obtained with a fit using Eq. \eqref{Fano_profile} are near to the theoretical values indicated in the plots. In particular, said $\delta \epsilon$ the absolute value of the distance between the fitted and theorertical value of the zero, we obtain $\delta \epsilon = 1.8 \times10^{-4}$ fot $\tilde{\nu}=0$, and $\delta\epsilon = 9.0 \times 10^{-5}$ for $\tilde{\nu}=0.7$.
}
\label{fig:1_impurity}
\end{figure*}
We apply this framework to the setup described in Fig.~\ref{fig:1}, considering only the impurity at $x=0$ (i.e. $\gamma_2=0$, $\nu_2=0$, $\gamma_{1}=\gamma$ and $\nu_{2}=\nu$). 
In a regime of  weak coupling which means $\tilde{\gamma}\lesssim 0.7$, the transmission probability is well-described by a two-channel approximation involving the open channel $n=0$ (continuum) and the first closed channel $n=-1$ (discrete), only. The channel $n=-1$ remains closed as long as its wavenumber $k_{-1}$ is imaginary, a condition that holds for incident energies $\tilde{E} < 1$.
An analytical expression for the transmission amplitude $t_0$ in the two-channel approximation can be derived as:
\begin{equation}
    \label{t_0_1_imp}
    t_{0} = \frac{2i\tilde{k}_{0}}{2i\tilde{k}_{0} + \frac{\tilde{\gamma}^2}{4}\frac{1}{2i|\tilde{k}_{-1}|-\tilde{\nu}}}.
\end{equation}
This expression of $t_{0}$ was obtained in Ref.~\cite{kim1998coherent} for three channels ($n=-1,0,1$).
The total transmission probability for a single impurity is given by $T=|t_{0}|^{2}$ and an example of the numerical determination is  shown in Fig.~\ref{fig:1_impurity}. We distinguish in the following the case with and without the static potential $\nu$. In both cases a Fano resonance is found at low driving amplitudes. 
In the absence of a static potential, shown in  Fig. \ref{fig:1_impurity}(a), the Fano resonance originates purely from the driving field. For small drive amplitudes, the resonance occurs at $\tilde{E} \simeq 1$. As the drive strength $\tilde{\gamma}$ increases, the resonance shifts. We extend previous works using an asymptotic expansion.
The asymptotic expansion of Eq.~\eqref{t_0_1_imp} is performed around the singularity $\tilde{E}=1$, up to fourth order in $\tilde{\gamma}$, following the procedure detailed in Appendix~\ref{app_cal} for the general case of two driven impurities without static potential $\nu=0$. In the case of a single driven impurity it leads to the zero transmission energy
\begin{equation}
\label{one_zero_nostatic}
    \tilde{E}_{zt} \simeq 1 - \frac{\tilde{\gamma}^4}{256},
\end{equation}
and the complex pole at
\begin{equation}
\label{pole_dynamic_1_imp}
    \tilde{E}_{p} \simeq \tilde{E}_{zt} - i\frac{\tilde{\gamma}^4}{256}.
\end{equation}
Comparing the real part of the pole with the zero, we observe that $\tilde{E}_{r}=\text{Re}(\tilde{E}_p) = \tilde{E}_{zt}$. According to Eq.~\eqref{zero_Fano}, this implies $q=0$. Consequently, the transmission spectrum is expected to exhibit a symmetric antiresonance, mirroring the ideal profile in Fig.~\ref{fig:fano_pedagogical}. The numerical result, plotted as the solid red curve in Fig.~\ref{fig:1_impurity}(c), confirms a dip located near $\tilde{E}=1$ with a small shift scaling as $\tilde{\gamma}^4$.
However, the observed profile deviates slightly from perfect symmetry. This arises from two main factors. First, $q$ is only approximately zero ($q \approx 0$) due to higher-order corrections in $\tilde{\gamma}$ neglected in the asymptotic expansion. Second, the $n=-1$ channel becomes an open channel at $\tilde{E}=1$, leading to a discontinuity in the transmission slope at that energy. This effect, known as threshold anomaly has been discussed for the single impurity in Ref.~\cite{martinez2001transmission}.
When a static potential is present ($\tilde{\nu}\ne0$), as shown in  Fig.~\ref{fig:1_impurity}(b), the physics is fundamentally modified by the existence of a bound state inherent to the static potential well. This state is located at energy $\tilde{E}_{b}=-\tilde{\nu}^{2}/4$~\cite{griffiths2018introduction}. In the Floquet picture, this bound state appears in the closed channel with $n=-1$, $\tilde{E}_{-1} = \tilde{E} - 1= -\frac{1}{4}\tilde{\nu}^2$. Thus, the zero of the resonance appears when the energy of the incoming particle meets the condition 
\begin{equation}
    \label{zero_1_imp}
    \tilde{E}_{zt,\nu} = 1 - \frac{1}{4}\tilde{\nu}^{2}.
\end{equation}
This can be extracted directly from  Eq.~\eqref{t_0_1_imp}, with the condition $2i|\tilde{k}_{-1}| - \tilde{\nu} = 0$, which leads to $t_{0}=0$.
To determine the complex pole, we apply a second-order perturbative approach in $\tilde{\gamma}$  around the static bound state energy $\tilde{E}_{b}$. This is explained in the general case of a two impurity-system with $\nu \ne 0$ in Appendix~\ref{app_cal}. For the single impurity case, the expansion yields the real part of the resonance energy:
\begin{equation}
    \label{pole_1_imp}
    \tilde{E}_{r,\nu} \simeq \tilde{E}_{zt,\nu} + \frac{1}{64}\frac{\tilde{\gamma}^2\tilde{\nu}^2}{\frac{\tilde{\nu}^2}{4} + \tilde{E}_{zt,\nu}},
\end{equation}
and the resonance width:
\begin{equation}
    \label{gamma_pole_1_imp}
    \tilde{\Gamma} \simeq \frac{\tilde{\gamma}^{2}\tilde{\nu}\sqrt{\tilde{E}_{zt,\nu}}}{16\left(\frac{\tilde{\nu}^2}{4} + \tilde{E}_{zt,\nu} \right)}.
\end{equation}
Crucially, unlike the case $\tilde{\nu}=0$, here the real part of the pole differs from the energy zero ($\tilde{E}_{r,\nu} \neq \tilde{E}_{zt,\nu}$). According to Eq.~\eqref{zero_Fano}, this implies a non-zero Fano asymmetry parameter ($q \neq 0$), resulting in the asymmetric profile observed in the solid blue curve in Fig. \ref{fig:1_impurity}(c).

\subsection{Quasi-BICs and Experimental Realization}

Equation~\eqref{gamma_pole_1_imp} demonstrates that the resonance width $\tilde{\Gamma}$ scales with $\tilde{\gamma}^2$ and remains finite for any non-zero coupling ($\gamma \ne 0$). Consequently, since $\tilde{\Gamma} \neq 0$, a single driven impurity cannot support a BIC. The formation of true BICs requires destructive interference mechanisms typically provided by spatially separated impurities or multifrequency driving schemes.
However, even in the absence of perfect BICs, the narrow Fano resonances detected in this system [see Fig.~\ref{fig:1_impurity}(c)] manifest as quasi-BICs. These are quasi-bound states weakly coupled to the continuum, characterized by a small resonance width $\Gamma$ and, consequently, a long lifetime $\tau = \hbar/\Gamma$. 
Physically, this corresponds to a scattering process with a significant time delay: the incoming particle interacts with the oscillating impurity and remains localized in its vicinity for a duration $\tau$ before escaping. 
Using the parameters corresponding to the red curve in Fig.~\ref{fig:1_impurity}(c), we evaluate the width of the Fano resonance via Eq.~\eqref{pole_dynamic_1_imp}. 
In dimensionless units, this yields $\tilde{\Gamma}  \simeq 4.9 \times 10^{-4}$, resulting in a dimensionless time delay $\tilde{\tau} = 1/\tilde{\Gamma} \simeq 2000$. 
To contextualize this result, we consider possible experimental platforms based on ultracold atoms in mesoscopic channels. 
Over the last decade, several experiments have investigated transport in such systems in the ballistic regime, specifically in channels without an underlying optical lattice~\cite{brantut2012conduction,krinner2015observation}. 
Localized impurities can be introduced into these one-dimensional channels using projected optical barriers~\cite{lebrat2018band}. 
By introducing one of these optical barriers and modulating them periodically with frequency $\omega$, one can realize the system described by the Hamiltonian in Eq.~\eqref{Hamiltonian} with $\nu_2=0=\gamma_2$. 
Considering typical drive frequencies used in these experiments in the range $\omega/2\pi \in [1, 20]$\,kHz, \cite{lebrat2018band}, the calculated dimensionless delay translates into physical localization times $\tau \simeq 35 - 680$\,ms. 
These timescales are well within the duration of typical transport experiments, which can extend up to $4$\,s~\cite{lebrat2018band}, in order to avoid detrimental heating effect due to the driving. 
Consequently, the single impurity system acts as a perfect energy filter: atoms with energy matching the energy of the Fano resonance $E=E_{r}=E_{zt}$ are delayed for a time $\tau$ before being reflected. This time is macroscopic but shorter than the experimental observation window.

\section{Overview of Transport Properties of the two-impurity system}
\label{5-overview}

\begin{figure*}[t]
    \centering
    % --- Blocco Sinistro ---
    \begin{minipage}[b]{\textwidth}
        \centering
        \includegraphics[width=0.8\textwidth]{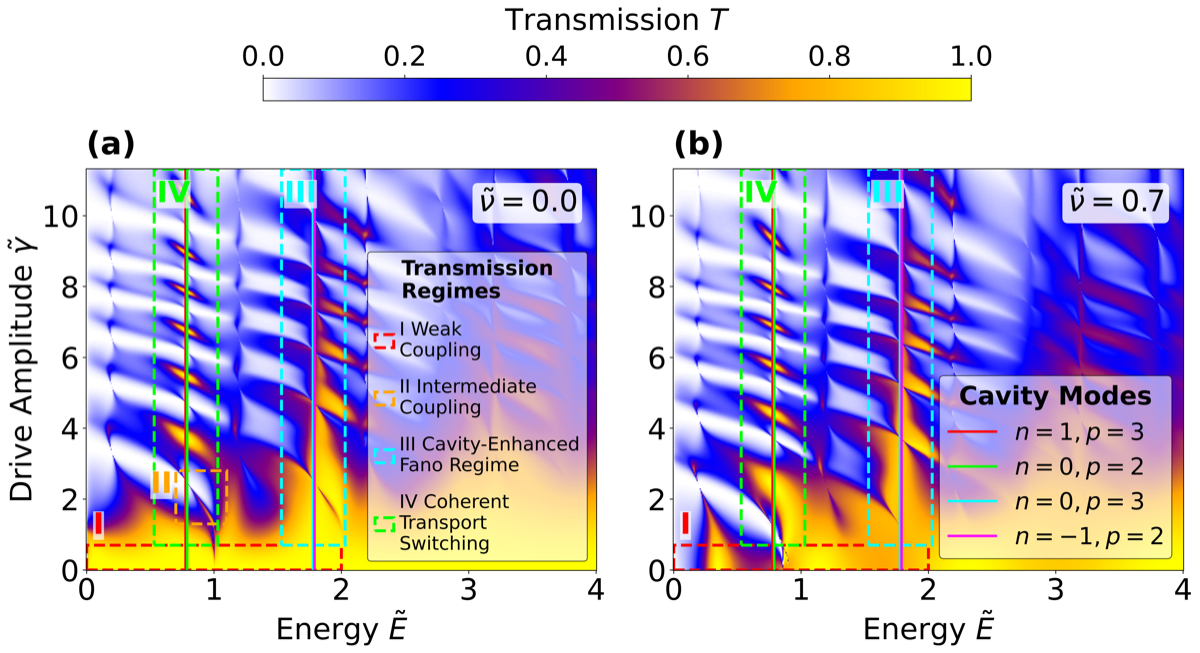}
        \vspace{1ex}  
    \end{minipage}
    \caption{Transmission probability  $T$ for the symmetric two-impurity system as a function of energy of the incoming particle $\tilde{E}$ and driving amplitude $\tilde{\gamma}$, with an inter-impurity distance $\tilde{l}=7.07$.
    \textbf{(a)} Case without static potential ($\tilde{\nu}=0$). The solid vertical lines indicate the theoretical positions of the Fabry-Pérot cavity modes [Eq.~\eqref{cavity_modes_E}] as indicated in the legend in panel (b). The squared dashed rectangles represent the regions of the transmisison spectrum I, II, III and IV characterized by different regimes.
    \textbf{(b)} Case with static potential ($\tilde{\nu}=0.7$).}
    \label{fig:overview}
\end{figure*}

While the previous section summarized and detailed the single-impurity case, in this section we present an overview of our results of the transport properties for the two-impurity system, and what we add with respect to the previous studies \cite{kim1998dynamic,kim1999coherent}. We restrict our analysis to the symmetric case ($\gamma_1=\gamma_2=\gamma, \nu_1=\nu_2=\nu$).  Introducing a weak asymmetry in the drive amplitudes leads to a small perturbation to the symmetric case, whereas strongly asymmetric drive amplitudes result in a behaviour converging toward the single-impurity limit.
Figure~\ref{fig:overview} displays a typical transmission probability $T(\tilde{E}, \tilde{\gamma})$ for two symmetric impurities, calculated numerically using 41 Floquet channels ($n_{max}=20$). We distinguish  between the case without a static potential ($\tilde{\nu}=0$) and the case with a static potential ($\tilde{\nu} > 0$) in Fig.~\ref{fig:overview}(a) and Fig.~\ref{fig:overview}(b), respectively.
Comparing these with the single-impurity results in Fig.~\ref{fig:1_impurity}, it is evident that the two-impurity system exhibits a significantly more complex transmission structure. The spectrum is characterized by a rich pattern of resonance features, showing rapid variations between regions of high transmission and (complete) reflection. 
Intuitively, this complexity can be understood by considering that the transport phenomenology is no longer determined only by Fano interference between discrete and continuum channels, but also by the fact that the system acts as a one-dimensional dynamical Fabry-Pérot cavity, with the two oscillating impurities acting as the cavity's semitransparent mirrors. 
Indeed, a significant part of the structures observed in Fig.~\ref{fig:overview} emerges from the interplay between these Fano interferences and cavity effects, which are absent in the single-impurity case. 
In the following, we highlight some of the most interesting regimes, which will be discussed in more detail later on.

\paragraph*{Case without static potential [$\tilde{\nu}_1 = \tilde{\nu}_2 =\tilde{\nu}=0$, Fig.~\ref{fig:overview}(a)]:} At low drive amplitudes ($\tilde{\gamma} \lesssim 0.7$) and low energies $(\tilde{E} < 2)$, highlighted in Fig. \ref{fig:overview}(a) with the red dashed rectangle and the number $I$ as weak-coupling regime, the spectrum is dominated by a single Fano resonance structure at $\tilde{E}\simeq1$. The Fano resonance arises from the interference between the open channel ($n=0$) and the first closed channel ($n=-1$). We detail in Sec.~\ref{6-wcr} how, in the weak-couping regime, these two channels are sufficient to explain the structure of the transmission.
As the drive amplitude increases, beyond the weak-coupling regime, the structure of the transmission becomes richer. An intermediate regime ($1.8 \lesssim \tilde{\gamma} \lesssim 2.5$) of intermidiate coupling (region II of the transmisison spectrum) emerges where a second Fano structure appears due to the involvement of the $n=-2$ channel (Sec \ref{7-bwcr}). At even stronger coupling ($\tilde{\gamma} \gtrsim 2.5$), we identify two persistent vertical features, around $\tilde{E} \simeq 0.78$ and $\tilde{E} \simeq 1.78$. At these energies Fabry-Pérot cavity modes form within the open channels between the impurities, as indicated by the vertical dashed lines in Fig.~\ref{fig:overview}, whose labels are given in the legend of Fig.~\ref{fig:overview}(b). The resonance condition for the $p$-th cavity mode in the $n$-th channel is given by $\tilde{k}_{n} \tilde{l} = p\pi$, which predicts specific resonant energies:
\begin{equation}
    \label{cavity_modes_E}
    \tilde{E}_{n,p} = \frac{\pi^{2}}{\tilde{l}^{2}}p^{2} - n.
\end{equation}
The solid vertical lines in Fig.~\ref{fig:overview}(a) correspond to these predicted energies. They mark well the novel complex structures arising: in particular, the one indicated as III shows the formation of nodes, while the one indicated as IV shows alternating regions of almost perfect transmission and almost perfect reflection. These structures, caused by an interplay between the cavity modes and Fano interference, will be analyzed in Secs.~\ref{7-bwcr} and \ref{8-localization}.

\paragraph*{Case with static potential ($\tilde{\nu}_{1}=\tilde{\nu}_{2}=\tilde{\nu} > 0$, Fig.~\ref{fig:overview}(b)):} When a static potential is active ($\tilde{\nu}=0.7$), the most significant change with respect to the $\tilde{\nu}=0$ case occurs in the weak-coupling regime. Instead of a single Fano resonance, we observe a double Fano structure. This splitting indicates that the static potential introduces new bound states within the closed channel $n=-1$, each generating its own interference path with the continuum spectrum (channel $n=0$). This mechanism will be the focus of Sec.~(\ref{6-wcr}). The structures at higher driving amplitudes or energies ($III$ and $IV$) seem similar to the $\tilde{\nu}=0$ case and will be analyzed in Secs.~\ref{7-bwcr} and \ref{8-localization}.

\begin{figure}[h]
    \centering
    \includegraphics[width=0.45\textwidth]{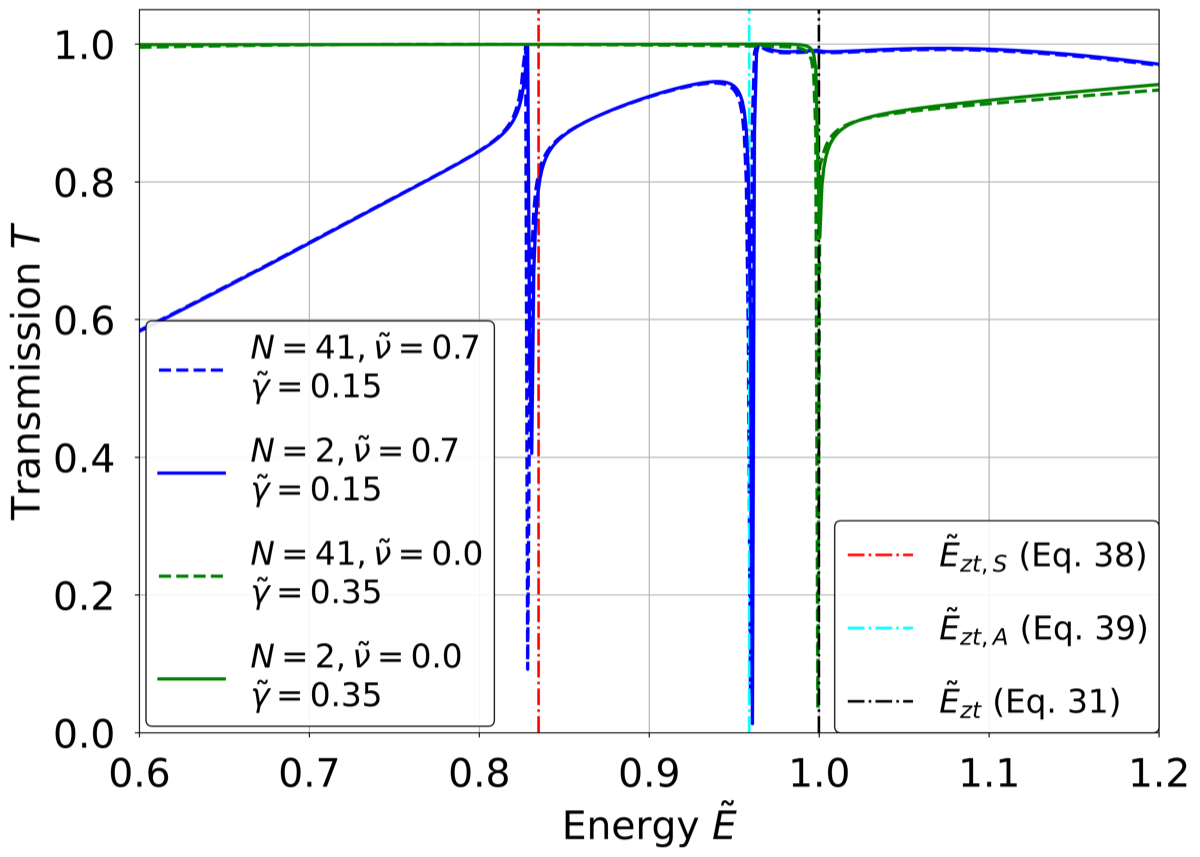}
    \caption{Transmission probability $T$ as a function of the incident particle energy $\tilde{E}$ for a system with two impurities separated by a distance $\tilde{l}=4.24$.
The green curves correspond to the case without static potential ($\tilde{\nu}=0$ and $\tilde{\gamma}=0.35$), exhibiting a single dip. The blue curves correspond to the case with a static potential ($\tilde{\nu}=0.7$ and $\tilde{\gamma}=0.15$), revealing a double Fano resonance structure.
Solid curves represent the results obtained within the 2-channel approximation, while dashed curves correspond to the 41-channel numerical simulation. The vertical lines are the theoretical prediction of the zeros, given by Eq.~\ref{zero_no_static}, Eq.~\ref{eq:zero_nu_simm} and Eq.~\ref{eq:zero_nu_anti}. The main features of the transmission probability are well described by the two-channel approximation.}
    \label{fig:2_imp_wcr}
\end{figure}

\section{Weak-Coupling Regime}
\label{6-wcr}

In this section we consider the weak-coupling regime. First, we examine the system without a static potential ($\tilde{\nu}=0$), where the coupling between channels is purely dynamical, in Subsec.~\ref{6-wcr}A. Second, we consider the case with a static potential ($\tilde{\nu} > 0$), in Subsec.~\ref{6-wcr}B, implying the existence of pre-existing bound states arising from the static double-delta potential problem, as known from standard quantum mechanics~\cite{griffiths2018introduction, book}.
Figure~\ref{fig:2_imp_wcr} provides the key differences of these two scenarios. For $\tilde{\nu}=0$ (green curve), the spectrum exhibits a single Fano resonance with $q\approx0$, qualitatively similar to the single-impurity result shown in Fig.~\ref{fig:1_impurity}(c) and discussed in the previous section. Conversely, activating the static potential ($\tilde{\nu} > 0$, blue curve) drastically alters the spectral response: a structure characterized by two distinct Fano profiles with asymmetry parameters of opposite sign arizes ($q<0$ corresponding to a peak-dip shape, and $q>0$ to a dip-peak shape).
To analyze the behavior of the Fano resonances in detail, we find that the essential features of the weak-coupling regime can be accurately described by reducing the multi-channel framework to a two-channel model, involving one open channel ($n=0$) and the first closed channel ($n=-1$). This approximation allows us to derive analytical insights and to identify the Fano resonances unambiguously. We verified that the two-channel model describes very well the structures seen in the transmission probability for driving amplitudes $0 \le \tilde{\gamma} \lesssim 0.7$ (weak-coupling regime) within the energy window $0 \le \tilde{E} < 2$. This  is demonstrated in Fig.~\ref{fig:2_imp_wcr}, where the solid curves represent the results of the 2-channel model, while the dashed curves correspond to the 41-channel model. Both agree very well. Within the two-channel approximation, which holds for small driving amplitudes,  one can carry out an asymptotical expansion in the driving amplitude $\tilde{\gamma}$ to obtain an approximate expression of the zero of the Fano resoances and compare them to the numerical solution. In the following, we utilize the two-channel approximation to derive analytical insights into the system's transport properties within the weak-coupling regime. We then validate these predictions by comparing them to full numerical simulations performed with 41 Floquet channels. A discussion of the agreement between the two models, as well as of observed deviations, is provided below.

\begin{figure}[h]
    \centering
    \includegraphics[width=0.45\textwidth]{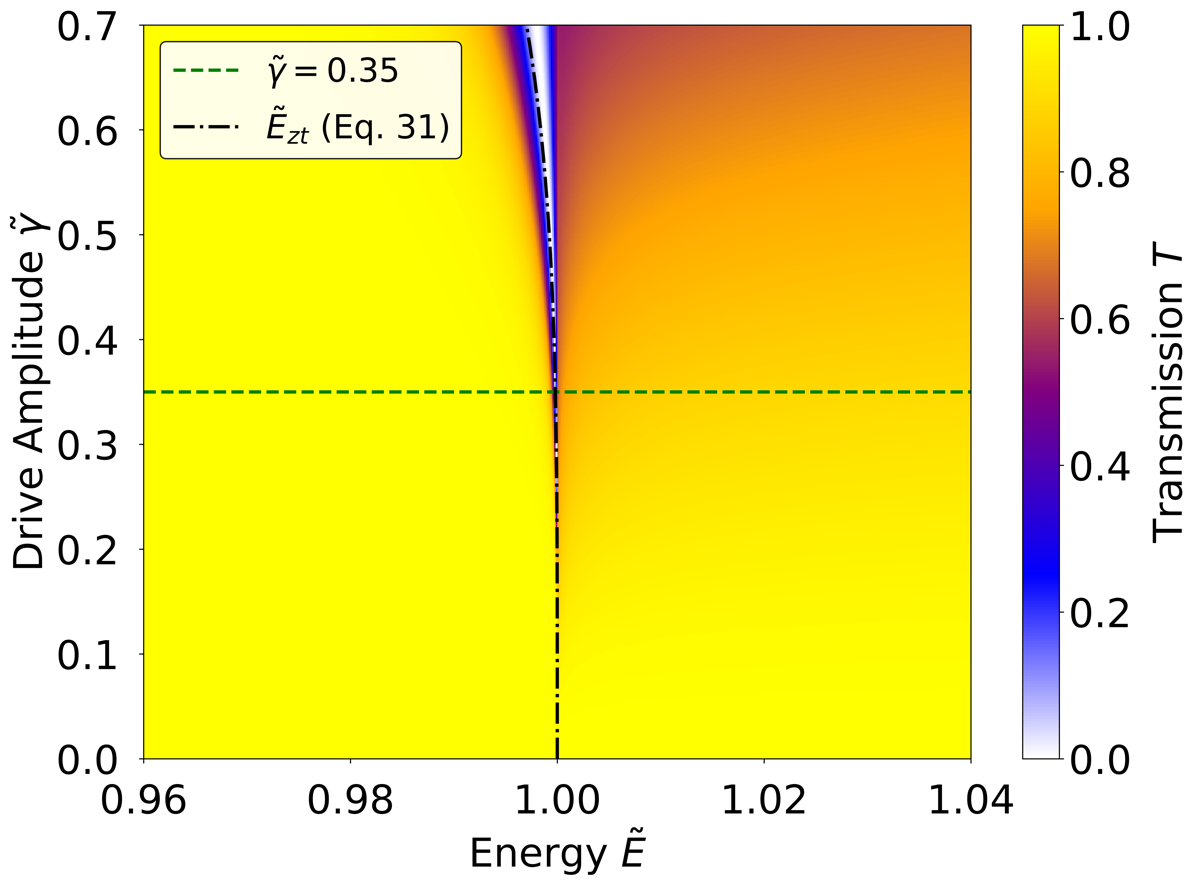}
    \caption{Transmission probability as a function of the incident energy $\tilde{E}$ and drive amplitude $\tilde{\gamma}$, calculated in the two-channel approximation for $\tilde{l}=4.24$ and $\tilde{\nu}=0$. The green dashed line indicates $\tilde{\gamma}=0.35$, corresponding to the drive amplitude of the green curve shown in Fig. \ref{fig:2_imp_wcr}. The numerical solution is obtained using 2 channels ($n=0$ and $n=-1$). The black dash-dotted line represents the analytical prediction of the transmission zero obtained through Eq.~\eqref{zero_no_static}.}
    \label{fig:T_map_nostatic}  
\end{figure}

\subsection{Two impurities in the absence of a static potential}

\begin{figure*}[t]
    \centering
    
    % --- PRIMA RIGA (2 Canali) ---
    \begin{minipage}[b]{\textwidth}
        \centering
        \includegraphics[width=0.8\textwidth]{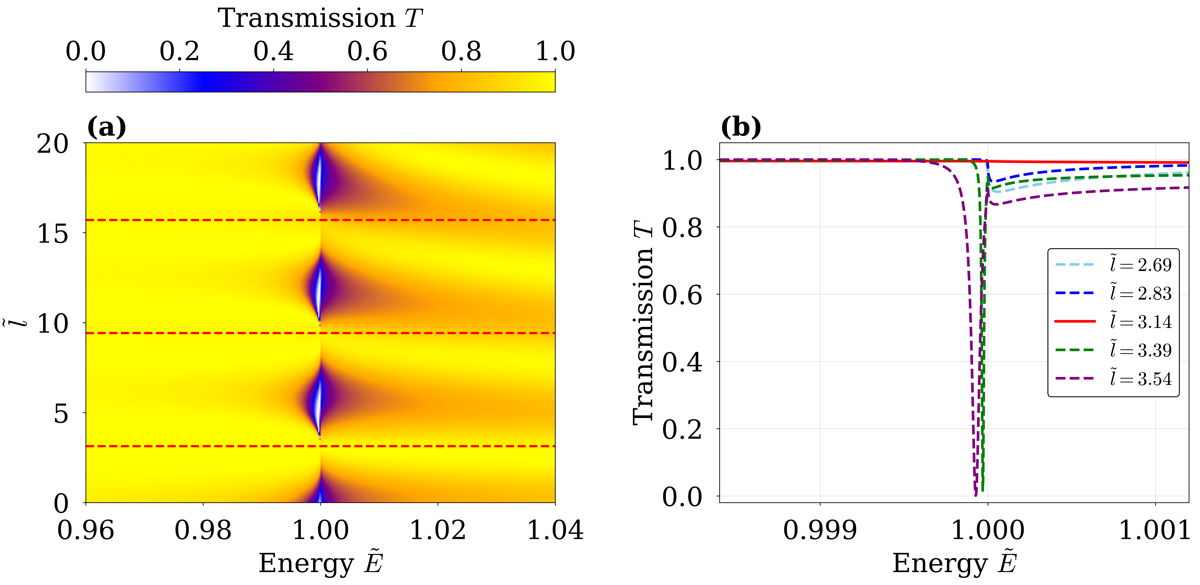}
    \end{minipage}
    
    % --- SECONDA RIGA (41 Canali) ---
    \begin{minipage}[b]{0.8\textwidth}
        \centering
        \includegraphics[width=\textwidth]{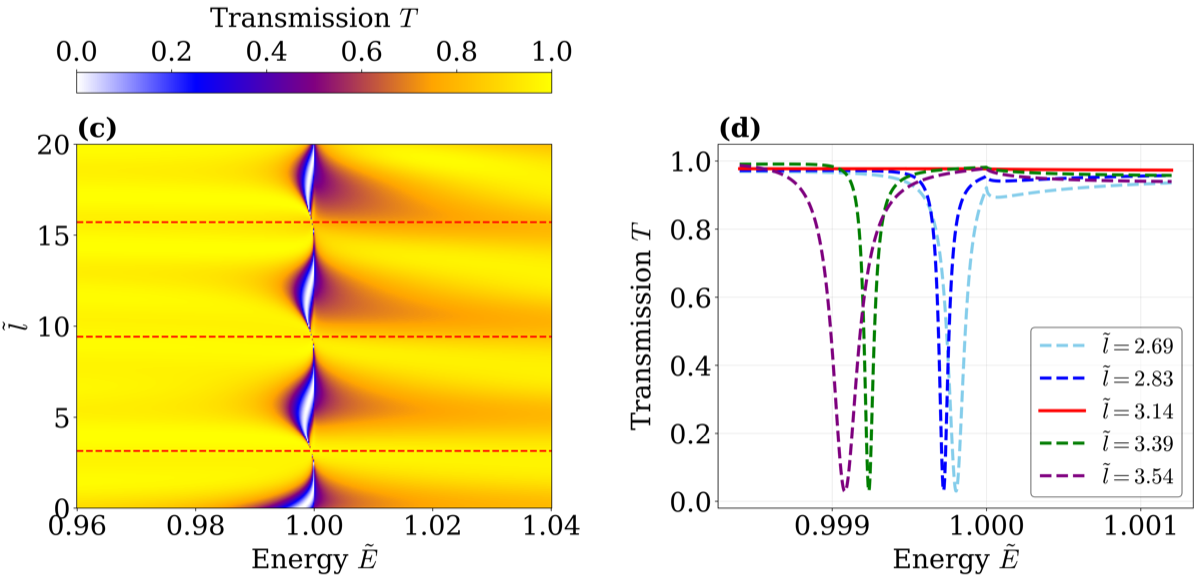}
        \textbf{(c)}
    \end{minipage}
    
    % --- CAPTION UNICA ---
    \caption{Transmission probability $T$ as a function of the incoming particle energy $\tilde{E}$ and inter-impurity distance $\tilde{l}$ for the  case $\tilde{\nu}=0$ and $\tilde{\gamma}=0.424$.
\textbf{[(a), (b)]:} Results obtained within the two-channel approximation.
\textbf{[(c), (d)]:} Numerical solution with 41 channels. The right column [(b), (d)] provides cuts around the resonance region for $\tilde{\gamma}=0.424$.
Note that the analytical prediction of the critical distances, represented by the horizontal red dashed lines, derived within the two-channel approximation, given by Eq.~(\eqref{l_critical_no_static}), is in excellent agreement with the full 41-channel numerical solution. The cuts at fixed $\tilde{l}$ (b)-(d) clearly show the complete disappearance of the Fano resonance at these critical distances (red flat curves), which correspond to the formation of Bound States in the Continuum (BICs).}
    \label{fig:nostatic} 
\end{figure*}

In this subsection, we analyze the transport properties in the absence of a static potential ($\tilde{\nu}=0$).
Figure \ref{fig:T_map_nostatic} displays a plot of the transmission probability as a function of  the drive amplitude $\tilde{\gamma}$ and the energy $\tilde{E}$ of the incoming particle calculated within the two-channel approximation. We observe a single sharp dip in the transmission which we identify as a Fano resonance. The position and width of the Fano resonance evolve with the driving amplitude.
In order to understand the origin of the dip in the transmission, we use the two-channel model to obtain analytical insights. 
The dip originates from the interference between the Floquet modes in the two relevant channels: the continuum spectrum ($n=0$) and the discrete spectrum ($n=-1$), where the coupling is purely dynamical and controlled by the drive amplitude $\gamma$. In this weak-coupling regime (see, e.g., $\tilde{\gamma}=0.35$ in Fig.~\ref{fig:T_map_nostatic}), a two-channel model is sufficient to capture this phenomenon. This is evident from the comparison between the green solid and dashed curves in Fig.~\ref{fig:2_imp_wcr}, representing the 2-channel and 41-channel simulations, respectively: the observed deviations are negligible, confirming that in this regime the physics is dominated by the Fano interference between the open channel $n=0$ and the closed channel $n=-1$. The resulting resonance is almost symmetric with a single dip and no peak ($q \approx 0$), similar to the behavior of the single impurity case without a static potential shown in Fig.~\ref{fig:1_impurity}.
Analytical expressions for the transmission zeros and poles can be derived using an asymptotic expansion near the channel threshold $\tilde{E} \simeq 1$, as detailed in Appendix \ref{app_cal}. The position of the transmission zero is given by:
\begin{equation}
    \label{zero_no_static}
    \tilde{E}_{zt} \simeq 1 - \left[ \frac{\tilde{\gamma}^{2}\sin(\tilde{l})}{8} \right]^{2} .
\end{equation}
This equation indicates that the zero always lies in the energy interval $\tilde{E} <1$ and approaches its upper bound quartically in $\tilde{\gamma}$ as long as the coupling strength $\tilde{\gamma}$ is small. The expression differs from the single impurity case, given by Eq.~(\ref{one_zero_nostatic}), by the factor $\sin(\tilde{l})$ in which the distance of the impurities becomes important. From Fig.~\ref{fig:T_map_nostatic} it is evident that the analytical results for the transmission zeros, obtained within the perturbation approach, are in good agreement with the numerical results across the entire weak-coupling regime.
For the resonance pole, a closed-form expression is not possible to achieve; instead, the complex energy $\tilde{E}_p$ is determined by the transcendental equation:
\begin{equation}
    \label{pole_no_static}
    \tilde{k}_0 (1 + e^{-i\tilde{k}_0 \tilde{l}}) = \frac{\tilde{\gamma}^2\tilde{l}}{8} \sin(\tilde{k}_0 \tilde{l}) ,
\end{equation}
where $\tilde{k}_0 = \sqrt{\tilde{E}_p}$ represents the complex wave vector in the open channel.
A crucial difference compared to the single-impurity case, given by Eq.~(\ref{pole_no_static}), is the dependence on the inter-impurity distance $\tilde{l}$. This distance acts as a control parameter. As previously discussed, the system behaves as a dynamical Fabry-Pérot cavity; in the Floquet picture, this corresponds to an infinite set of coupled static cavities of length $\tilde{l}$. These cavities support modes at energies given by condition \eqref{cavity_modes_E}. Consequently, one expects an interplay between the cavity effects and the Fano interference, which allows for a control of the Fano resonance that would not be possible with a single impurity.
In this regard, from  Eq. \eqref{pole_no_static}, we can identify critical distances $\tilde{l}_{c,p'}$ at which the resonance width vanishes ($\tilde{\Gamma} \to 0$). As we have explained in Sec. \ref{4-single}, a resonance with $\tilde{\Gamma}=0$ is characterized by the fact that the pole is real and coincides with the zero, $\tilde{E}_{p}=\tilde{E}_{zt}=\tilde{E}_{r}$. From Eq. \eqref{pole_no_static} one obtains that it happens if and only if $\cos(\tilde{k}_0 \tilde{l}) = -1$. This condition leads to the critical lengths:
\begin{equation}
    \label{l_critical_no_static}
    \tilde{l}_{c,p'} = (2p'+1)\pi,
\end{equation}
with $p' \in \mathbb{N}$. 
The critical lengths $\tilde{l}_{c,p'}$ correspond to the geometric conditions under which the open channel ($n=0$) supports cavity modes, as follows directly from Eq.~\eqref{cavity_modes_E}. 
For odd integers $p$, the incident particle fails to couple to these cavity modes. Since such a mode is completely decoupled from the continuum, it remains inaccessible in a standard scattering experiment, where the particle is by definition prepared in a continuum state. 
As will be discussed in detail in Subsection~C, this condition implies that the scattering state is degenerate with a BIC. Consequently, in the scattering problem, the BIC appears effectively transparent, leaving no signature in the transmission spectrum due to its complete decoupling from the open channel. This mechanism is unique to the two-impurity configuration, as it strictly requires a spatial separation $l'$ to allow for the phase cancellation between the two scattering events.
At the critical values, Eqs. \eqref{zero_no_static} and \eqref{pole_no_static} predict that both the zero $\tilde{E}_{zt}$ and the real part of the pole $\tilde{E}_r$ collapse onto the energy $\tilde{E} = 1$. Consequently, the pole and zero coincide, leading to a cancellation of the resonant features in the transmission spectrum. Away from these critical points, the zero and the pole remain close but distinct, resulting in a Fano dip ($q \approx 0$), as observed looking at the green solid curve in Fig.  \ref{fig:2_imp_wcr}.
The analytical predictions about the zeros and the poles are confirmed by the numerical simulations presented in Fig. \ref{fig:nostatic}. Figures \ref{fig:nostatic}(a) and \ref{fig:nostatic}(b) show the transmission spectrum in the two-channel approximation and a cut for $\tilde{\gamma}=0.424$. The Fano dip vanishes completely at the predicted critical distances given by Eq.~(\ref{l_critical_no_static}), restoring perfect transmission.
To validate the two channel approximation, we compare these results with the 41 Floquet channels solution, shown in Fig. \ref{fig:nostatic}(c) and \ref{fig:nostatic}(d). The main features of the transmission probability, including the location of the BICs, are well preserved, confirming that the two-channel model captures the essential physics in the weak-coupling regime. The primary deviation is an underestimation of the resonance width in the approximate model; as a result, the Fano resonance re-emerges more gradually in the 41-channel numerical solution as the distance deviates from $\tilde{l}_{c,p'}$.

\subsection{Two impurities with static potential}

\begin{figure*}[t]
    \centering
    
    % --- RIGA SUPERIORE ---
    
    % Pannello (a)
    \begin{minipage}[t]{\textwidth}
        \centering
        % Le doppie graffe {{...}} servono per proteggere le virgole nel nome file
        \includegraphics[width=\textwidth]{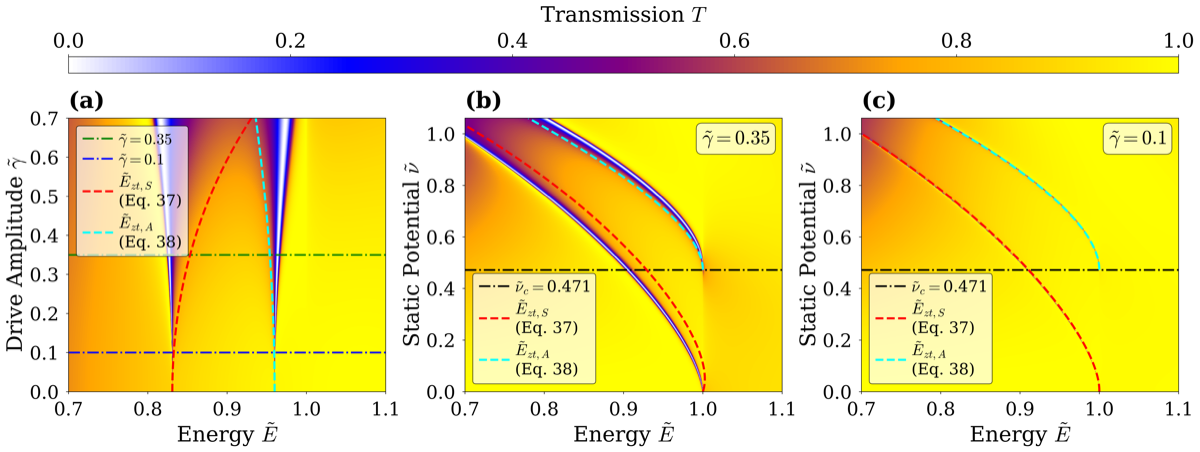}
    \end{minipage}
    
    \begin{minipage}[t]{\textwidth}
        \centering
        \includegraphics[width=\textwidth]{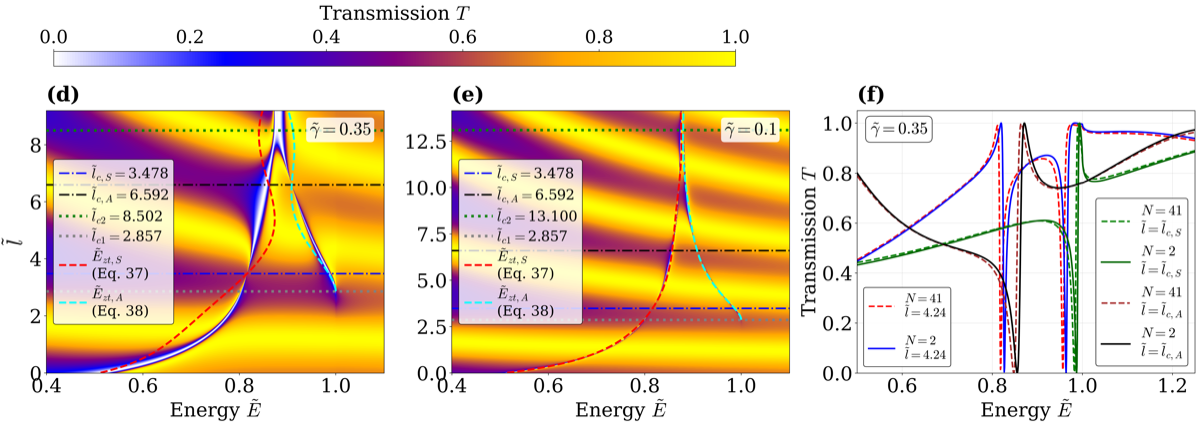}
    \end{minipage}

    \caption{Numerical solution for the system with a static potential ($\tilde{\nu} >0$) in the weak-coupling regime. The solutions in \textbf{(a)}-\textbf{(e)} are obtained within the 2-channel approximation. In \textbf{(f)} the main results are compared to the solution with 41 channels, showing that in the regime considered the 2-channel approximation describes well the main properties of transport. The dashed red and cyan lines indicate the analytical prediction for the Fano resonance zeros, obtained with Eq. \eqref{eq:zero_nu_simm} and \eqref{eq:zero_nu_anti}, respectively. Panel \textbf{(a)} represents the transmission probability  $T$  as a function of drive amplitude $\tilde{\gamma}$ and energy $\tilde{E}$ of the incoming particle with $\tilde{l}=4.24$ and $\tilde{\nu}=0.7$. Horizontal green and black dash-dotted lines indicate selected drive amplitudes representing two distinct regimes: one showing a discrepancy between numerical and analytical results (green) and another where they exhibit good agreement (black).
Panels \textbf{(b)}-\textbf{(c)} represent the transmission probability $T$ as a function of static potential $\tilde{\nu}$ and energy of the incoming particle $\tilde{E}$ for fixed distance $\tilde{l}=4.24$ and drive amplitudes $\tilde{\gamma}=0.35$ for \textbf{(b)} and $\tilde{\gamma}=0.1$ for \textbf{(c)}. The  black horizontal dash-dotted line represents the critical value under which the channel $n=-1$ hosts only one bound state, obtained from \eqref{eq:critical_static}, $\tilde{\nu}_{c}=2/\tilde{l}$. This value of the static potential marks the disappearance of the Fano resonance associated with the vanishing bound state. Panels \textbf{(d)}-\textbf{(e)} represent the transmission probability as function of the distance between the impurities $\tilde{l}$ and energy of the incoming particle $\tilde{E}$, at fixed static potential $\tilde{\nu}=0.7$ and drive amplitude $\tilde{\gamma}=0.35$ (for \textbf{d}) and $\tilde{\gamma}=0.1$ (for \textbf{e}). The dotted horizontal lines represents the lower and upper critical distances. The lower is given by  \eqref{eq:critical_static} $\tilde{l}_{c1}=2/\tilde{\nu}$, the upper is given by Eq. \eqref{eq:upper_critical_l} with $\tilde{\epsilon}=0.025$ for \textbf{(d)} and $\tilde{\epsilon}=0.005$ for \textbf{(e)}. The black horizontal dash-dotted lines represent the distances $\tilde{l}_{c,S}=3.478$ and $\tilde{l}_{c,A}=6.592$ at which the Fano resonances disappear and the BICs form. Panel \textbf{(f)} shows some cuts of the transmission probability at fixed $\tilde{\gamma}, \tilde{\nu}$ and $\tilde{l}$.}
\label{fig:static}
\end{figure*}
In this section, we extend our analysis to the case where a static potential is present ($\tilde{\nu} > 0$). To understand the structure of the transmission probability of this system, it is instructive to first consider the static limit ($\tilde{\gamma} = 0$). In this regime, the Hamiltonian describes a particle interacting with a double potential well formed by two attractive Dirac delta functions.
Such a configuration supports up to two bound states \cite{griffiths2018introduction, ahmed2016revisiting, erman2018scattering}: a ground state with a symmetric wavefunction and a first excited state with an antisymmetric wavefunction (with respect to the center of the cavity $\tilde{l}/2$).
The symmetric bound state exists for any non-vanishing attractive potential $\tilde{\nu}$. In contrast, the antisymmetric bound state exists only if the impurities are sufficiently separated or the potential is sufficiently strong. The condition for the existence of the antisymmetric state is given by:
\begin{equation}
\label{eq:critical_static}
\tilde{l} \tilde{\nu} \ge 2.
\end{equation}
When the periodic drive is activated ($\tilde{\gamma} \neq 0$), in the two channel approximation the discrete bound states associated with the closed channel ($n=-1$) become quasi-bound states coupled to the continuum of the open channel ($n=0$).
Since the channel $n=-1$ supports two static bound states, the resonant scattering pathway in energy splits into two, resulting in a transmission spectrum characterized by a double resonance structure with distinct Fano profiles. This feature is accurately captured by the two-channel approximation, as illustrated in Fig.~\ref{fig:static}(a). A comparison of transmission cuts at  fixed driving amplitude against the full 41-channel solution Fig.~\ref{fig:static}(f) confirms the robustness of the two-channel approximation in this regime. Thus, the physics in the weak-coupling regime is dominated by the Fano interference between the channels $n=0$ and $n=-1$.
The condition given by Eq.~\eqref{eq:critical_static} defines a sharp boundary in the parameter space between the single- and double-resonance regimes. At a fixed distance $\tilde{l}$, we identify a critical potential strength $\tilde{\nu}_c = 2/\tilde{l}$; for $\tilde{\nu} \le \tilde{\nu}_c$, the resonance associated with the antisymmetric state vanishes, and only the symmetric Fano resonance is observed [see Fig.~\ref{fig:static}(b)-(c)]. Equivalently, for a fixed potential $\tilde{\nu}$, this condition defines a lower critical distance $\tilde{l}_{c_1} = 2/\tilde{\nu}$, below which only the symmetric resonance exists, as shown in Fig.~\ref{fig:static}(d)-(e). We define an upper critical distance scale $\tilde{l}_{c_2}$, above which the two Fano resonances effectively merge. Physically, this merging occurs because, at large separations, the two potential wells act as independent scatterers, each supporting a single bound state at the same energy.
It is important to distinguish the nature of these thresholds: while $\tilde{\nu}_c$ and $\tilde{l}_{c_1}$ are exact bounds derived from the existence of the static bound states (which persist in the weak-coupling driven limit), $\tilde{l}_{c_2}$ is an asymptotic scale defined where the energy splitting becomes negligible:
\begin{equation}
\label{eq:upper_critical_l}
    \tilde{l}_{c_2} = \frac{2}{\tilde{\nu}} \ln\left(\frac{\tilde{\nu}^2}{\tilde{\epsilon}}\right).
\end{equation}
In principle, $\tilde{\epsilon}$ represents the energy splitting between the two Fano resonances arising from the symmetric and antisymmetric states. Here, we set this separation equal to the width of the single Fano resonance that emerges from their merging, namely $\tilde{\epsilon} = \Delta \tilde{E} = \tilde{E}_A - \tilde{E}_S$. This quantity is evaluated numerically. 
More information about the Fano resonances observed in Fig. \ref{fig:static} can be analytically obtained within the two-channel approximation by treating the drive as a perturbation on the static bound states, as detailed in Appendix \ref{app_cal}. The energies of the static bound states are $\tilde{E}_{S,A} = -\kappa^{2}_{S,A}$. In the two-channel approximation, as a starting point for the perturbation theory in $\tilde{\gamma}$ the zeros, corresponding also to the resonance energies, and to the real part of the pole, $\tilde{E}_{zt,\alpha}^{0}=\tilde{E^{0}}_{r,\alpha}$ with $\alpha=S,A$, can be written as the energies corresponding to the static bound states shifted in the channel $n=-1$, using the fact that $\tilde{E}_{-1} = \tilde{E} - 1$ one obtains:

\begin{equation}
    \tilde{E}_{r,\alpha}^{0} = 1 - \kappa_{\alpha}^2 \quad (\alpha = S, A),
\end{equation}
It is worth noting that, as explained in appendix \ref{app_cal}  $\kappa_{\alpha}^2$ depends on the inter-impurity distance, and so also does $\tilde{E}_{r,\alpha}^{0}$. This dependence of the energy on the distance $\tilde{l}$ explains the curved trajectories of the Fano resonances in the $(\tilde{E}, \tilde{l})$ plane.
As explained in Appendix (\ref{app_cal}), at the second order in $\tilde{\gamma}$ one obtains:
\begin{align}
\label{eq:zero_nu_simm}
    \tilde{E}_{zt, S} &= \tilde{E}^{0}_{r,S} - \frac{\tilde{\gamma}^2 \mathcal{A}_S}{\tilde{k}_0} \sin(\tilde{k}_0 \tilde{l}), \\
\label{eq:zero_nu_anti}
    \tilde{E}_{zt, A} &= \tilde{E}^{0}_{r,A} + \frac{\tilde{\gamma}^2 |\mathcal{A}_A|}{\tilde{k}_0} \sin(\tilde{k}_0 \tilde{l}),
\end{align}
where $\mathcal{A}_S$ and $\mathcal{A}_A$ are positive amplitude factors related to the static wavefunctions and depending on $\tilde{l}$, whose explicit form is given in appendix \ref{app_cal}. These analytical results corroborate the numerical observations: the first Fano resonance (symmetric state) exhibits a dip-peak structure ($q>0$), since $\tilde{E}_{zt,S} < \tilde{E}^{0}_{r,S}$, while the second (antisymmetric state) exhibits a peak-dip structure ($q<0$) since since $\tilde{E}_{zt,A} > \tilde{E}^{0}_{r,A}$. From Fig.~\ref{fig:static}(a) one can notice that the results obtained within the perturbation approach hold only for small values of the drive amplitude $\tilde{\gamma}$. In particular, the analytical prediction of the Fano resonance zeros are in good agreement with the numerical solution only for $\tilde{\gamma}\lesssim 0.12$, beyond this limit there is a deviation, as it is shown in Fig.~\ref{fig:static}. Unlike the case without static potential in Sec.~\ref{6-wcr}A, one does not have an agreement between the analytical and numerical results across all the weak-coupling regime, since here the first non-zero term of the perturbation is the one with the second power of $\tilde{\gamma}$, while for the case with no static potential the fourth power. Here we expect to have a better agreement considering higher order corrections in the perturbation theory.

The lifetime of these resonances is governed by the interference between the decay channels. We derive, using a perturbation approach in $\tilde{\gamma}$, within the two-channel approximation, compact expressions for the coefficients of the imaginary part of the poles (more details are provided in appendix \ref{app_cal}):
\begin{align}
    \tilde{\Gamma}_S &= 4\tilde{\gamma}^2\frac{|\psi_{S}(0)|^2}{\tilde{k}_{0}} \cos^2\left(\frac{\tilde{k}_0 \tilde{l}}{2}\right), \\
    \tilde{\Gamma}_A &= 4 \tilde{\gamma}^2\frac{|\psi_{A}(0)|^2}{\tilde{k}_0} \sin^2\left(\frac{\tilde{k}_0 \tilde{l}}{2}\right),
\end{align}
where $\tilde{k}_0 = \sqrt{\tilde{E}_{0}}$ is the wave vector in the open channel ($n=0$).
The derived expressions for $\tilde{\Gamma}_{S,A}$ directly determine the existence conditions for BICs, defined by a vanishing decay width. As previously discussed, these are non-decaying states with infinite lifetimes: their complete decoupling from the continuum ensures that the escape probability is zero, effectively trapping the particle within the system. One can find the conditions for these critical distances at which the Fano resonances disappear, for the symmetric state $\tilde{l}_{c,S}$ and the antisymmetric one $\tilde{l}_{c,A}$. 
 For the symmetric state, the condition $\tilde{\Gamma}_S = 0$ implies $\tilde{k}_0 \tilde{l}_{c,S} = (2p'+1)\pi$ with $p' \in \mathbb{N}$. Conversely, for the antisymmetric state, the condition $\tilde{\Gamma}_A = 0$ requires $\tilde{k}_0 \tilde{l}_{c,A} = 2p'\pi$ with $p' \in \mathbb{N}$. 
These conditions correspond precisely to the points in Fig.~\ref{fig:static}(d)-(e) where the Fano resonances disappear on the two branches originating from the symmetric and antisymmetric states. In Fig.~\ref{fig:static} cuts of the transmission probability at those critical distances show that one Fano resonance disappears (see the black and green curve in Fig.~\ref{fig:static}(f), where only one Fano structure is present).
We will give more details on the BICs in the next subsection.

\subsection{Bound States in the Continuum (BICs)}

\begin{figure*}[t]
    \centering
    
    % --- Pannello Sinistro (a) ---
    \begin{minipage}[b]{\textwidth}
        \centering
        \includegraphics[width=\textwidth]{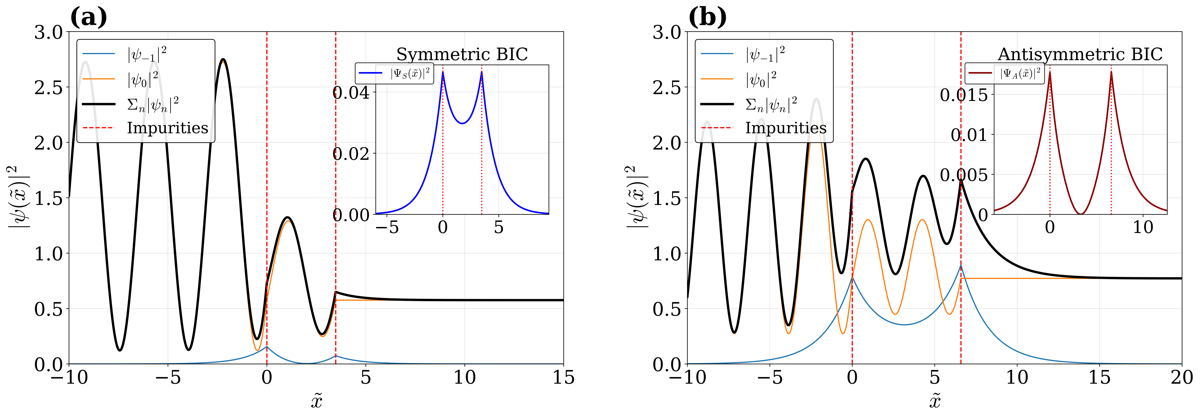}
    \end{minipage}

    % --- Caption Principale ---
    \caption{Probability density $|\psi(x)|^2$ computed within the two-channel approximation. The solid black lines represent the time-averaged probability density for the scattering problem, obtained from the solution to the Schrödinger equation given by Eq.~\eqref{Floquet_solution}. The blue and orange solid lines denote the probability densities associated with the individual Floquet modes in channels $n=-1$ and $n=0$, respectively. The insets display the probability density of the solutions for the problem without an incoming particle. The static potential strength is set to $\tilde{\nu}=0.7$, and the system is evaluated at the critical parameters corresponding to the emergence of bound states in the continuum (BICs). 
\textbf{(a)} Profiles evaluated at the critical distance $\tilde{l}_{c,S}\simeq 3.48$ and energy $\tilde{E}_{c,S} \simeq 0.82$. At these specific parameters, the Fano resonance originating from the symmetric bound state of the static potential completely disappears. The inset illustrates the corresponding perfectly symmetric BIC. 
\textbf{(b)} Profiles evaluated at the critical distance $\tilde{l}_{c,A}\simeq 6.59$ and energy $\tilde{E}_{c,A} \simeq 0.91$. At this configuration, the Fano resonance associated with the antisymmetric static bound state vanishes. The inset shows the corresponding purely antisymmetric BIC.
In the scattering problem, the solutions at these critical distances are running waves with transmission probabilities $T \simeq 0.57$ (symmetric case) and $T \simeq 0.77$ (antisymmetric case). The  transmission zeros vanish with the Fano resonances because the bound states in the $n=-1$ channel become completely decoupled from the continuum ($\Gamma=0$), preventing the resonant capture of the wave in channel $n=0$.} 
    \label{fig:BICs}
\end{figure*}
While the existence of BICs is well-established in the literature~\cite{vonNeumann1929bic,friedrich1985interfering, longhi2013floquet,wang2024optical,tang2025dynamically}, their direct visualization remains challenging~\cite{plotnik2011experimental,hsu2016bound,liu2025high}. These are states embedded within the continuum spectrum but remain completely decoupled from it and are spatially localized. In a scattering problem, one must look for these states where the system admits a resonance with a purely real pole. This implies that the decay rate $\Gamma$ of the resonant state into the continuum is zero, and thus the state acquires an infinite lifetime $\tau$. When such state with an infinite lifetime is strictly localized (i.e., its wavefunction is square-integrable), it is classified as BIC. 
Looking at the Fano resonance formulas in Eqs.~\eqref{zero_Fano} and \eqref{pole_Fano}, one can see that when $\Gamma=0$, the pole and the zero of the Fano resonance exactly coincide; they both occur at the energy $E_{r}$. As a result, the resonance profile completely disappears from the transmission spectrum in the scattering problem. As shown in Figs.~\ref{fig:nostatic} and \ref{fig:static}, we have identified specific conditions for the distance between the two impurities in the weak-coupling regime, both numerically and via the perturbative approach, that lead to the complete disappearance of the Fano line shape. 
These states are therefore strong candidates to be BICs. The crucial next step is to confirm that they are indeed spatially localized. Since states perfectly decoupled from the continuum are inherently invisible in the scattering framework studied previously, one must proceed by solving the homogeneous Schrödinger equation, without an incoming scattered particle. For the Floquet channels, this requires adopting an ansatz containing exclusively outgoing waves in the external regions, with no incident term:

\begin{equation}
\label{ansatz_noincoming}
    \psi_n(x) = \begin{cases} 
    r_n e^{-ik_n x} & x < 0  \\
    A_{n} e^{ik_n x} + B_{n} e^{-ik_n x} & 0 \le x \le l \\
    t_n e^{ik_n x} & x > l 
    \end{cases}
\end{equation}
By imposing the boundary conditions at the impurity sites $x=0$ and $x=l$, as for the scattering problem in Sec.~\ref{3-T}, one obtains the same equations \eqref{eq:LinearSystem}(a)-(d) without the term related to the incoming particle $\delta_{n,0}e^{ik_{n}x}$. This leads to a linear homogeneous system for the unknown coefficients 
$\mathbf{x} = [ \{r_n\}, \{A_n\}, \{B_n\}, \{t_n\} ]^T$ 
for $n \in [-n_{max}, n_{max}]$. Where $n_{max} \in \mathbb{N}$ is the cut-off for the Floquet channel to include in the numerical solution, that can be written in a matrix formulation:
\begin{equation}
    M(E) \cdot \mathbf{x} = 0,
\end{equation}
This system admits non-trivial solutions as long as the determinant of $M(E)$ is non-zero. Crucially, while this matrix is structurally identical to the coefficient matrix of the scattering problem. The poles of the transmision probability of the problem with the incoming particle of Sec. \ref{2-model}, as explained in Appendix \ref{app_cal} are solutions to the equation
\begin{equation}
    \det[M(E_p)] = 0.
\end{equation}
As discussed in Sec.~\ref{6-wcr}B, for the system with a static potential, the two Fano resonances originate from the pre-existing bound states of the undriven problem. We demonstrated the disappearance of these Fano resonances at specific critical distances within the weak-coupling regime [shown in Figs.~\ref{fig:static}(d) and \ref{fig:static}(f)].
Interestingly, at these critical distances, the scattering and homogeneous problems exhibit different solutions to the Schrödinger equation. Specifically, at the critical length $\tilde{l}_{c,S}$, the scattering problem yields a running wave with a transmission probability $T \simeq 0.57$, as depicted in Fig.~\ref{fig:BICs}(a). Conversely, solving the problem without an incoming particle for the exact same energy, drive parameters, and distance reveals a symmetric localized state. This perfectly confined state, representing a BIC, is shown in the inset of Fig.~\ref{fig:BICs}(a). Similarly, at the critical length $\tilde{l}_{c,A}$, the scattering problem exhibits a running wave with a transmission probability $T \simeq 0.77$ [Fig.~\ref{fig:BICs}(b)], whereas removing the incoming particle reveals a localized state with an antisymmetric profile, which again corresponds to a BIC. 
Although these states are strictly invisible in the transport problem, their existence is indirectly proven by the disappearance of the Fano resonance. These BICs, inheriting the parity of the underlying static bound states, must not be confounded with the probability density observed in the Floquet channel ($n=-1$). To understand why this channel exhibits a non-zero probability density even when the Fano resonance is completely suppressed, as shown in Fig.~\ref{fig:BICs}, it is crucial to distinguish between two concurrent physical mechanisms: resonant capture and direct scattering.
The mechanism of resonant capture takes place when the energy of the incoming particle matches the energy of the bound states of the static problem, $\tilde{E}_{r,\alpha}$. It leads to the Fano resonance, which is characterized by a zero and a pole related to the resonant energy. Since the pole is generally complex, its imaginary part, $\Gamma$, represents the decay rate of these bound states into the continuum. As long as this coupling $\Gamma$ is non-zero, the capture can take place and is temporary, because the state eventually decays back into the continuum. However, at the critical distances, these bound states completely decouple from the continuum ($\Gamma = 0$), meaning the resonant capture cannot take place. Consequently, the Fano resonance vanishes entirely from the scattering problem. This phenomenon conclusively identifies the BIC as a perfect spectral dark state, fully decoupled from the continuum \cite{hsu2016bound}. 
On the other hand, when an incident wave in the continuous channel ($n=0$) impinges on the oscillating impurities, a fraction of the probability amplitude is scattered into the $n=-1$ channel directly upon impact. This generates the probability density associated with $\psi_{-1}(x)$ seen in Fig.~\ref{fig:BICs}(a)-(b), which exists simply because the time-dependent potential is continuously acting on the particle. Therefore, the residual, low-level probability density observed in the $n=-1$ channel during scattering is not the BIC itself, but merely the unavoidable footprint of the direct scattering process.
Finally, we note that the study in this weak-coupling regime has been carried out using the 2-channel approximation. As established earlier, this truncation guarantees numerical convergence, a robustness that extends to the evaluation of the BICs.

\begin{figure}[h!]
    \centering
    \includegraphics[width=0.45\textwidth]{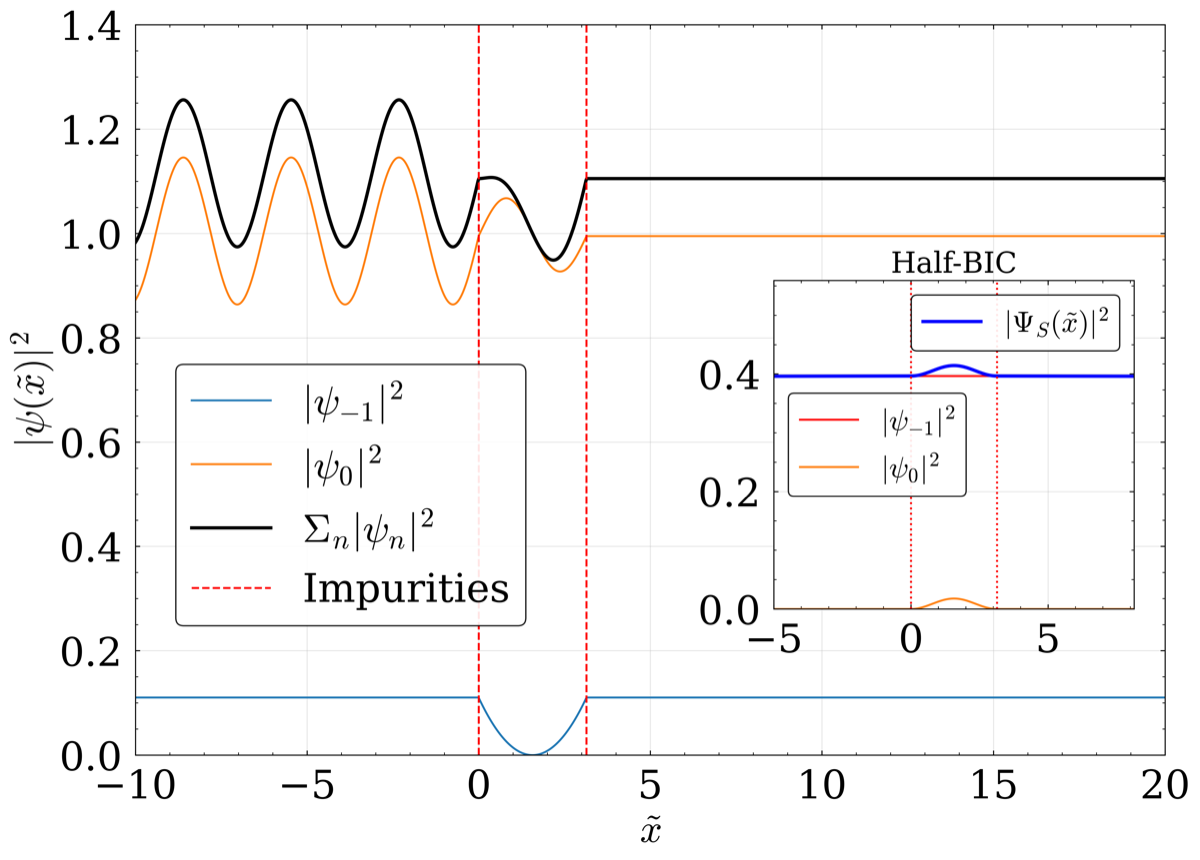}
    \caption{Probability density $|\psi(x)|^2$ computed within the two-channel approximation. The solid black lines represent the time-averaged probability density for the scattering problem, obtained from the solution to the Schrödinger equation given by Eq.~\eqref{Floquet_solution}. The blue and orange solid lines denote the probability densities associated with the individual Floquet modes in channels $n=-1$ and $n=0$, respectively. The insets display the probability density of the solutions for the problem without an incoming particle. The static potential strength is set to $\tilde{\nu}=0$, and the system is evaluated at the critical parameters $\tilde{E}=1$ and $\tilde{l}=\tilde{l}_{c,1}=\pi$ at which the Fano resonace disappears in the problem with no static potential, being pushed on the threshold. In the scattering problem there is a running wave that is transmitted in region III almost perfectly $T \simeq 1$. In channel $n=-1$ (light blue curve) there is zero energy state or half bound state due to threshold $k_{-1}=0$ completely decoupled from the channel $n=0$ ($\Gamma=0$), while in the problem without incoming particle one finds the Half-BIC.}
    \label{fig:half_bic}
\end{figure}

For the system without a static potential, we similarly find conditions leading to the disappearance of the Fano resonance at specific critical distances, specifically when $\tilde{l}$ is an odd multiple of $\pi$, as shown in Fig.~\ref{fig:static}. However, the nature of the underlying state in this regime differs fundamentally from the static potential case. Here, the collapse of the imaginary part of the Fano pole ($\Gamma \to 0$) occurs exactly at the energy $\tilde{E}=1$. This corresponds to the threshold at which the $n=-1$ channel opens up, meaning its wave number vanishes ($k_{-1}=0$). Consequently, this channel does not contribute to the probability current but yields a constant probability density in regions I and III, as seen in Fig.~\ref{fig:half_bic}. In the equivalent homogeneous problem (without an incoming particle), shown in the inset of Fig.~\ref{fig:half_bic}, the system supports a state with an infinite lifetime. Unlike the previously discussed BICs, this state is not spatially localized; rather, it is a Half-bound state \cite{neufeld2022asymptotically} embedded in the continuum, yet completely decoupled from it like a standard BIC—therefore acting as a 'Half-BIC'.
According to standard 1D scattering theory, total reflection ($R = 1$) is expected for incident quantum particles with energies approaching a channel threshold, where the momentum vanishes 
\cite{ senn1988threshold}. However, this 
paradigm breaks down in the presence of half-bound states: zero-energy 
resonances characterized by a non-vanishing, constant asymptotic 
probability amplitude and a zero spatial derivative. Rather than 
reflecting, an incident wave can perfectly couple to these flat, 
non-square-integrable states, which induces macroscopic transport 
anomalies and gives rise to the so-called ``zero-reflection paradox'' 
($R=0$) \cite{kiers1996scattering, ahmed2016paradoxical}. While these states are typically studied through the ad hoc engineering of static potentials \cite{hinton1991embedded, patient2021supersymmetry, neufeld2022asymptotically}, we demonstrate here that they can be dynamically generated using local periodic drives. 
Similar to the static potential scenario, the constant probability density observed in the $n=-1$ channel during the scattering process [Fig.~\ref{fig:half_bic}] is merely the footprint of the direct scattering mechanism. Because the Fano resonance is suppressed, the Half-BIC remains completely decoupled from the continuum. The incident wave is not captured by the half bound state since its coupling to the continuum spectrum is zero ($\Gamma=0$), which leads to an almost perfect transmission ($T \simeq 0.995$) at the threshold in the scattering problem.

\section{Fano Resonances Beyond the Weak-Coupling Regime}

\label{7-bwcr}

Beyond the weak-coupling regime ($\tilde{\gamma} > 0.7$), the two-channel approximation proves insufficient to fully capture the complexity of the transmission spectrum. As the drive amplitude increases, the enhanced coupling between Floquet modes brings higher-order channels into play, giving rise to richer spectral features.  Due to the complexity of the multi-channel coupling, an analytical treatment is no longer feasible. Here we focus our attention on the region II and III of the transmission spectrum indicated in Fig.~\ref{fig:overview}.

\subsection{Intermediate-Coupling Regime}

\begin{figure}[h]
    \centering
    
        \includegraphics[width=0.5\textwidth]{{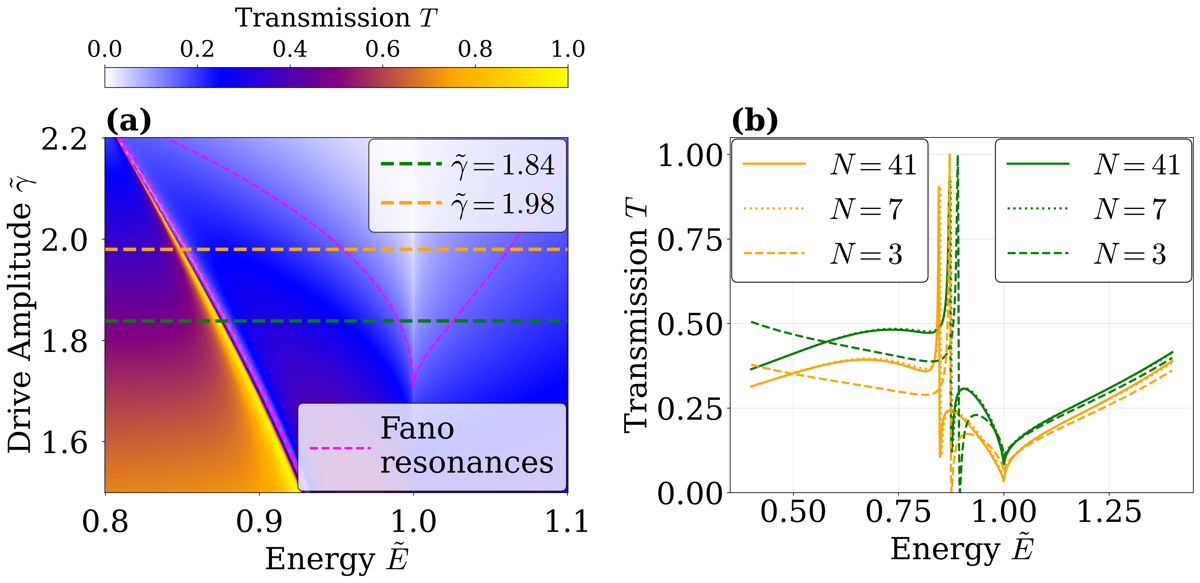}}

    \caption{Transmission probability in the intermediate-coupling regime. The system is evaluated for an inter-impurity distance $\tilde{l}=3.54$ in the absence of a static potential ($\tilde{\nu}=0$).
\textbf{(a)} Transmission probability $T$ as a function of the incident particle energy $\tilde{E}$ and the drive amplitude $\tilde{\gamma}$. The horizontal dashed lines indicate the specific drive amplitudes extracted for the cuts in panel (b). The magenta dashed contours, numerically obtained by tracing the boundary where $T \le 0.15$, explicitly highlight the regions of probability drop associated with the two Fano resonances.
\textbf{(b)} Transmission probability cuts extracted at $\tilde{\gamma}=1.84$ (green lines) and $\tilde{\gamma}=1.98$ (orange lines). The panel compares the minimal model restricted to three non symmetric channels ($n=-2, -1, 0$, dashed lines), the 7-channel approximation (dotted lines), and the 41-channel solution (solid lines). The good agreement validates the robustness of the 7-channel approximation.}
\label{fig:restoredDouble}
\end{figure}

A double resonance structure emerges in the absence of a static potential ($\tilde{\nu}=0$), as shown in Fig.~(\ref{fig:restoredDouble}), which corresponds to the region II of the transmission spectrum highlighted in Fig.~\ref{fig:overview}(a) of Sec.~\ref{5-overview}. This behavior marks a qualitative difference from the single-impurity case, where only single resonances are present, even beyond the weak-coupling regime.
The origin of this double structure lies in the specific Floquet channels participating in the scattering process. In this intermediate- coupling regime ($1.8 \le \tilde{\gamma} \le 2.5$), the continuum of the open channel ($n=0$) couples significantly not only to the first closed channel ($n=-1$) but also to the second closed channel ($n=-2$). Consequently, the incoming particle interacts simultaneously with two distinct bound states, corresponding to the two Floquet modes in the channels $n=-1$ and $n=-2$.
From a physical perspective, the inclusion of these three channels ($n=0, -1, -2$) is sufficient to qualitatively capture the double resonance, as shown in Fig. \ref{fig:restoredDouble}(b). However, in this regime, the numerical convergence of the solution is achieved by increasing the number of Floquet channels to seven (from $n=-3$ to $n=3$), as shown in Fig.~\ref{fig:restoredDouble}(b).
This stands in contrast to the single-impurity problem, where the transmission spectrum in the same region is characterized by a single Fano resonance and is still physically described by the two-channel approximation ($n=0, -1$), with higher-order channels contributing only to numerical refinement rather than new physical features.
This scenario, for the two-impurity problem without static potential, is analogous to the case with a static potential offset ($\tilde{\nu} \ne 0$), where the system supported two pre-existing bound states (symmetric and antisymmetric). Here, however, the two discrete states are dynamically generated by different Floquet modes. The Fano interference between the single continuum pathway ($n=0$) and these two discrete pathways ($n=-1,-2$) leads to a double Fano resonance profile. A notable difference from the weak-coupling limit is that the transmission minima do not reach exactly zero; which indicates that higher-order coupling effects generate a non-negligible background that prevents perfect destructive interference.

\subsection{Cavity-enhanced Fano Regime}

\begin{figure*}[t]
    \centering

     \begin{minipage}[t]{\textwidth}
        \centering
            \includegraphics[width=0.65\textwidth]{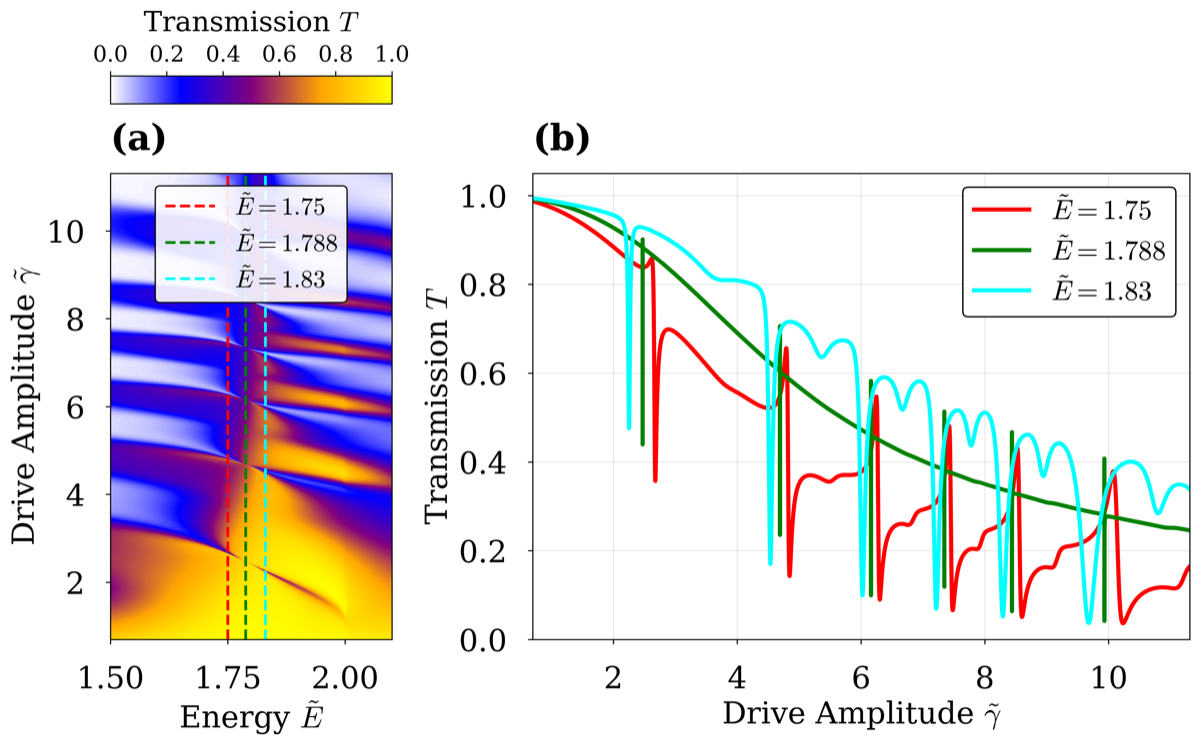}
     \end{minipage}
          \begin{minipage}[t]{\textwidth}
        \centering
            \includegraphics[width=\textwidth]{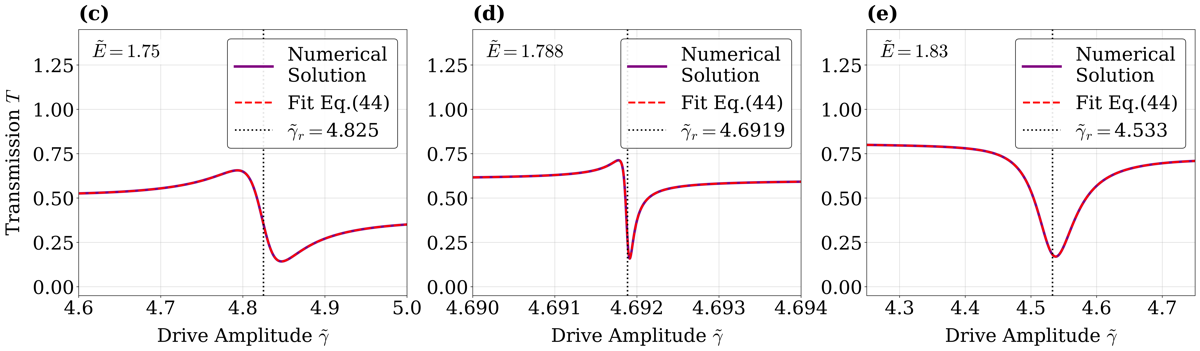}
     \end{minipage}
    
    \begin{minipage}[t]{\textwidth}
        \centering
            \includegraphics[width=\textwidth]{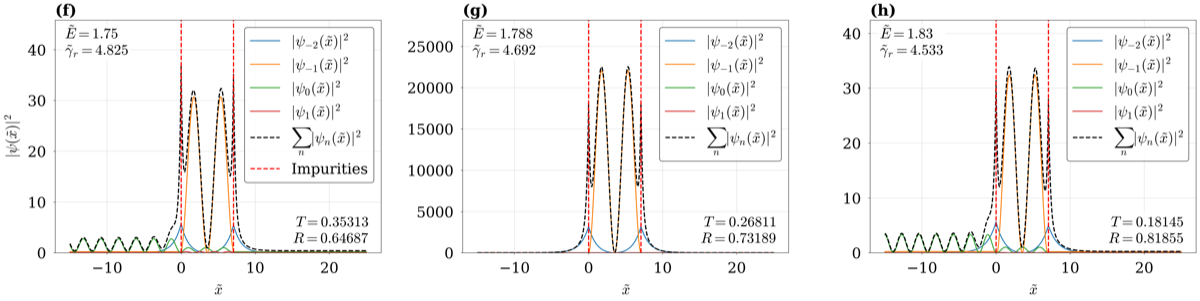}
     \end{minipage}

    \caption{Numerical solution with 41 Floquet channels in region III of the transmission spectrum [Fig.~\ref{fig:overview}(a)], calculated for an inter-impurity distance $\tilde{l}=7.07$ without a static potential ($\tilde{\nu}=0$). \textbf{(a)} Transmission probability as a function of the incoming particle energy $\tilde{E}$ and the drive amplitude $\tilde{\gamma}$. The vertical dashed lines indicate the energies $\tilde{E}=1.75$, $\tilde{E}=1.83$, and the cavity mode energy $\tilde{E}_{-1,2}=1.788$. \textbf{(b)} Cuts of the transmission probability as a function of the drive amplitude $\tilde{\gamma}$, evaluated at the energies indicated by the vertical lines in panel (a). \textbf{(c)-(e)} Fits of the Fano resonances for the transmission probability as a function of $\tilde{\gamma}$ at the three selected energies of panel (a), using Eq.~\eqref{fano_fit}. \textbf{(f)-(h)} Corresponding probability densities and time-averaged total probability density for the same energies.}
    \label{fig:RegionIII}
\end{figure*}

An important aspect emerging beyond the weak-coupling regime—particularly within region III of the transmission spectrum in Fig.~\ref{fig:overview}(a)—is the onset of cavity effects. These effects compete with the Fano interference, leading to novel features in the transmission profile. Region III is characterized by the condition that cavity modes, given by Eq.~(\ref{cavity_modes_E}) are supported by two different propagating channels, specifically $n=0$ and $n=-1$. The energies of these cavity modes are $\tilde{E}_{0,3} \simeq 1.774 $ and $\tilde{E}_{-1,2} \simeq1.788$. 
As shown in Fig.~\ref{fig:RegionIII}(a), an interesting interplay occurs when a cavity mode overlaps with the Fano interference pattern. Large Fano resonances arise in this area of the transmission spectrum, far from the energy of the cavity modes. However, as the incident particle energy approaches the energy of the cavity modes, in particular of the channel $n=-1$, the spectral profile of the Fano resonances is heavily affected. Notably, exactly at the cavity mode energy (highlighted by the green dashed line at $\tilde{E} \approx 1.788$), the Fano resonances undergo a drastic narrowing, becoming remarkably sharp.
Figure~\ref{fig:RegionIII}(b) presents three cuts of the transmission probability as a function of the drive amplitude. The transmission profile is governed by two combined effects: a slow background decay of $T$ with increasing $\tilde{\gamma}$, upon which several sharp Fano resonances are superimposed. As previously observed, the sharpest resonances emerge precisely at the energy of the cavity mode associated with the $n = -1$ Floquet channel. 

To quantitatively analyze this behavior, we isolate a single Fano resonance for each energy cut—specifically, which consistently falls within the drive amplitude interval $\tilde{\gamma} \in [4.5, 4.9]$. We perform a non-linear fit of these isolated resonances, shown in Figs.~\ref{fig:RegionIII}(c)-(e), using a superposition of a  Fano profile and a second-order polynomial function:
\begin{equation}
    \label{fano_fit}
    \begin{split}
        T_{\text{Fano}}(\tilde{\gamma}) &= A\frac{(q \tilde{\Gamma}/2 + \tilde{\gamma} - \tilde{\gamma}_{r})^2}{(\tilde{\Gamma}/2)^2 + (\tilde{\gamma} - \tilde{\gamma}_{r})^2} \\
        &\quad + a_{0} + a_{1}(\tilde{\gamma}-\tilde{\gamma}_{r}) + a_{2}(\tilde{\gamma}-\tilde{\gamma}_{r})^{2}.
    \end{split}
\end{equation}
This specific functional form is employed to cleanly decouple the intrinsic resonance features from the transmission baseline. While the first term captures the pure Fano lineshape, the polynomial tail acts as a local Taylor expansion of the background around the resonance point. This procedure ensures a highly accurate extraction of the key physical parameters: the asymmetry parameter $q$, the resonance position $\tilde{\gamma}_r$, and the width $\tilde{\Gamma}$.
The results of the fit are detailed in the caption of Fig.~\ref{fig:RegionIII}. Notably, the asymmetry parameter $q$ is negative for all the analyzed Fano resonances, resulting in a characteristic peak-dip lineshape. The resonance corresponding to the cavity mode energy ($\tilde{E}_{-1,2}$) exhibits the narrowest width, $\tilde{\Gamma} \simeq 1.1 \times 10^{-4}$. This value is two orders of magnitude smaller than the widths extracted at the other two energies, which are detuned respectively below and above the cavity mode. Consequently, the associated time delay at this specific energy is remarkably enhanced ($\tilde{\tau} \simeq 9200$), being roughly two orders of magnitude larger than the time delays observed at the other energies ($\tilde{\tau} \simeq 20$ for $\tilde{E}=1.75$, and $\tilde{\tau} \simeq 14$ for $\tilde{E}=1.83$). 

The physical mechanisms underlying the difference in resonance width of two orders of magnitude (and consequently in time delay) is the effect of the cavity modes in the open Floquet channels in capturing and delaying the particle, together with the bound states in the closed Floquet channels. An incoming particle in the $n=0$ Floquet channel can be scattered in all the others Floquet channels, since these are coupled at the impurity sites. Fano interference typically arises from capture of the particle into bound states hosted within channels with $n \le -2$.
However, in the specific energy interval considered, the trapping mechanism is enhanced by coupling to cavity modes lying within the open channels $n=-1$ and $n=0$. This is visually demonstrated in Figs.~\ref{fig:RegionIII}(f)-(h), which display the spatial probability densities associated with dominant individual Floquet modes $|\psi_n(x)|^2$. They also show the time-averaged total probability density, $\langle |\psi(x,t)|^2 \rangle_T$, of the full scattering solution derived from Eq.~\ref{psi_tot}. This average, calculated over one driving period $T$, is mathematically equivalent to the sum of the individual modes densities:
\begin{equation}
\label{Time-av-density}
    \langle |\psi(x,t)|^2 \rangle_T = \frac{1}{T}\int_{0}^{T}|\psi(x,t)|^{2} dt = \sum_{n}|\psi_{n}(x)|^{2}.
\end{equation}
Across all three energy cases falling near the cavity-mode supporting ranges for channels $0$ and $-1$, the plots reveal clear cavity-mode structures—specifically, a second-order cavity mode ($p=2$) in channel $n=-1$ and a third-order mode ($p=3$) in $n=0$. Crucially, we observe that the peak probability density in channel $n=-1$ is substantially higher than in other channels, including the contributing discrete channel ($n=-2$). This high modal density enhancement indicates that channel $n=-1$ plays the dominant role in particle trapping.
While the numerical simulation incorporates a large number of channels (41) in order to describe the general shape of the transmission probability (with all the channels giving a non-negligible contribution), the spatial profile of the time-averaged total wavefunction density at the Fano resonance energy is governed almost exclusively by the spatial structure of these three key channels ($n=-1$, $n=0$, and the contributing $n \le -2$). We have explicitly included the bound state in channel $n=-2$ in the plots as it generally exhibits the highest density enhancement among channels belonging to the discrete spectrum. Alongside it, we plot the two open channels supporting cavity modes ($n=-1$ and $n=0$). Density enhancement is most pronounced in channel $n=-1$ and less so in $n=0$, yet both display the characteristic cavity-mode signature: strong localization in the region between the two impurities, decaying outside that region to constant asymptotic values. Contribution from the remaining channels at the Fano resonances energies and drive amplitudes, both discrete and within the continuum, only becomes significant at substantially higher values of the drive amplitude.
The main result here is that at the specific energy matching the cavity mode $\tilde{E}_{-1,2}$, shown in Fig.~\ref{fig:RegionIII}(g), we observe the highest values of total density enhancement, much greater than the cases of the energies in Fig.~\ref{fig:RegionIII}(f) and Fig.~\ref{fig:RegionIII}(h). This means that the intrinsic Fano interference (capture into discrete bound states) is locally assisted and enhanced by coupling to the cavity mode supported within the open channel $n=-1$. This resonant cooperative effect drastically increases the effective time delay, leading directly to the corresponding sharp narrowing observed in the spectral width of the Fano resonance.

The interplay of Fano interference with the cavity modes enables the engineering of states with significantly enhanced lifetimes, which manifest as remarkably narrow Fano resonances even at high drive amplitudes. As discussed in Sec.~\ref{4-single} for the single-impurity case, these long-lived states act as quasi-BICs. However, a fundamental limitation of the single-impurity system is that such quasi-BICs—yielding time delays on the order of $\tilde{\tau} \simeq 10^{3}$—are strictly confined to the weak-coupling regime. As shown in Fig.~\ref{fig:1_impurity}(a), increasing the drive amplitude for a single impurity inevitably broadens the Fano resonances, leading to a drastic reduction in the time delay. In contrast, the two-impurity configuration leverages cavity-enhanced Fano interference to overcome this limitation. This system provides a robust mechanism to engineer even longer-lived states at significantly larger drive amplitudes and across a broader range of incident energies.

\begin{figure}[h]
    \centering
    \includegraphics[width=0.48\textwidth]{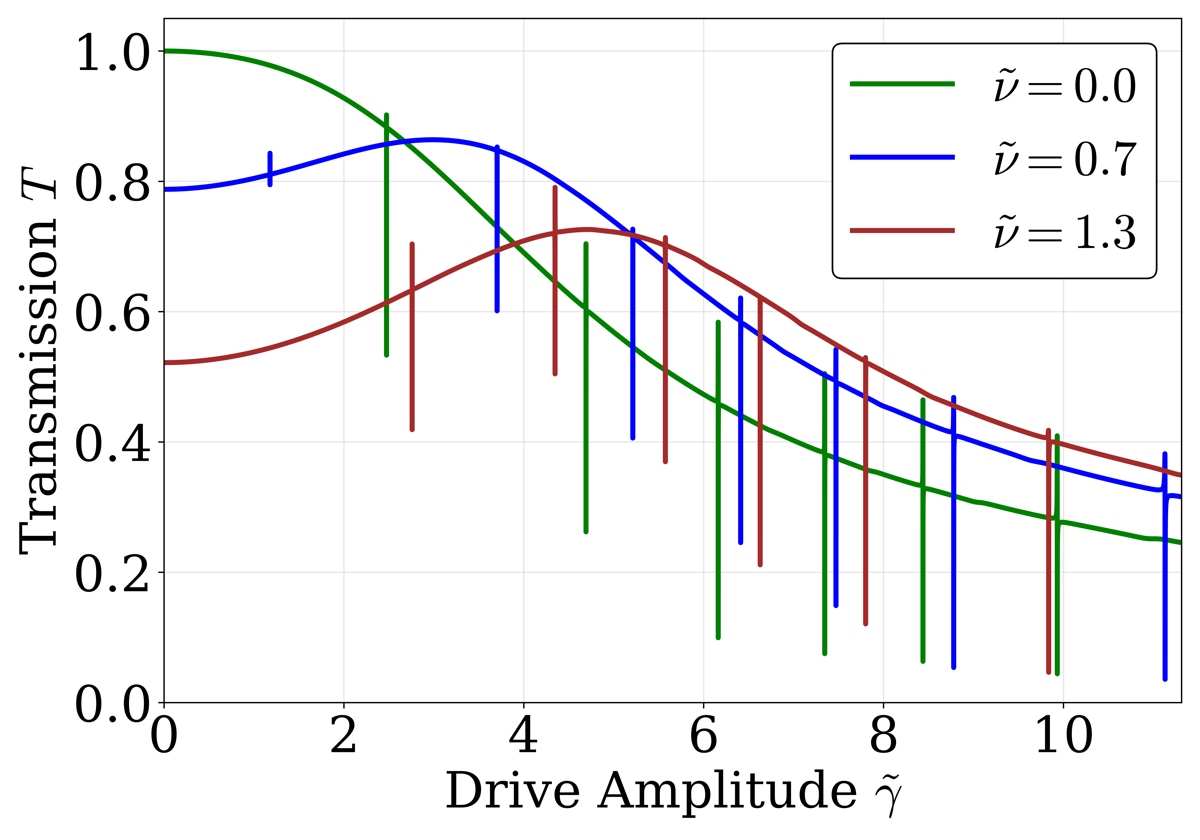} 
    \caption{Transmission probability as a function of drive amplitude $\tilde{\gamma}$ for fixed energy $\tilde{E}=1.774$ and distance $\tilde{l}=7.07$. The green curve represents the case without static potential ($\tilde{\nu}=0$), while the blue curve shows the case with a static potential $\tilde{\nu}=0.7$, the brown curve the case with $\tilde{\nu}=1.3$ .}
    \label{fig:ComparisonregionIII}
\end{figure}

Finally, in Fig. \ref{fig:ComparisonregionIII}, we investigate the interplay between a static potential and the periodic drive by comparing the case ($\tilde{\nu}=0$) with the cases including static potential ($\tilde{\nu}=0.7$ and $\tilde{\nu}=1.3$).
For the case without a static potential ($\tilde{\nu}=0$), the system exhibits two main features. First, the background transmission monotonically decreases starting from $T=1$ at zero drive amplitude. This occurs because, as the drive amplitude increases, the incident particle is scattered into different Floquet channels. These include closed channels, where the particle can only be reflected, and open channels, which allow both transmission and reflection. The participation of more Floquet channels at higher drive amplitudes inevitably leads to a decrease of the transmission probability. Second, on top of this decreasing background, sharp Fano resonances appear. At fixed incident energies, these dips occur exactly at the drive amplitudes for which the system admits quasi-bound states.
When the static potential is non-zero ($\tilde{\nu}=0.7$ and $\tilde{\nu}=1.3$), we observe a clear breakdown of the background monotonicity and a shift in the position of the Fano resonances. The shift of the resonances occurs because the static potential acts as an attractive well that modifies the positions of the underlying quasi-bound states.
Furthermore, at zero drive amplitude ($\tilde{\gamma}=0$), the static potential reduces the initial transmission probability from $T=1$ to $T < 1$. As the drive is activated, the transmission is initially assisted by the oscillation of the two impurities. In this intermediate window, the scattering in the Floquet channels effectively help the particle overcome the static obstacle, leading to a temporary increase in $T$. However, for higher values of the drive amplitude relative to the static potential, the scattering into other channels becomes dominant, and the transmission exhibits a decreasing behavior again. As clearly shown in Fig.~\ref{fig:ComparisonregionIII}, if $\tilde{\nu}$ increases, the initial transmission at zero drive is lower, and a larger drive amplitude is required to reach the transmission maximum before the decreasing behavior eventually takes over.

\section{Dynamical Switching between Localization and Perfect Transparency}

\label{8-localization}

In this final section, we explore an application of coherent control over transport properties, arising from the interplay between dynamical cavity effects and Fano interference. Specifically, we demonstrate that the system can operate as a tunable quantum switch. By varying the drive amplitude at a fixed energy, it is possible to drive the system from  localization, with the incoming particle trapped between the two impurities for a giant time delay, to perfect transparency, where the particle is fully transmitted.

\subsection{Localization and transparency}

\begin{figure*}[t]
    \centering
    \begin{minipage}[t]{0.75\textwidth}
        \centering
        \includegraphics[width=\textwidth]{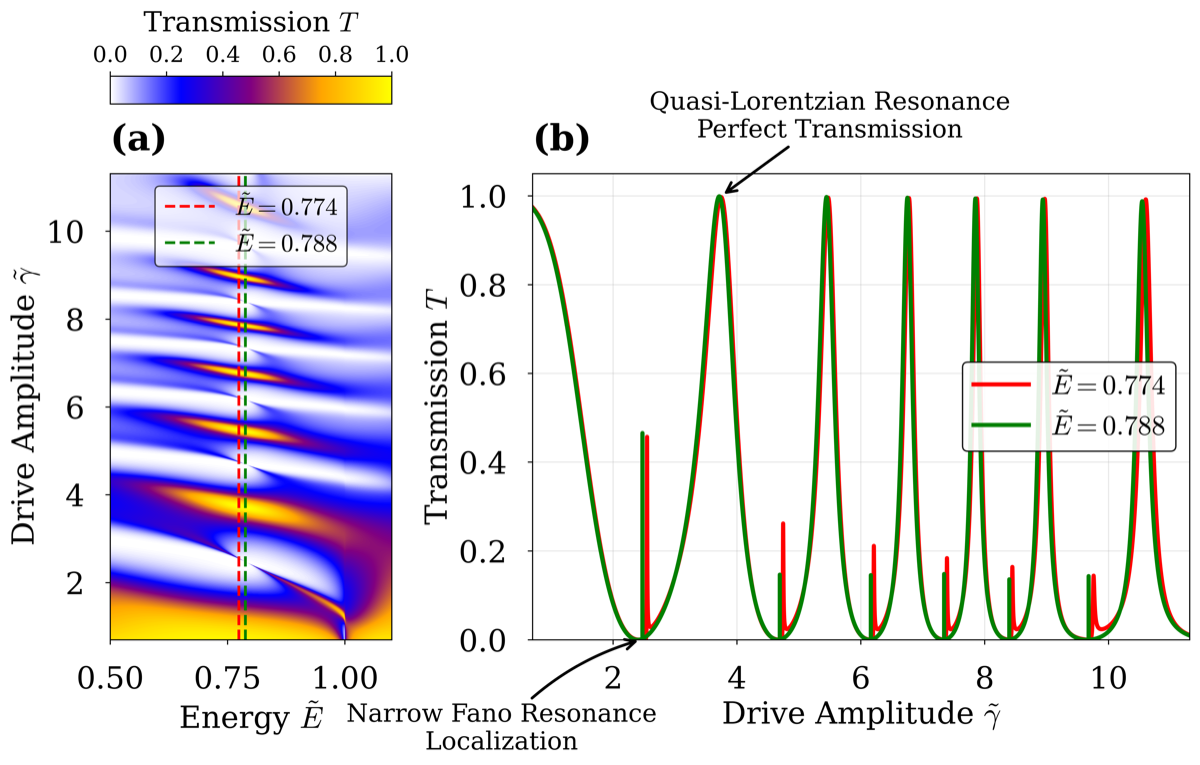} 
    \end{minipage}

    \caption{\textbf{(a)} Transmission probanility as function of th eenergy of the incoming particle $\tilde{E}$ and drive amplitude $\tilde{\gamma}$, numerical solution with 41 Floquet channels in region $III$ of the transmission spectrum, calculated for an inter-impurity distance $\tilde{l}=7.07$ without a static potential ($\tilde{\nu}=0$). The dashed vertical lines represent the energies of the third cavity mode of the channel $n=1$, $\tilde{E}_{1,3}=0.774$ and the energy of the second cavity mode of the channel $n=0$, $\tilde{E}_{0,2}=0.788$.
    \textbf{(b)} Cuts of the transmission probability $T$ as a function of drive amplitude $\tilde{\gamma}$ at the two specific cavity mode energies identified in (a). The profiles reveal a recurring pattern of narrow Fano resonances and broad quasi-Lorentzian peaks of (almost) perfect transmission.}
    \label{fig:regionIV}
\end{figure*}
\begin{figure*}[t]
    \centering
    
    % --- Pannello (a) ---
    \begin{minipage}[t]{\textwidth}
        \centering
        \includegraphics[width=\textwidth]{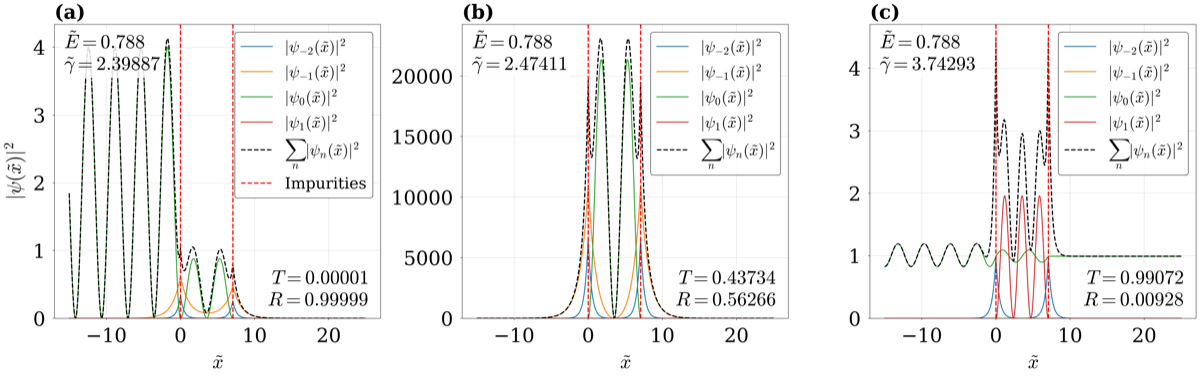}
    \end{minipage}
    \caption{Probability density of the Floquet modes and time-averaged probability density for $\tilde{l}=7.07$ at the energy $\tilde{E} = \tilde{E}_{0,2}$=0.788
    \textbf{(a)} at the zero of the narrow Fano resonance ($\tilde{\gamma}\simeq2.39$). The system is in a state of perfect reflection ($R\simeq 1, T\simeq0$);
    \textbf{(b)} at the narrow Fano resonance drive amplitude ($\tilde{\gamma}\simeq2.47$). Here the particle is trapped mainly in the region between the two impurities for long time; \textbf{(c)} at the quasi-Lorentzian peak of perfect transmission ($\tilde{\gamma}\simeq3.74$). Here the particle is transmitted with $T \simeq 1$. }
    \label{fig:wavefunctionsIV}
\end{figure*}

We focus our analysis on the transmission spectrum region labeled IV in Fig.~\ref{fig:overview}. Figure~\ref{fig:regionIV}(a) displays a zoom of this area. This region is characterized by the close proximity in energy of two distinct cavity modes belonging to different propagating channels: the third cavity mode of the $n=1$ channel ($\tilde{E}_{1,3}=0.774$) and the second cavity mode of the $n=0$ channel ($\tilde{E}_{0,2}=0.788$).
In Fig.~\ref{fig:regionIV}(b), we plot the transmission probability $T$ as a function of the drive amplitude $\tilde{\gamma}$ at these two specific energies. These cuts reveal a recurring pattern consisting of an alternating sequence of narrow Fano resonances and quasi-Lorentzian peaks of perfect or almost perfect transmission.
A difference compared to the regime analyzed in Sec.~\ref{7-bwcr}B concerns the behavior of the background transmission. In that section, for $\nu=0$, the transmission decreased monotonically as the drive amplitude increased, with the Fano resonances on top of these background. Here, one observes a slow suppression of the transmission peaks with increasing drive amplitude. In particular, within the investigated range ($\tilde{\gamma} \in [0, 11]$), the first two Lorentzian peaks reach perfect transmission, while the subsequent ones decrease slightly but maintain values close to unity. Observing the decay of the transmission probability in this region would require exploring much higher values of $\tilde{\gamma}$, which implies the need for a significantly larger number of Floquet channels ($N > 41$) to ensure numerical convergence.
We now focus our analysis on the energy $\tilde{E} = \tilde{E}_{0,2}$, although analogous behavior is observed for $\tilde{E} = \tilde{E}_{1,3}$, as shown in Fig.~\ref{fig:regionIV}(b). By keeping the inter-impurity distance and the incident energy fixed, we tune the drive amplitude $\tilde{\gamma}$ to coherently control the transport characteristics of the system.

Figure \ref{fig:wavefunctionsIV} illustrates the behavior of the Floquet modes in three distinct transport regimes.
At the Fano resonance zero [see Fig. \ref{fig:wavefunctionsIV}(a)], we observe a condition of strong reflection ($R \simeq 1$). Among the propagating channels, only in the one with $n=0$—which hosts the second cavity mode—the probability density is non-negligible. Conversely, the closed channels belonging to the discrete spectrum are populated, with a probability that decreases as $|n|$ increases.
The time-averaged probability density, $\langle |\psi(x,t)|^2 \rangle_T$ given by Eq.~\ref{Time-av-density}, reveals a localized structure between the two impurities. This localization mimics the profile of the second cavity mode of the $n=0$ channel but is enhanced at the boundaries due to the contribution of the bound states in the closed channels. In region I (for $x < 0$), the wavefunction corresponds to a  running wave resulting from total reflection.
This implies that at the specific drive amplitude corresponding to the zero of the Fano resonance, the particle is completely reflected after a short time delay. At the peak of perfect transmission [see Fig. \ref{fig:wavefunctionsIV}(c)] we observe that two propagating channels are active. Crucially, the $n=1$ channel exhibits a cavity mode profile, while the mode in the $n=0$ channel appears deformed. This configuration results in a propagating wave emerging beyond the second impurity, signaling perfect transparency. Therefore, by tuning the drive amplitude between the Fano zero and the perfect transmission peak, the system operates as an efficient energy filter, capable of fully reflecting or transmitting the particle based only on the driving amplitude. 
At the Fano resonance drive amplitude $\tilde{\gamma}=\tilde{\gamma}_{r}\simeq2.474$ indicated in Fig.~\ref{fig:wavefunctionsIV}(b), the system exhibits a different behavior: the particle is localized in the region between the two impurities for a long time, since the time-averaged probability density reaches very high values, as shown in Fig.~\ref{fig:wavefunctionsIV}. Thus, in this regime the two impurities act as a cavity, localizing the particle in the region between them for long time.
However, since these Fano resonances are extremely narrow, performing a reliable numerical fit (as done in Sec.~\ref{7-bwcr}) proves difficult. To address this, a more robust approach based on the Wigner time delay will be employed, as shown in Sec.~\ref{8-localization}B. 
On the other hand, for the broad quasi-Lorentzian peaks, the lifetime is significantly shorter than that of the narrow Fano resonances, and can be straightforwardly estimated from the width of the resonance itself.

As evident in Fig.~(\ref{fig:regionIVComparison}), the introduction of a static potential does not significantly alter the underlying physics. The characteristic spectral structure, featuring alternating narrow-Fano resonances and quasi-Lorentzian peaks, is qualitatively preserved. The primary effect of the static potential is to shift the resonance positions and modify their linewidths, without disrupting the overall transport mechanism.

\begin{figure}[h]
        \centering
        \includegraphics[width=0.48\textwidth]{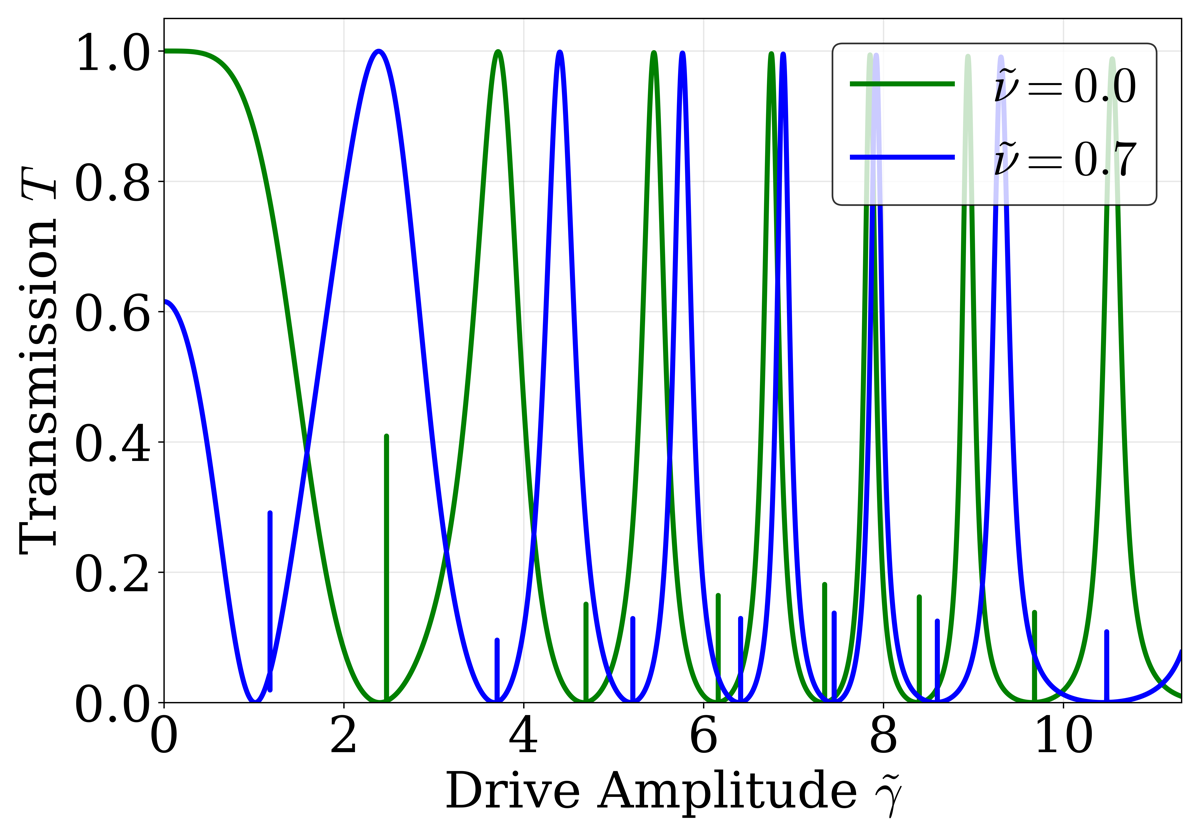}
        \caption{Transmission probability as a function of drive amplitude $\tilde{\gamma}$ for fixed energy $\tilde{E}=0.788$ and distance $\tilde{l}=7.07$. The green curve represents the case without static potential ($\tilde{\nu}=0$), while the blue curve shows the case with a static potential $\tilde{\nu}=0.7$.}
        \label{fig:regionIVComparison}
\end{figure}

\subsection{Quantifying Localization: Giant Wigner Time Delay}

By tuning the drive amplitude $\tilde{\gamma}$, one is able to control coherently the transport properties of the system, switching from a state of strong localization  to a state of perfect transparency or one of strong reflection. To quantitatively evaluate the duration of the particle's interaction with the scattering center—and thus the localization time of the quasi-bound states corrensponding to the Fano resonance drive amplitude $\tilde{\gamma}_{r}$—we employ the concept of the Wigner time delay $\tau_W$ \cite{texier2016wigner, wigner1955lower}.
In its original formulation for single-channel scattering, this quantity is defined as the energy derivative of the scattering phase shift $\eta(E)$:
\begin{equation}
    \label{eq:Wigner_scalar}
    \tau_W(E) = \hbar \frac{d\eta}{dE}.
\end{equation}
Physically, $\tau_W$ corresponds to the time lag of a particle interacting with the scattering potential relative to a free particle \cite{Smith1960}. In our specific setup, formed by two periodically driven impurities, this delay quantifies the excess time the particle remains localized within the scattering region. Crucially, this region is not strictly confined to the space between the impurities but extends slightly outwards, reflecting the exponential decay of the probability density characteristic of the evanescent Floquet modes [see the curves decaying from the two impurity positions in Fig.~\ref{fig:wavefunctionsIV}(b)]. However, since the periodic driving maps the system onto a set of coupled Floquet sidebands, our problem is inherently multichannel. Consequently, a generalized multichannel scattering formalism must be employed to properly evaluate the time delay.
The transport properties of the system are fully described by the multichannel scattering matrix $S$ \cite{li1999floquet}. For a system with $N_o$ open scattering channels, $S$ is a $2N_o \times 2N_o$ unitary matrix relating the incoming asymptotic amplitudes to the outgoing ones. It is organized in a block structure describing reflection and transmission processes from the left ($L$) and right ($R$) leads:
\begin{equation}
    S = 
    \begin{pmatrix}
    r & t' \\
    t & r'
    \end{pmatrix},
\end{equation}
where $r$ and $t$ are the $N_o \times N_o$ reflection and transmission sub-matrices for incidence from the left, while $r'$ and $t'$ correspond to incidence from the right.
Crucially, to ensure the unitarity of $S$ (flux conservation), the matrix elements are normalized by the square root of the ratio of the group velocities (which are proportional to the wavevectors $k$). For a transition from an input channel $m$ to an output channel $n$, the scattering element is defined as:
\begin{equation}
    S_{nm} = A_{nm} \sqrt{\frac{k_n}{k_m}},
\end{equation}
Here, $A_{nm}$ represents the raw amplitude of the wavefunction component in channel $n$. Generally, $A_{nm}$ corresponds to the transmission coefficient $t_{nm}$ (for forward scattering) or the reflection coefficient $r_{nm}$ (for backward scattering) resulting from a unit-amplitude incident wave in channel $m$. In our specific case, since the incoming particle originates from the channel $m=0$, these amplitudes map directly onto the notation used in previous sections, i.e., $A_{n0}$ corresponds to $t_n$ or $r_n$.
Considering the incoming particle form the channel $n=0$ witho probability 1, the matrix $S$ is normalized. This normalization ensures that $S$ is unitary, implying that  $|\det(S)|=1$. We can thus define the total scattering phase shift $\Phi_{\text{tot}}(E)$ via the relation $\det S = e^{i\Phi_{\text{tot}}(E)}$ (it takes into account the contribution from all the propagating channels, summing them up).
To analyze the temporal dynamics of the scattering process, we utilize the lifetime matrix $Q$, introduced by Smith to generalize Wigner's delay time to multichannel systems \cite{Smith1960}:
\begin{equation}
    \label{eq:Smith_Matrix}
    Q = -i \hbar S^\dagger \frac{dS}{dE}.
\end{equation}
This Hermitian matrix encodes the time delay properties of the system. The eigenvalues of $Q$ correspond to the specific time delays of the scattering eigenchannels \cite{li1999floquet}. 
The global observable of interest is the Wigner time delay $\tau_W$, defined as the trace of the lifetime matrix $Q$ introduced by Smith \cite{Smith1960}. This quantity represents the sum of the proper time delays of all scattering channels and is explicitly given by:
\begin{equation}
    \label{eq:Wigner_Trace}
    \tau_W(E) = \text{Tr}[Q] = -i\hbar \text{Tr}\left[ S^\dagger \frac{dS}{dE} \right].
\end{equation}Using the identity $\text{Tr}(S^\dagger dS) = d(\ln \det S)$, $\tau_W$ can be directly expressed as the energy derivative of the total phase shift:
\begin{equation}
    \label{eq:wigner_def}
    \tau_W(E)  = \hbar \frac{d}{d E} \text{Im} \left[ \ln (\det S) \right].
\end{equation}
This quantity provides a measure of the probability density localized within the scattering region.
However, the numerical evaluation of this derivative near extremely narrow resonances, such as the quasi-BICs observed in Fig.~\ref{fig:regionIV}(b), presents significant challenges. Standard finite-difference methods applied directly to the total scattering phase $\Phi_{\text{tot}}(E)$ are prone to severe numerical artifacts. 
First, the numerical extraction of the phase from the S-matrix typically yields values wrapped within the principal branch $(-\pi, \pi]$. Consequently, the phase exhibits artificial discontinuous jumps of $2\pi$ (phase wrapping), which, upon differentiation, result in non-physical singularities in the time delay. 
Second, when the resonance linewidth $\Gamma$ is comparable to or narrower than the energy sampling step, the rapid phase variation characterizing the resonance is undersampled (aliasing), leading to a significant underestimation or complete loss of the derivative signal.
To overcome this limitation and ensure numerical stability, we employed a high-precision direct difference method based on the logarithmic derivative of the determinant.

\begin{figure}[h]
    \centering
    \includegraphics[width=0.48\textwidth]{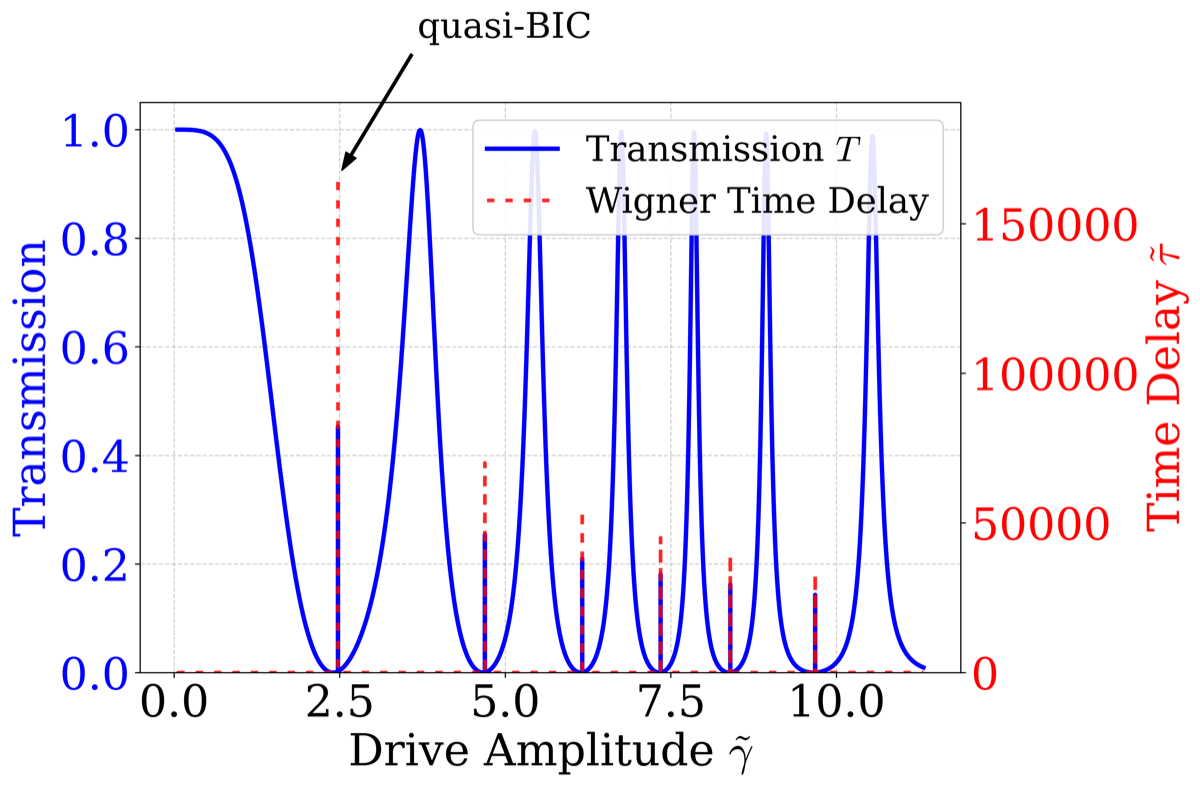}
    \caption{Transmission probability $T$ (blue solid line, left axis) and the Wigner time delay $\tau_W$ (red dashed spikes, right axis) as a function of the drive amplitude $\tilde{\gamma}$. The system parameters are set to $\tilde{l} = 7.07$, $\tilde{\nu} = 0$, and the incident energy is fixed at $\tilde{E} = 0.788$. The plot reveals a strict correspondence between the Fano resonances and the divergences in the time delay. The magnitude of $\tau_W$ reaches values of the order of $\times 10^5$, indicating the extreme temporal localization of the particle within the scattering region matching the quasi-BIC conditions. The number of open channels used in the simulation is $N_{o}=20$}
    \label{fig:WignerTimedelay}
\end{figure}

To fully resolve the ultra-narrow linewidths characterizing the quasi-BICs, the parameter space was sampled with extremely high resolution. We utilized a dense grid of $N_{\gamma} = 5 \times 10^5$ points over the investigated range, corresponding to a step size $\Delta \tilde{\gamma} \approx 1.6 \times 10^{-5}$, ensuring that even the sharpest resonance peaks are accurately captured.
Specifically, for each value of the drive amplitude $\tilde{\gamma}$, we compute the scattering matrix $S$ at the target energy $\tilde{E}$ and at an infinitesimally shifted energy $\tilde{E} + \delta \tilde{E}$. The energy step was set to $\delta \tilde{E} = 10^{-9}$ (in natural units), a value chosen to be sufficiently small to approximate the derivative ($\delta \tilde{E} \ll \tilde{\Gamma}$) yet large enough to avoid machine precision errors. The time delay is then evaluated as:
\begin{equation}
    \label{eq:robust_wigner}
    \tau_W(E) \approx \hbar \frac{\text{Im}\left[ \ln \left( \frac{\det S(E+\delta E)}{\det S(E)} \right) \right]}{\delta E}.
\end{equation}
By computing the imaginary part of the logarithm of the ratio of the determinants, this formulation automatically handles the branch cuts of the complex phase. This numerical approach effectively eliminates the artifacts associated with phase unwrapping algorithms, allowing for the accurate resolution of the trapping time $\tau_W$. As illustrated in Fig.~\ref{fig:WignerTimedelay}, $\tau_W$ manifests as large positive peaks exactly coinciding with the transmission dips of the Fano resonances. Notably, the magnitude of the Wigner time delay reaches the order of $10^{5}$, confirming the extreme temporal localization associated with the quasi-BIC states. This indicates that the dwell time of the particle within the scattering region matching the quasi-BIC condition is three to five orders of magnitude larger than the time spent during the broad quasi-Lorentzian resonances associated with perfect transparency. While the delay in the latter regime is negligible and could be estimated via standard Fano fitting with a background, the quasi-BICs represent a regime of extreme temporal localization, where the particle is effectively trapped for timescales far exceeding the transit time.

\begin{figure}[h]
        \centering
        \includegraphics[width=0.48\textwidth]{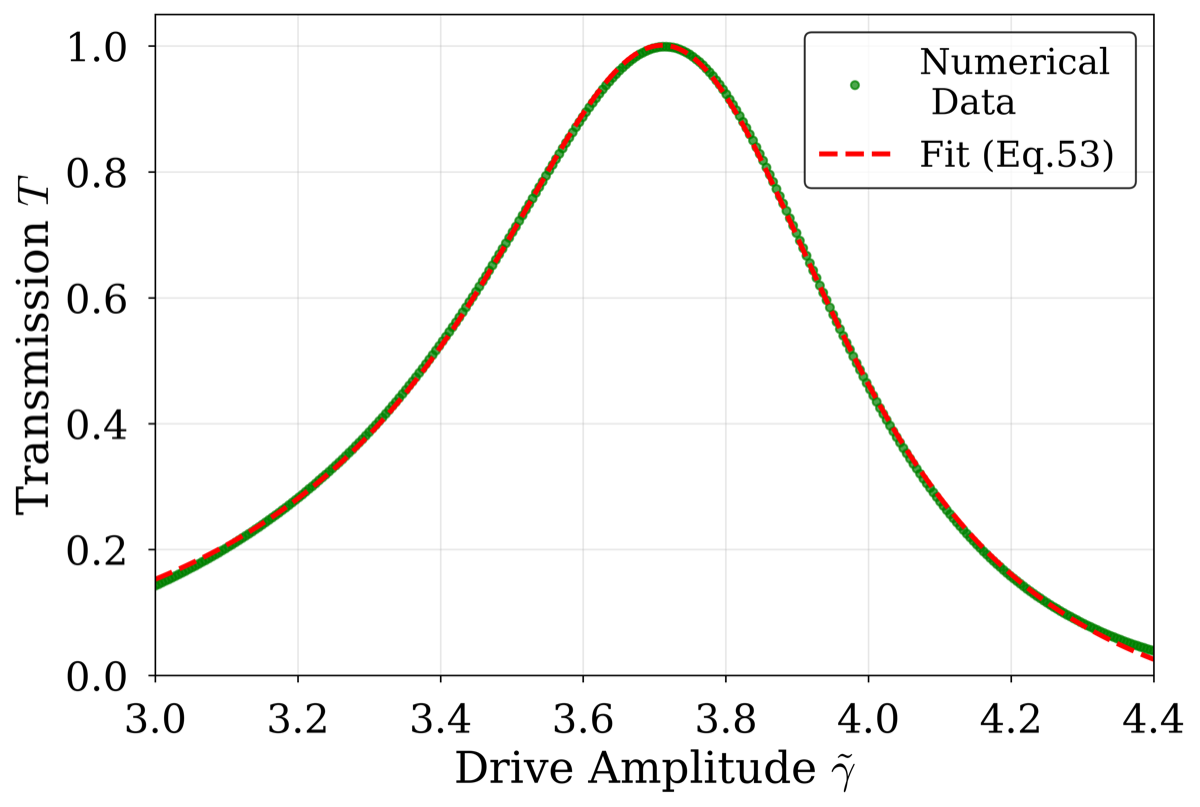}
        \caption{Fano-Lorentzian fit of the numerical transmission data (green dotted curve, $\tilde{E} = 0.788$) from Fig.~\ref{fig:regionIV}(b) using the model described by Eq.~\eqref{quasilorentzian}. The dashed red line represents the best fit, yielding a resonance center $\tilde{\gamma}_0 \simeq 3.74$, a linewidth $\tilde{\Gamma} \simeq 0.65$ and an asymmerry parameter $q\simeq-10.1$. The background offset is $T_{\text{bg}} \simeq -0.112$.}
        \label{fig:FanoLorentzianFit}
\end{figure}

The fitting equation used for the broad quasi-Lorentzian resonance in Fig.~(\ref{fig:FanoLorentzianFit}) is based on an asymmetric Fano profile. The transmission probability $T$ as a function of the drive amplitude $\tilde{\gamma}$ is given by
\begin{equation}
\label{quasilorentzian}
    T(\tilde{\gamma}) = A \frac{\left(1 + \frac{\epsilon}{q}\right)^2}{1 + \epsilon^2} + T_{\text{bg}},
\end{equation}
where $\epsilon = 2(\tilde{\gamma} - \tilde{\gamma}_0)/\tilde{\Gamma}$ represents the dimensionless detuning. In this expression, $A$ is the resonance amplitude, $\tilde{\gamma}_0$ is the exact resonance center (where perfect transmission is achieved), $\tilde{\Gamma}$ denotes the resonance linewidth, and $q$ is the Fano asymmetry parameter. Finally, $T_{\text{bg}}$ is a small constant offset introduced to capture the non-resonant background transmission.
The comparison highlights that while the broad quasi-Lorentzian resonance allows for fast particle transmission, the quasi-BIC states effectively freeze the particle dynamics, demonstrating the extreme tunability of the system.

\subsection{Experimental Realization}
\label{sec:Outlook}

In order to appreciate the coherent trapping mechanism that we have described in this section and the experimental relevance of these timescales, we map our dimensionless results onto physical parameters for the specific platform that we have introduced in Sec.~\ref{4-single}(C), consisting of ultracold atoms in mesoscopic ballistic channels, engineered with two oscillating impurities with frequencies in the range $\omega/2\pi \in [1,20]$~kHz. For the dimensionless time delay of $1.5 \times 10^{5}$, one obtains localization times $\tau \simeq 1.2 - 23.9$~s. These are of the same order of magnitude, or up to one order of magnitude larger, than the duration of a typical transport experiment, which can last up to about $4$~s, and are much larger than the typical time delay of a single-impurity system discussed in Sec.~\ref{4-single}(C).
Beyond the observation of long time delays, our results suggest a protocol for the active engineering of quantum devices. The extreme sensitivity of the time delay to the drive parameters ($\tilde{\gamma}$ and $\tilde{E}$) allows for the implementation of a tunable mechanism for trapping and detrapping the particle. One can even engineer the system to achieve perfect reflection or perfect transmission, allowing it to act as a trap or as an energy filter.

By tuning the drive amplitude $\tilde{\gamma}$ towards the resonance condition of a quasi-BIC, an incoming particle can be effectively captured and localized mainly within the regions between the two impurities for a duration $\tau_W$. Conversely, a slight detuning of $\tilde{\gamma}$ can lead to perfect reflection or perfect transmission of the particle. 
The quasi-BIC states represent protected islands in the parameter space where leakage to the continuum is heavily suppressed. This isolation makes them ideal candidates for quantum memory applications, as information encoded in the localized state is protected from the environment for timescales $\tau \gg t_{\text{transit}}$. 
In conclusion, this tunable storage capability marks a qualitative departure from the single-impurity case. While a single driven impurity functions primarily as an energy filter via  Fano resonances transmission zeros and peaks, the double-impurity system exploits cavity effects and their interplay with Fano interference to engineer robust particle traps. Consequently, it acts as a programmable delay line where the storage time can be dynamically tuned over several orders of magnitude simply by adjusting the drive amplitude.

\section{Conclusion and Outlook}
\label{sec:conclusion}

In this work, we have investigated the quantum transport properties of a one-dimensional ballistic channel  engineered with two periodically driven impurities. By employing the Floquet formalism, we mapped this time-dependent problem into an equivalent stationary multichannel scattering problem.
Our analysis highlights the rich physics emerging from the introduction of a second impurity compared to the widely studied single-impurity case. Crucially, the inter-impurity distance $l$ acts as an additional control parameter. The rich physics arises from the interplay between the cavity effects in the Floquet sidebands and the Fano interference.
Working in the weak-coupling regime we have shown that the properties of the system can be explained considering only two Floquet channels, and in this regime we have carried out an analytical calculation to determine the Fano resonances parameters. 

We show how one can engineer Bound States in the continuum (BICs) in the two-impurity system, and their signature is the disapperence of the Fnao resonance. 
In the scattering picture, these states are invisible to the incident flux due to the lack of coupling; however, in the associated eigenvalue problem, without the incoming particle, they appear as localized eigenstates degenerate with the continuum.
We provided an analytical description of these phenomena within the two-channel approximation, discussing its range of validity and its limitations, particularly concerning the physics at higher drive amplitudes and the inability to capture threshold anomalies associated with higher-order channel openings.
To overcome these limitations, we performed a numerical analysis beyond the weak-coupling regime. We emphasized the role of the region between the oscillating impurities as a dynamical Fabry-Pérot cavity. The interference between the continuum pathways and the discrete quasi-bound states gives rise to a complex landscape of Fano resonances.

Unlike true BICs, which are perfectly decoupled, we focused on Quasi-BICs, representing long-lived states that remain weakly coupled to the continuum. We showed that these states play a crucial role in coherent transport control.
Specifically, we demonstrated a coherent trapping mechanism based on the tunable switching between narrow Fano resonances (localization) and Lorentzian peaks of perfect transmission. By tuning the drive amplitude when cavity effects are predominant, one can switch the system from a highly transmissive state to a regime of strong localization to one of perfect reflection.
We connected these theoretical findings to relevant experimental platforms, such as ultracold atoms in mesoscopic channels. Through the calculation of the Wigner time delay, we provided a quantitative analysis of the trapping efficiency, finding time delays orders of magnitude larger than the natural transit time. This suggests concrete applications in quantum technologies, such as tunable delay lines, quantum memories, or dynamical switches.

Future investigations can expand in several promising directions, exploiting the versatility of this two-impurity setup.
First, introducing a phase difference between the two driven impurities would add a further degree of coherent control. This could be employed to study coherent caloritronics effects, such as generating heat currents in the absence of a thermal gradient (quantum pumping) or, conversely, applying a thermal gradient to investigate enhanced Seebeck effects driven by the threshold anomalies discussed in this work. Having a second impurity, and more coherent parameters can lead to thermoelectric effects better to control.
A second direction involves extending this model to discrete lattice systems, such as the tight-binding or Bose-Hubbard models, which are directly relevant to optical lattice experiments.
Finally, a natural and necessary step forward is to include particle-particle interactions. Investigating how strong correlations modify the Fano resonances and the stability of the Floquet states would allow probing the interplay between many-body physics and coherent driving, potentially leading to novel phenomena.

\begin{acknowledgments}
We thank  S. Eggert, T. Giamarchi,  F. Hemman, H. Ott, E. Demler for fruitful discussions. We
acknowledge support by the Deutsche Forschungsgemeinschaft (DFG, German Research Foundation) under Project No. 277625399-TRR 185 OSCAR (“Open
System Control of Atomic and Photonic Matter”, B3, B4),
No. 277146847-CRC 1238 (“Control and dynamics of
quantum materials”, C05), No. 511713970-CRC 1639
NuMeriQS (“Numerical Methods for Dynamics and
Structure Formation in Quantum Systems”), and under
Germany’s Excellence Strategy – Cluster of Excellence
Matter and Light for Quantum Computing (ML4Q)
EXC 2004/1 – 390534769. This research was supported by the Erasmus+ programme.
R.C. acknowledges support by the PNRR MUR project
PE0000023 “National Quantum Science and Technology Institute” (NQSTI), through the cascade funding
projects TOPQIN and SPUNTO.
\end{acknowledgments}

\section*{Data Availability}
The Julia source code used to generate the numerical data and plots for this study is openly available at Zenodo \cite{zenodo_code}.

\appendix

\section{Quasienergy Operator}
\label{App:Floquet_Structure}

In this Appendix, we explicitly derive the Fourier components of the Hamiltonian and the matrix structure of the Floquet operator $\hat{Q}$.
The time-dependent Hamiltonian of the system is given by:
\begin{equation}
    \hat{H}(t) = -\frac{\hbar^2}{2m}\frac{\partial^2}{\partial x^2} + V(x,t),
\end{equation}
where the potential term describes the two driven impurities:
\begin{equation}
    V(x,t) = -\sum_{j=1,2} [\nu_j + \gamma_j \cos(\omega t)]\delta(x-x_j),
\end{equation}
with $x_1=0$ and $x_2=l$. Using the Euler relation $\cos(\omega t) = \frac{1}{2}(e^{i\omega t} + e^{-i\omega t})$, we can expand the Hamiltonian into its Fourier series components $\hat{H}_n$ defined by $\hat{H}(t) = \sum_n e^{-in\omega t} \hat{H}_n$.

The only non-vanishing terms are the ones with $n=0$ and $n=\pm 1$. The diagonal component ($n=0$) contains the kinetic energy and the static part of the potential:
\begin{equation}
    \hat{H}_0 = -\frac{\hbar^2}{2m}\frac{\partial^2}{\partial x^2} - \sum_{j=1,2} \nu_j \delta(x-x_j).
\end{equation}
The off-diagonal components ($n=\pm 1$) are determined solely by the driving amplitude $\gamma$:
\begin{equation}
    \hat{H}_{1} = \hat{H}_{-1} = - \sum_{j=1,2} \frac{\gamma_j}{2} \delta(x-x_j).
\end{equation}
All other components $\hat{H}_n$ for $|n|>1$ are zero.
These explicit forms highlight that the static potential $\nu$ governs the intra-channel scattering (within the same $n$), while the drive amplitude $\gamma$ is responsible for the inter-channel coupling, connecting channel $n$ with $n\pm 1$.

Substituting these components into the definition of the Floquet operator matrix elements $\hat{Q}_{n,m} = \hat{H}_{n-m} + n\hbar\omega \delta_{n,m}$, we obtain a block-tridiagonal structure for the quasi-energy operator $\hat{Q}$ in the extended Hilbert space $\mathcal{F}$:
\begin{equation}
    \hat{Q} = 
    \begin{pmatrix}
    \ddots & \vdots & \vdots & \vdots & \vdots & \iddots \\
    \dots & \hat{H}_0 + \hbar\omega & \hat{H}_{-1} & 0 & 0 & \dots \\
    \dots & \hat{H}_{1} & \hat{H}_0 & \hat{H}_{-1} & 0 & \dots \\
    \dots & 0 & \hat{H}_{1} & \hat{H}_0 - \hbar\omega & \hat{H}_{-1} & \dots \\
    \dots & 0 & 0 & \hat{H}_{1} & \hat{H}_0 - 2\hbar\omega & \dots \\
    \iddots & \vdots & \vdots & \vdots & \vdots & \ddots
    \end{pmatrix}.
\end{equation}
This matrix representation clearly demonstrates that the problem is equivalent to a tight-binding-like model in the frequency domain (Floquet lattice), where $\hat{H}_0$ represents the on-site energy (shifted by the photon energy quanta $n\hbar\omega$) and $\hat{H}_{\pm 1}$ acts as the hopping term between adjacent sites.

\section{Transmission Probability}
\label{Transmission}

In this section, we provide more details about the calculation of the transmission probability \( T \). This probability is derived from the conservation of the time-averaged probability current. The fundamental definition of the probability current depends on the total, time-dependent wavefunction \( \Psi(x,t) \) \cite{griffiths2018introduction}:

\begin{equation}
\label{eq:probability_current}
\begin{split}
j(x,t) = \frac{\hbar}{2mi} \biggl[ & \Psi^*(x,t) \frac{\partial \Psi(x,t)}{\partial x} + \\
         & -\Psi(x,t) \frac{\partial \Psi^*(x,t)}{\partial x} \biggr].
\end{split}
\end{equation}

Our total wavefunction is the Floquet expansion
\begin{equation}
\Psi(x,t) = \sum_n \psi_n(x) e^{-iE_n t / \hbar},
\end{equation}
where \( E_n = E + n\hbar\omega \).  
When this sum is inserted into Eq.~(\ref{eq:probability_current}), we have the products of two sums, one over $n$ and the other over $m$. The resulting current \( j(x,t) \) contains both static terms (for \( n = m \)) and oscillating terms (for \( n \neq m \)). For a stationary scattering problem, we are interested in the time-averaged current,
\begin{equation}
j(x) = \langle j(x,t) \rangle_t.
\end{equation}
When averaging over time, all oscillating cross-terms (\( n \neq m \)) vanish.  
The total time-averaged current is thus simply the sum of the currents from each Floquet channel calculated independently:
\begin{equation}
\begin{split}
j(x) = & \sum_n j_n(x) = \\
 = &\sum_n \frac{\hbar}{2mi} \left[ \psi_n^*(x) \psi_n'(x) - \psi_n(x) \psi_n^{*'}(x) \right].
\end{split}
\end{equation}

By demanding current conservation, the time-averaged flux must be equal in the asymptotic regions:
\[
j(x<0) = j(x>l).
\]
In region I, where $x<0$ we have
\begin{equation}
\psi_n(x) = \delta_{n0} e^{ik_0 x} + r_n e^{-ik_n x}.
\end{equation}
Plugging this into the formula for \( j(x) \) gives:
\begin{equation}
j_{\text{I}} = \underbrace{\frac{\hbar k_0}{m}}_{\text{incident}} 
 - \sum_{n} \underbrace{\frac{\hbar k_n}{m} |r_n|^2}_{\text{reflected}}.
\end{equation}
Here, the sum is only over open (propagating) channels, as evanescent (closed) channels have imaginary \( k_n \) and carry zero net current.

In region III, where $x>l$:
\begin{equation}
\psi_n(x) = t_n e^{ik_n x}.
\end{equation}
The flux is purely transmitted:
\begin{equation}
j_{\text{III}} = \sum_{n} \underbrace{\frac{\hbar k_n}{m} |t_n|^2}_{\text{transmitted}}.
\end{equation}
The total transmission probability \( T \) is defined as the ratio of the total transmitted flux \( j_{\text{trans}} = j_{III} \) to the incident flux \( j_{\text{in}} = \frac{\hbar k_0}{m} \):
\begin{equation}
T = \frac{j_{\text{trans}}}{j_{\text{in}}} 
  = \frac{\displaystyle \sum_{n} \frac{\hbar k_n}{m} |t_n|^2}{\frac{\hbar k_0}{m}}.
\end{equation}
This simplifies to the final expression for the total transmission:
\begin{equation}
T = \sum_{n} \frac{k_n}{k_0} |t_n|^2.
\label{eq:transmission_final}
\end{equation}

\section{Derivation of Zeros and Poles}
\label{app_cal}

In this Appendix, we detail the analytical derivation of the transmission zeros and  poles discussed in the main text.
The mathematical treatment differs between the case without static potential ($\tilde{\nu}=0$) and the case with static potential ($\tilde{\nu} \ge 0$). This distinction is necessary due to the different physical origins of the resonances:
For $\tilde{\nu}=0$, the resonances are dynamically induced by the driving field near the channel threshold opening ($\tilde{E}\lesssim1$). Since the effective coupling diverges at the threshold, we must employ an asymptotic expansion of the exact scattering equations around the singularity $\tilde{E} = 1$. For $\tilde{\nu} \ge 0$, the system already supports discrete bound states in the static limit ($\tilde{\gamma}=0$). What one does in this case is a perturbative expansion up to the second order in the drive amplitude $\tilde{\gamma}$ to find correction to the static bound states. 

In the framework of the two-channel approximation, in the energy interval $0\le\tilde{E}<2$ the system dynamics is characterized by an open channel with wavevector $\tilde{k}_0 = \sqrt{\tilde{E}}$ and a closed channel for $0<\tilde{E}<1$ with wavevector $\tilde{k}_{-1} = \sqrt{1 - \tilde{E}}=i\kappa$. The equations \ref{BC_Explicit_0}(a)-(b) and \ref{BC_Explicit_l}(a)-(b) can be written only fopr the channel $n=0$ and include the coupling between it and the channel $n=-1$ in the effective potential given by $V(\tilde{E}) = (\tilde{\gamma})^2 /8\kappa$. Thus, one obtains only four equations for the wavefunction $\psi_{0}(x)$, and the effect of the closed Floquet channel ($n=-1$) is fully taken into account within the effective potential $V(\tilde{E})$.

The transmission zero occurs at a real energy $\tilde{E}_{zt}$ where the transmission amplitude $t_0$ vanishes. This requires the wavefunction in the open channel to vanish for $\tilde{x} > \tilde{l}$, and by continuity, $\psi_0(\tilde{l})=0$.
Solving the Schrödinger equation in the  region $0 < \tilde{x} < \tilde{l}$ with the boundary condition $\psi_0(\tilde{l})=0$ yields the standing wave 
\begin{equation}
  \psi_0(x) = A \sin(\tilde{k}_0(\tilde{x}-\tilde{l})).  
\end{equation}
Applying the boundary condition on the derivative jump at $\tilde{x}=\tilde{l}$ induced by the effective potential:
\begin{equation}
    \psi'_0(\tilde{l}^+) - \psi'_0(\tilde{l}^-) = -V(\tilde{E})\left[\psi_0(\tilde{l}) - e^{-\kappa \tilde{l}}\psi_0(0)\right].
\end{equation}
Since $\psi_0(\tilde{l})=0$ and $\psi'_0(\tilde{l}^+) = 0$, this simplifies to:
\begin{equation}
    -\tilde{k}_0 = -V(\tilde{E}) e^{-\kappa \tilde{l}} \sin(-\tilde{k}_0 \tilde{l}).
\end{equation}
In the limit $\tilde{E} \to 1$, one has $\kappa \to 0$. One can approximate the non-singular terms $\tilde{k}_0 \simeq 1$ and $e^{-\kappa \tilde{l}} \approx 1$, obtaining the equation
\begin{equation}
    1 \approx \frac{\tilde{\gamma}^{2}}{8\kappa} \sin( \tilde{l}).
\end{equation}
Solving for $\kappa$ and using the relation $\tilde{E} = 1 - \kappa^2$, we obtain the explicit formula for the zero:
\begin{equation}
    \tilde{E}_{zt} \approx 1-\left[ \frac{\tilde{\gamma^{2}}\sin(\tilde{l})}{8} \right]^{2}.
\end{equation}

The poles correspond to the complex energies $\tilde{E}_p$ where the system without incoming particle admits non-trivial solutions. The wavefunction in the open channel is written as an outgoing wave ansatz:
\begin{equation}
    \psi_0(x) = \begin{cases} 
    r_0 e^{-i\tilde{k}_0 \tilde{x}} & \tilde{x} < 0 \\
    A e^{i\tilde{k}_0 \tilde{x}} + B e^{-i\tilde{k}_0 \tilde{x}} & 0 < \tilde{x} < \tilde{l} \\
    t_0 e^{i\tilde{k}_0 \tilde{x}} & \tilde{x} > \tilde{l}.
    \end{cases}
\end{equation}
Using the continuity conditions as well as the boundary conditions on the derivative jumps at $\tilde{x}=0$ and $\tilde{x}=\tilde{l}$ can be recast as a matrix equation for the internal amplitudes $A$ and $B$:
\begin{equation}
    M \begin{pmatrix} A \\ B \end{pmatrix} = 0.
\end{equation}
The matrix $M$ governing the homogeneous system is explicitly given by:

\begin{equation}
    M = \begin{pmatrix}
        M_{11} & M_{12} \\
        M_{21} & M_{22}
    \end{pmatrix},
\label{eq:matrix_M}
\end{equation}
where the individual elements are defined as:
\begin{align*}
    M_{11} &= 2i\tilde{k}_0 + V(\tilde{E}_{p}) + V(\tilde{E}_{p})e^{-\kappa \tilde{l}} e^{i\tilde{k}_0 \tilde{l}}, \\
    M_{12} &= V(\tilde{E}_p)\left(1 + e^{-\kappa \tilde{l}}e^{-i\tilde{k}_{0}\tilde{l}}\right), \\
    M_{21} &= V(\tilde{E}_p)\left(1 + e^{-\kappa \tilde{l}}e^{i\tilde{k}_{0}\tilde{l}}\right), \\
    M_{22} &= 2i\tilde{k}_0 + V(\tilde{E}_{p})e^{-i\tilde{k}_0 \tilde{l}} + V(\tilde{E}_{p})e^{-\kappa \tilde{l}}.
\end{align*}
The condition for the existence of a pole is $\det(M) = 0$. Near the threshold ($\kappa \to 0$), the terms proportional to $V^2(\tilde{E}_{p}) \propto 1/\kappa^2$ cancel out in the determinant expansion.
By performing a Laurent expansion of $\det(M)$ with respect to the small parameter $\kappa$ and imposing the vanishing of the dominant remaining term (scaling as $1/\kappa$), the exact condition reduces to the following transcendental equation for the open channel wavevector $\tilde{k}_0$:
\begin{equation}
    \label{eq:pole_transcendental}
    \tilde{k}_0 \left( 1 + e^{-i\tilde{k}_0 \tilde{l}} \right) = \frac{\tilde{\gamma}^{2}\tilde{l}}{8} \sin(\tilde{k}_0 \tilde{l}).
\end{equation}
This equation governs the physics of the resonances. Specifically, for odd multiples of the critical length ($\tilde{k}_0 \tilde{l} = p\pi$, with $p$ odd), the LHS vanishes ($1+e^{-i\pi}=0$) and the RHS vanishes ($\sin(p\pi)=0$), allowing for a solution with a real wavevector, thus a real pole. For even multiples ($p$ even), the LHS is non-zero ($1+e^{-i2\pi}=2$), while the RHS vanishes, meaning that no real solution for $\tilde{k}_{0}$ is found.

When a static potential is present, the resonance poles, defined as $E_p = E_r - i\Gamma/2$, are derived from the bound states of the static double-well potential $V_{\text{stat}}(x) = -\nu [\delta(x) + \delta(x-\tilde{l})]$. In the weak-coupling regime ($\gamma <<\nu$), we assume that the resonance positions $E_r$ coincide with the energies of these static bound states shifted into the $n=-1$ Floquet channel. Subsequently, we determine the transmission zeros and the imaginary parts of the poles (the resonance widths) by applying perturbative corrections up to the second order in the drive amplitude $\tilde{\gamma}$.
 Let $\tilde{E}_{S,A}=-\kappa_{S,A}^2$ be the energies of the static bound states. The wavefunctions associeted to them are one symmetric ($\psi_S$) and the other antisymmetric ($\psi_A$) with respect to the center of the cavity.
Considering the static problem without drive and imposing the boundary conditions at the delta potentials, as expleined beforehand in Sec.~\ref{3-T}, leads to two transcendental equations for the decay constant $\kappa_{S,A}$:
\begin{align}
    \frac{2\kappa_S}{\tilde{\nu}} &= 1 + e^{-\kappa_S \tilde{l}} \quad (\text{Symmetric State}), \\
    \frac{2\kappa_A}{\tilde{\nu}} &= 1 - e^{-\kappa_A \tilde{l}} \quad (\text{Antisymmetric State}).
\end{align}
The energies of the resonances, coinciding with the real part of the poles, are givern by the energies of the static bound states in the channel $n=-1$:
\begin{equation}
    E_{r,\alpha} \simeq 1 - \kappa_{\alpha}^2 ~ (\alpha = S, A).
\end{equation}

The decay width $\tilde{\Gamma}$ is determined by the coupling of these localized states to the outgoing continuum waves $\frac{1}{\sqrt{2\pi}}e^{\pm i\tilde{k}_0 \tilde{x}}$ via the operator $\hat{V} = -\frac{\tilde{\gamma}}{2}[\delta(\tilde{x}) + \delta(\tilde{x}-\tilde{l})]$. Evaluating the transition rates according to Fermi's Golden Rule \cite{sakurai2020modern}, the amplitude $\mathcal{M}$ from the bound state in the closed channel ($n=-1$) to the open channel ($n=0$) is given by the coherent sum of the emissions from the two impurities and is given by:
\begin{equation}
    \tilde{\Gamma} = \frac{\tilde{\gamma}^2}{\tilde{k}_0} \left| \psi_{S,A}(0) + e^{i\tilde{k}_0 \tilde{l}} \psi_{S,A}(\tilde{l}) \right|^2,
\end{equation}
where the phase factor $e^{i\tilde{k}_0 \tilde{l}}$ accounts for the propagation distance between the impurities.

To fully understand the physical origin of the decay widths and the positions of the transmission zeros, it is necessary to explicitly calculate the probability amplitude of the static bound states at the impurity locations. 
The ground state is symmetric with respect to the cavity center, characterized by an unnormalized wavefunction of the form $\phi_S(\tilde{x}) = e^{-\tilde{\kappa}_S|\tilde{x}|} + e^{-\tilde{\kappa}_S|\tilde{x}-\tilde{l}|}$. The normalization consant $N_S$ is obtained integrating the square modulus over all space:
\begin{align}
    \frac{1}{N_S^2} &= \int_{-\infty}^{+\infty} |\phi_S(\tilde{x})|^2 d\tilde{x} \nonumber \\
    &= \frac{2}{\tilde{\kappa}_S} \left[ 1 + e^{-\tilde{\kappa}_S \tilde{l}}(1 + \tilde{\kappa}_S \tilde{l}) \right].
\end{align}
The normalized wavefunction is $\psi_S(\tilde{x}) = N_S \phi_S(\tilde{x})$. Evaluated at the impurity sites, it yields $\psi_S(0) = \psi_S(\tilde{l}) = N_S (1 + e^{-\tilde{\kappa}_S \tilde{l}})$. We define the strictly positive amplitude overlap factor $\mathcal{A}_S$ as the product of the wavefunction at the two interaction sites:
\begin{equation}
    \mathcal{A}_S \equiv \psi_S(0)\psi_S(\tilde{l}) = \frac{\tilde{\kappa}_S \left(1 + e^{-\tilde{\kappa}_S \tilde{l}}\right)^2}{2 \left[ 1 + e^{-\tilde{\kappa}_S \tilde{l}}(1 + \tilde{\kappa}_S \tilde{l}) \right]}.
\end{equation}

Conversely, the first excited state is antisymmetric, possessing the functional form $\phi_A(\tilde{x}) = e^{-\tilde{\kappa}_A|\tilde{x}|} - e^{-\tilde{\kappa}_A|\tilde{x}-\tilde{l}|}$. Following the same spatial integration, its normalization constant $N_A$ is given by:
\begin{align}
    \frac{1}{N_A^2} &= \int_{-\infty}^{+\infty} |\phi_A(\tilde{x})|^2 d\tilde{x} \nonumber \\
    &= \frac{2}{\tilde{\kappa}_A} \left[ 1 - e^{-\tilde{\kappa}_A \tilde{l}}(1 + \tilde{\kappa}_A \tilde{l}) \right].
\end{align}
Due to its odd parity, the normalized wavefunction at the two impurity sites evaluates to opposite values: $\psi_A(0) = N_A (1 - e^{-\tilde{\kappa}_A \tilde{l}})$ and $\psi_A(\tilde{l}) = -\psi_A(0)$. As a direct consequence, the overlap factor for the antisymmetric state is strictly negative:
\begin{equation}
    \mathcal{A}_A \equiv \psi_A(0)\psi_A(\tilde{l})- \frac{\tilde{\kappa}_A \left(1 - e^{-\tilde{\kappa}_A \tilde{l}}\right)^2}{2 \left[ 1 - e^{-\tilde{\kappa}_A \tilde{l}}(1 + \tilde{\kappa}_A \tilde{l}) \right]}.
\end{equation}

It is simple to notice that $\mathcal{A}_{S}=|\psi_{S}(0)|^2$ and $\mathcal{A}_{A}=|\psi_{A}(0)|^2$
These exact spatial overlap factors govern the coupling between the bound states (in the channel $n=-1$) and the open continuum (channel $n=0$). For the symmetric state, using the trigonometric identity $|1+e^{i\phi}|^2 = 4\cos^2(\phi/2)$, the decay width evaluates to:
\begin{equation}
    \tilde{\Gamma}_S = \frac{4\tilde{\gamma}^2 \mathcal{A}_S}{\tilde{k}_0} \cos^2\left(\frac{\tilde{k}_0 \tilde{l}}{2}\right).
\end{equation}
This width vanishes, signaling the formation of a BIC, when the destructive interference condition $\tilde{k}_0 \tilde{l} = (2n+1)\pi$ is met. Similarly, for the antisymmetric state, by factoring out the absolute value $|\mathcal{A}_A| = |\psi_A(0)|^2$ and using $|1-e^{i\phi}|^2 = 4\sin^2(\phi/2)$, the decay width becomes:
\begin{equation}
    \tilde{\Gamma}_A = \frac{4\tilde{\gamma}^2 |\mathcal{A}_A|}{\tilde{k}_0} \sin^2\left(\frac{\tilde{k}_0 \tilde{l}}{2}\right).
\end{equation}
In this latter case, the BIC formation occurs at the complementary geometric condition $\tilde{k}_0 \tilde{l} = 2n\pi$.

For the determination of the zero of the transmission probability $T$, one can proceed as in the case $\tilde{\nu}=0$, with the difference that here we have preexisting bound states in the channel $n=-1$. The equation for the closed channel ($n=-1$) depends on the static Hamiltonian $H_{0} = -\frac{d^2}{d\tilde{x}^2} - \tilde{\nu}[\delta(\tilde{x}) + \delta(\tilde{x}-\tilde{l})]$, reading:
\begin{equation}
    (H_{0} - \tilde{E}) \psi_{-1}(\tilde{x}) = -\tilde{\gamma}[\delta(\tilde{x}) + \delta(\tilde{x}-\tilde{l})]\psi_0(\tilde{x}).
\end{equation}
$H_{0}$ possesses unperturbed eigenstates $\phi_\alpha(\tilde{x})$ (with $\alpha \in \{S, A\}$) and corresponding resonance energies $\tilde{E}_{r,\alpha}^0$. Our goal is to find the energies $\tilde{E} = \tilde{E}_{zt,\alpha}$ where the transmission is identically zero. If there is a transmission zero, by definition, the wave cannot exist past the obstacle. Thus, for $\tilde{x} > \tilde{l}$, we have $\psi_0(\tilde{x}) = 0$. For the wavefunction to remain continuous, it must vanish at the second impurity, namely $\psi_0(\tilde{l}) = 0$.

Between the two impurities ($0 < \tilde{x} < \tilde{l}$), the wave obeys the free-particle equation, since the delta potentials act only at the boundaries. The general solution that vanishes at $\tilde{l}$ is necessarily a standing wave:
\begin{equation}
    \psi_0(\tilde{x}) = C \sin(\tilde{k}_0(\tilde{x} - \tilde{l})).
\end{equation}

We then evaluate the boundary condition exactly at the site $\tilde{x} = \tilde{l}$. We calculate the left ($\tilde{l}^-$) and right ($\tilde{l}^+$) spatial derivatives as $\psi_0'(\tilde{l}^-) = C \tilde{k}_0 \cos(0) = C \tilde{k}_0$ and $\psi_0'(\tilde{l}^+) = 0$. The derivative jump is therefore $\Delta\psi_0'(\tilde{l}) = 0 - C \tilde{k}_0 = -C \tilde{k}_0$. Physically, by integrating the open-channel equation, namely Eq.~\ref{BC_Explicit_l}(b), around $\tilde{l}$, one obtains:
\begin{equation}
    \Delta\psi_0'(\tilde{l}) = -\tilde{\nu}\psi_0(\tilde{l}) - \tilde{\gamma}\psi_{-1}(\tilde{l}).
\end{equation}
Since $\psi_0(\tilde{l}) = 0$, the static potential term vanishes completely, leading to $-C \tilde{k}_0 = -\tilde{\gamma}\psi_{-1}(\tilde{l})$. Isolating the amplitude constant $C$, one obtains:
\begin{equation}
    C = \frac{\tilde{\gamma}}{\tilde{k}_0} \psi_{-1}(\tilde{l}).
\end{equation}
Now that the amplitude $C$ is known, we can write the open wave throughout the internal space and evaluate it when it strikes the first impurity at $\tilde{x} = 0$, which gives $\psi_0(0) = C \sin(-\tilde{k}_0\tilde{l}) = -C \sin(\tilde{k}_0\tilde{l})$. Substituting the newly found value of $C$, one establishes the exact driving term at $\tilde{x}=0$:
\begin{equation}
    \psi_0(0) = -\frac{\tilde{\gamma}}{\tilde{k}_0} \psi_{-1}(\tilde{l}) \sin(\tilde{k}_0\tilde{l}).
\end{equation}

Near the resonance, one can approximate the internal wavefunction $\psi_{-1}$ at the transmission zero $\tilde{E}_{zt,\alpha}$ using the static bound state $\phi_\alpha$. Due to the narrow width of the Fano resonance, $\psi_{-1}$ is simply proportional to $\phi_\alpha$ by an amplitude factor $a$, such that $\psi_{-1}(\tilde{x}) \approx a \phi_\alpha(\tilde{x})$. Substituting this into the closed-channel equation, and recalling that $H_{0} \phi_\alpha = \tilde{E}_{r,\alpha}^0 \phi_\alpha$, gives:
\begin{equation}
    a (\tilde{E}_{r,\alpha}^0 - \tilde{E}_{zt,\alpha}) \phi_\alpha(\tilde{x}) = -\tilde{\gamma}[\delta(\tilde{x}) + \delta(\tilde{x}-\tilde{l})]\psi_0(\tilde{x}).
\end{equation}
We project this relation by multiplying the entire equation by $\phi_\alpha(\tilde{x})$ and integrating over space. On the left side, the static wavefunction is normalized to 1. On the right side, the delta functions sample the values at $0$ and $\tilde{l}$:
\begin{equation}
    a (\tilde{E}_{r,\alpha}^0 - \tilde{E}_{zt,\alpha}) = -\tilde{\gamma} \left[ \phi_\alpha(0)\psi_0(0) + \phi_\alpha(\tilde{l})\psi_0(\tilde{l}) \right].
\end{equation}
Once again, we apply the condition $\psi_0(\tilde{l}) = 0$. The second term on the right side vanishes entirely, reducing the projection to $a (\tilde{E}_{r,\alpha}^0 - \tilde{E}_{zt,\alpha}) = -\tilde{\gamma} \phi_\alpha(0)\psi_0(0)$. We now assemble all the derived components. From the approximation $\psi_{-1}(\tilde{l}) \simeq a \phi_\alpha(\tilde{l})$, we extract the amplitude as $a = \psi_{-1}(\tilde{l}) / \phi_\alpha(\tilde{l})$. We then insert the previously calculated expression for the entrance wave $\psi_0(0)$. The equation becomes:
\begin{equation}
    \left(\frac{\psi_{-1}(\tilde{l})}{\phi_\alpha(\tilde{l})}\right) (\tilde{E}_{r,\alpha}^0 - \tilde{E}_{zt,\alpha}) = -\tilde{\gamma} \phi_\alpha(0) \left[ -\frac{\tilde{\gamma}}{\tilde{k}_0} \psi_{-1}(\tilde{l}) \sin(\tilde{k}_0\tilde{l}) \right],
\end{equation}
which can be simplified to:
\begin{equation}
    \tilde{E}_{zt,\alpha} = \tilde{E}_{r,\alpha}^0 - \frac{\tilde{\gamma}^2}{\tilde{k}_0} [\phi_\alpha(0)\phi_\alpha(\tilde{l})] \sin(\tilde{k}_0\tilde{l}).
\end{equation}
Eventually, for the symmetric and antisymmetric bound states, one obtains:
\begin{equation}
    \tilde{E}_{zt,S} = \tilde{E}_{r,S}^0 - \frac{\tilde{\gamma}^2 \mathcal{A}_S}{\tilde{k}_0} \sin(\tilde{k}_0\tilde{l})
\end{equation}
and
\begin{equation}
    \tilde{E}_{zt,A} = \tilde{E}_{r,A}^0 + \frac{\tilde{\gamma}^2 |\mathcal{A}_A|}{\tilde{k}_0} \sin(\tilde{k}_0\tilde{l}).
\end{equation}
\begin{figure*}[t]
    \centering
    
    % --- PRIMA FIGURA (Sinistra) ---
    \begin{minipage}[t]{\textwidth}
        \centering
        \includegraphics[width=\textwidth]{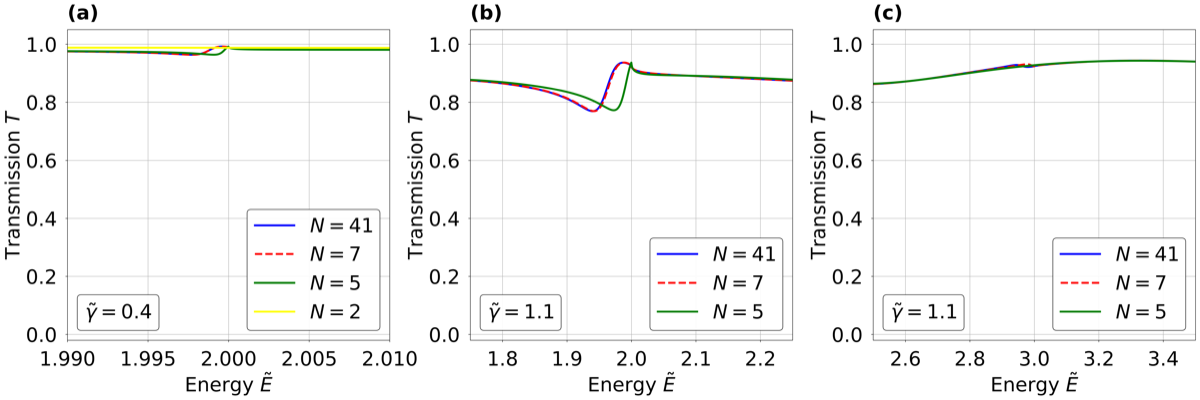}
    \end{minipage}
    \caption{Transmission probability $T$ as a function of the incident particle energy $\tilde{E}$, computed for different numbers of Floquet channels $N$ at a fixed distance $\tilde{l} = 4.5$. The panels display a zoom on the threshold anomalies: (a) the second threshold ($\tilde{E} \simeq 2.0$) in the weak-coupling regime ($\tilde{\gamma}=0.4$); (b) the second ($\tilde{E} \simeq 2.0$) and (c) the third ($\tilde{E} \simeq 3.0$) threshold beyond the weak-coupling regime ($\tilde{\gamma}=1.1$).}
    \label{fig:threshold}
\end{figure*}

\section{Threshold anomalies}
\label{threshold}

In Sec.~\ref{6-wcr}, we analyzed the system transport properties within the two-channel approximation. As discussed, this truncation is reliable only within a limited range of drive amplitudes and, crucially, captures the essential physics only for incident energies  $\tilde{E} < 2$.
The fundamental limitation of this approximation becomes evident as the energy increases. In the two-channel framework, the channel $n=-1$ is initially treated as a closed, evanescent mode ($\tilde{k}_{-1}$ is imaginary). As the energy reaches the threshold $\tilde{E} = 1$, the longitudinal wavenumber vanishes ($\tilde{k}_{-1} \to 0$), and for $\tilde{E} > 1$, the channel becomes open and propagating ($\tilde{k}_{-1}$ becomes real).
However, since the model restricts the spectrum to only two channels ($n=0$ and $n=-1$), it inherently fails to account for subsequent channel openings at higher integer multiples of the drive frequency 0 ($\tilde{E} = n$ with $n \in [2,+\infty]\cap\mathbb{N}$). Consequently, the two-channel approximation cannot reproduce the threshold anomalies associated with higher-order closed channels entering the continuum.
This transition corresponds to the opening of a new scattering pathway. In accordance with Wigner's threshold law~\cite{wigner1948behavior}, the abrupt availability of phase space for transmission into the $n=-1$ channel induces a non-analytic behavior in the scattering amplitudes of the fundamental channel ($n=0$). This manifests as a threshold anomaly—characteristically appearing as a cusp or a derivative discontinuity—in the transmission spectrum $T$ precisely at $\tilde{E}=1$. This feature is clearly resolved within the two-channel approximation, as evidenced by the distinct cusp at $\tilde{E}=1$ visible in Figs.~\ref{fig:nostatic}(b) and \ref{fig:static}(b). 
However, for higher energies ($\tilde{E} > 1$), the two-channel model fails to predict subsequent thresholds. Since the model is truncated, it cannot account for the opening of higher-order channels at multiples $n$. Generally, at energy $\tilde{E} = |n|$, the Floquet channel $-|n|$ opens as a propagating mode, inducing a threshold anomaly. To correctly capture these features and the associated redistribution of probability flux, the theoretical description must necessarily go beyond the two-channel approximation. 
To determine the effective number of Floquet channels required to capture the system dynamics, Fig.~\ref{fig:threshold} compares approximations with different truncation orders against a numerically converged benchmark. The reference solution is computed using a basis of 41 channels (spanning the indices $n \in [-20, 20]$), which ensures both numerical stability and precision.
In Fig.~\ref{fig:threshold}(a), we focus on the second threshold at $\tilde{E} = 2$ within the weak-coupling regime. The 5-channel approximation (restricted to $n \in [-2, 2]$) is sufficient to qualitatively resolve the anomaly, as it explicitly includes the channel $n=-2$ responsible for the threshold opening. However, achieving quantitative agreement with the converged solution necessitates expanding the basis to at least 7 channels ($n \in [-3, 3]$).
This effect persists beyond the weak-coupling regime. In Fig.~\ref{fig:threshold}(b), with a stronger drive ($\tilde{\gamma} = 1.1$), the anomaly at $\tilde{E} = 2$ remains visible with 5 channels, though 7 channels are needed for better accuracy.
Crucially, the necessity of including the specific opening channel is demonstrated in Fig. \ref{fig:threshold}(c), which focuses on the third threshold at $\tilde{E} = 3$. Here, the 5-channel approximation ($n \in [-2,+2]\cap \mathbb{Z}$) shows no sign of the anomaly because the relevant channel, $n=-3$, is excluded from the numerical simulation. The anomaly appears only when we consider at least  7 channels. Nevertheless, due to the stronger coupling, numerical stability at this higher energy requires further channels to be taken into account, achieving convergence only with 11 channels ($n \in [-5,+5]\cap \mathbb{Z}$).
Finally, it is worth noting that while threshold anomalies are a universal feature of multichannel scattering—present also in the single impurity case \cite{martinez2001transmission}—their manifestation in the double-impurity setup offers unique opportunities for spectral engineering.
From a fundamental perspective, the correct resolution of these cusps serves as a rigorous benchmark for our numerical approach, confirming the validity of the multi-channel expansion in capturing non-analytic behaviors.
From an applicative standpoint, the interplay between these threshold singularities and the cavity-induced resonances provides a distinct advantage over the single-impurity case.
A recent study on the single periodically driven impurity \cite{zhang2025enhanced} suggests that the diverging derivative of the transmission probability ($\partial T/\partial E$) at the threshold can be exploited to generate enahnced thermoelectric effects. The double-impurity geometry introduces the inter-impurity distance  as an additional control parameter, potentially amplifying the thermoelectric response.

\bibliographystyle{apsrev4-2}
\bibliography{ref}

\end{document}